\pdfoutput=1
\documentclass[aps,pr,twocolumn,superscriptaddress,preprintnumbers,showpacs,amsmath,amssymb,floatfix]{revtex4}

\usepackage{graphicx} 
\usepackage{dcolumn}  

\usepackage{epsfig}
\usepackage{epsbox}
\usepackage[T1]{fontenc}
\usepackage{wrapfloat}
\usepackage{subfigure}
\usepackage{here}
\usepackage{array}
\usepackage{amsmath, amssymb, hhline, multirow, hangcaption, subfigure}
\usepackage{graphics}
\usepackage{graphicx}
\usepackage{lscape}
\usepackage{fancyhdr}
\usepackage{longtable}
\usepackage{supertabular}
\usepackage{refmerge}
\usepackage{dcolumn}
\usepackage{color}

\def \vec #1{\mbox{{\boldmath $#1$}}}

\def \lr #1{\left( #1 \right)}

\def \B {{\cal B}}

\def \GeV {{\rm GeV}}
\def \MeV {{\rm MeV}}

\def \eV {{\rm eV}}
\def \simlt {\stackrel{<}{\sim}}

\def \vec #1{\mbox{{\boldmath $#1$}}}
\def \ks {K^0_S}

\begin{document}
\preprint{\vbox{ \hbox{   }
                \hbox{Belle Preprint 2013-18}
                 \hbox{KEK Preprint 2013-28}
                \hbox{July 2013, November 2013 Revised}
}}
\title{ 
High-statistics study of $\ks$ pair production in 
two-photon collisions\\
}

\begin{abstract}
We report a high-statistics measurement of the differential cross section
of the process $\gamma \gamma \to \ks \ks$ in the range 
$1.05~\GeV  \leq W \leq 4.00$~GeV, where $W$ is the center-of-mass
energy of the colliding photons,
using 972~fb$^{-1}$ of data
collected with the Belle detector at the KEKB asymmetric-energy 
$e^+ e^-$ collider operated at and near the $\Upsilon$-resonance 
region.
The differential cross section is fitted by parameterized
S-, D$_0$-, D$_2$-, G$_0$- and G$_2$-wave amplitudes.
In the D$_2$ wave, the $f_2(1270)$, $a_2(1320)$ and $f_2'(1525)$
are dominant and a resonance, the $f_2(2200)$, is also present.
The $f_0(1710)$ and possibly the $f_0(2500)$ are seen in the S wave.
The mass, total width and 
product of the two-photon partial decay width and decay branching
fraction to the $K \bar{K}$ state
$\Gamma_{\gamma \gamma}\B(K \bar{K})$ are 
extracted for the $f_2'(1525)$, $f_0(1710)$, $f_2(2200)$ and $f_0(2500)$.
The destructive interference between the $f_2(1270)$ 
and $a_2(1320)$ is confirmed by measuring their relative phase.
The parameters of the charmonium states $\chi_{c0}$ and $\chi_{c2}$ 
are updated.
Possible contributions from the $\chi_{c0}(2P)$ and  $\chi_{c2}(2P)$
states are discussed.
A new upper limit for the branching fraction of the
$P$- and $CP$-violating decay channel $\eta_c \to \ks \ks$ is 
reported.
The detailed behavior of the cross section 
is updated and compared with QCD-based calculations.
\end{abstract}


\noaffiliation
\affiliation{University of the Basque Country UPV/EHU, 48080 Bilbao}
\affiliation{Beihang University, Beijing 100191} 
\affiliation{University of Bonn, 53115 Bonn}
\affiliation{Budker Institute of Nuclear Physics SB RAS and Novosibirsk State University, Novosibirsk 630090}
\affiliation{Faculty of Mathematics and Physics, Charles University, 121 16 Prague}
\affiliation{Chiba University, Chiba 263-8522}
\affiliation{Deutsches Elektronen--Synchrotron, 22607 Hamburg}
\affiliation{Justus-Liebig-Universit\"at Gie\ss{}en, 35392 Gie\ss{}en}
\affiliation{Gifu University, Gifu 501-1193}
\affiliation{II. Physikalisches Institut, Georg-August-Universit\"at G\"ottingen, 37073 G\"ottingen}
\affiliation{Gyeongsang National University, Chinju 660-701}
\affiliation{Hanyang University, Seoul 133-791}
\affiliation{University of Hawaii, Honolulu, Hawaii 96822}
\affiliation{High Energy Accelerator Research Organization (KEK), Tsukuba 305-0801}
\affiliation{Hiroshima Institute of Technology, Hiroshima 731-5193}
\affiliation{Ikerbasque, 48011 Bilbao}
\affiliation{University of Illinois at Urbana-Champaign, Urbana, Illinois 61801}
\affiliation{Indian Institute of Technology Guwahati, Assam 781039}
\affiliation{Institute of High Energy Physics, Chinese Academy of Sciences, Beijing 100049}
\affiliation{Institute of High Energy Physics, Vienna 1050}
\affiliation{Institute for High Energy Physics, Protvino 142281}
\affiliation{INFN - Sezione di Torino, 10125 Torino}
\affiliation{Institute for Theoretical and Experimental Physics, Moscow 117218}
\affiliation{J. Stefan Institute, 1000 Ljubljana}
\affiliation{Kanagawa University, Yokohama 221-8686}
\affiliation{Institut f\"ur Experimentelle Kernphysik, Karlsruher Institut f\"ur Technologie, 76131 Karlsruhe}
\affiliation{Korea Institute of Science and Technology Information, Daejeon 305-806}
\affiliation{Korea University, Seoul 136-713}
\affiliation{Kyungpook National University, Daegu 702-701}
\affiliation{\'Ecole Polytechnique F\'ed\'erale de Lausanne (EPFL), Lausanne 1015}
\affiliation{Faculty of Mathematics and Physics, University of Ljubljana, 1000 Ljubljana}
\affiliation{Luther College, Decorah, Iowa 52101}
\affiliation{University of Maribor, 2000 Maribor}
\affiliation{Max-Planck-Institut f\"ur Physik, 80805 M\"unchen}
\affiliation{School of Physics, University of Melbourne, Victoria 3010}
\affiliation{Moscow Physical Engineering Institute, Moscow 115409}
\affiliation{Moscow Institute of Physics and Technology, Moscow Region 141700}
\affiliation{Graduate School of Science, Nagoya University, Nagoya 464-8602}
\affiliation{Kobayashi-Maskawa Institute, Nagoya University, Nagoya 464-8602}
\affiliation{Nara Women's University, Nara 630-8506}
\affiliation{National Central University, Chung-li 32054}
\affiliation{National United University, Miao Li 36003}
\affiliation{Department of Physics, National Taiwan University, Taipei 10617}
\affiliation{H. Niewodniczanski Institute of Nuclear Physics, Krakow 31-342}
\affiliation{Nippon Dental University, Niigata 951-8580}
\affiliation{Niigata University, Niigata 950-2181}
\affiliation{Osaka City University, Osaka 558-8585}
\affiliation{Pacific Northwest National Laboratory, Richland, Washington 99352}
\affiliation{Panjab University, Chandigarh 160014}
\affiliation{University of Pittsburgh, Pittsburgh, Pennsylvania 15260}
\affiliation{Research Center for Electron Photon Science, Tohoku University, Sendai 980-8578}
\affiliation{RIKEN BNL Research Center, Upton, New York 11973}
\affiliation{University of Science and Technology of China, Hefei 230026}
\affiliation{Seoul National University, Seoul 151-742}
\affiliation{Soongsil University, Seoul 156-743}
\affiliation{Sungkyunkwan University, Suwon 440-746}
\affiliation{School of Physics, University of Sydney, NSW 2006}
\affiliation{Tata Institute of Fundamental Research, Mumbai 400005}
\affiliation{Excellence Cluster Universe, Technische Universit\"at M\"unchen, 85748 Garching}
\affiliation{Toho University, Funabashi 274-8510}
\affiliation{Tohoku Gakuin University, Tagajo 985-8537}
\affiliation{Tohoku University, Sendai 980-8578}
\affiliation{Department of Physics, University of Tokyo, Tokyo 113-0033}
\affiliation{Tokyo Institute of Technology, Tokyo 152-8550}
\affiliation{Tokyo Metropolitan University, Tokyo 192-0397}
\affiliation{Tokyo University of Agriculture and Technology, Tokyo 184-8588}
\affiliation{University of Torino, 10124 Torino}
\affiliation{CNP, Virginia Polytechnic Institute and State University, Blacksburg, Virginia 24061}
\affiliation{Wayne State University, Detroit, Michigan 48202}
\affiliation{Yamagata University, Yamagata 990-8560}
\affiliation{Yonsei University, Seoul 120-749}
  \author{S.~Uehara}\affiliation{High Energy Accelerator Research Organization (KEK), Tsukuba 305-0801} 
  \author{Y.~Watanabe}\affiliation{Kanagawa University, Yokohama 221-8686} 
  \author{H.~Nakazawa}\affiliation{National Central University, Chung-li 32054} 
  \author{I.~Adachi}\affiliation{High Energy Accelerator Research Organization (KEK), Tsukuba 305-0801} 
  \author{H.~Aihara}\affiliation{Department of Physics, University of Tokyo, Tokyo 113-0033} 
  \author{D.~M.~Asner}\affiliation{Pacific Northwest National Laboratory, Richland, Washington 99352} 
  \author{V.~Aulchenko}\affiliation{Budker Institute of Nuclear Physics SB RAS and Novosibirsk State University, Novosibirsk 630090} 
  \author{T.~Aushev}\affiliation{Institute for Theoretical and Experimental Physics, Moscow 117218} 
  \author{A.~M.~Bakich}\affiliation{School of Physics, University of Sydney, NSW 2006} 
  \author{A.~Bala}\affiliation{Panjab University, Chandigarh 160014} 
% ...
\collaboration{The Belle Collaboration}

\normalsize



\maketitle

\tighten


\section{Introduction}
\label{sec:intro}
We present a high-statistics study of the
cross section for the process 
$\gamma \gamma \to \ks \ks$,
through the measurement of 
$e^+ e^- \to (e^+ e^-) \ks \ks$ where neither
a scattered electron nor positron is detected (zero-tag mode),
in the $W$ region
from close to its threshold to $4.0~\GeV$ 
and in the angular range $|\cos \theta^*| \le 0.8$,
where $W$ is the total energy of the parent photons
and $\theta^*$ is the scattering angle of the $\ks$
in their center-of-mass (c.m.) reference frame.
Measurements of exclusive hadronic final states in two-photon
collisions provide valuable information concerning the physics of 
light- and heavy-quark resonances, perturbative and non-perturbative 
QCD and hadron-production mechanisms.
The Belle collaboration has measured the production cross sections 
for charged-pion pairs~\cite{mori1,mori2,nkzw},
charged and neutral-kaon pairs~\cite{nkzw,kabe,chen},
and proton-antiproton pairs~\cite{kuo}.
Belle has also analyzed $D$-meson-pair production and observed a new
charmonium state identified as the $\chi_{c2}(2P)$~\cite{uehara}.
In addition, Belle has measured the production cross section
for  the $\pi^0 \pi^0$, $\eta \pi^0$ and $\eta \eta$
final states~\cite{pi0pi0,pi0pi02, etapi0, etaeta}.
The statistics of these measurements are two to three orders of
magnitude higher than in pre-$B$-factory measurements~\cite{past_exp}, 
opening a new era in two-photon physics.

The $f_J$ and $a_J$ mesons (with even spin $J$) both
contribute to the process of $\gamma \gamma \to K \bar{K}$.
The almost degenerate $f_J$ and $a_J$ that are predominantly
$u \bar{u}$ and $d \bar{d}$
are predicted to interfere destructively 
in $\gamma \gamma \to K^0 \bar{K}^0$ 
and constructively in $\gamma \gamma \to K^+ K^-$~\cite{lipkin}.
This is due to the Okubo-Zweig-Iizuka rule~\cite{ozirule}
where the $d \bar{d}$ ($u \bar{u}$) initial state dominates
in $K^0 \bar{K}^0$ ($K^+K^-$) production.
To the extent that the $s \bar{s}$ component is ignored,
the $d \bar{d}$ ($u \bar{u}$) state can be expressed as
$(f_J - a_J)/\sqrt{2}$ ($(f_J + a_J)/\sqrt{2}$) by the isospin 
consideration.

In the $\gamma \gamma \to \ks \ks$ reaction near the threshold,
Refs.~\cite{achasov, achasov2} predict a destructive interference between 
the $f_0(980)$ and $a_0(980)$, irrespective of their nature,
that suppresses the production cross section
to below 1~nb.
They consider the $\ks \ks$ production to be 
dominated by the rescattering process of $K^+ K^- \to K^0 \bar{K}^0$
near the threshold.
There have been no further data to shed light on this.

The destructive interference between the $f_2(1270)$ and $a_2(1320)$
was confirmed and the parameters of the $f_2'(1525)$ were measured
in many experiments~\cite{tasso1, pluto, cello, tasso2, L3}.
More recently, the process $\gamma \gamma \to \ks \ks$ has been 
investigated by L3~\cite{L3}, where
prominent peaks were observed around 1.3, 1.5 and 1.8~GeV.
Two peaks were interpreted to be due to 
$f_2(1270)$/$a_2(1320)$ interference and the $f_2'(1525)$,
respectively.
The third was attributed to the $f_J(1710) \; (J=2)$~\cite{L3}.
The limited statistics of these experiments ({\it e.g.}, 
0.588~fb$^{-1}$ for the L3 results~\cite{L3})
were insufficient to resolve and to study higher mass resonances.
Although these experiments operated at higher $e^+ e^-$ c.m. energies,
the cross section of each two-photon production process 
in a specific $W$ range rises only logarithmically with the $e^+ e^-$ 
c.m. energy.

The CLEO collaboration published the distribution of the
invariant mass for $\gamma\gamma \to \ks \ks$
in a search for $\eta(1440) \to \ks K^{\pm} \pi^{\mp}$
based on  13.8~fb$^{-1}$ of data~\cite{cleokskpi};
the $\ks\ks$ measurement was  used solely for the calibration of the $\ks$ 
efficiency, but no physics results were extracted.
Intriguingly, several resonant structures can be observed clearly
in their $\ks \ks$ mass spectrum.

In the previous Belle study of the $\gamma \gamma \to K^+ K^-$ reaction,
enhancements near 1.75~GeV, 2.0~GeV and 2.3~GeV were reported and
attributed to the $a_2(1700)$, $f_2(2010)$ and 
$f_2(2300)$, respectively~\cite{kabe, pdg2012}.

In this article, we present a high-statistics study of the
cross section for $\gamma \gamma \to 
\ks \ks$ from close to its threshold to $W = 4.0~\GeV$.
The data are based on an integrated luminosity of 
972~fb$^{-1}$.
This significantly extends our previous study~\cite{chen}, 
where the measurement of this process was reported for 
$2.4~\GeV \leq W \leq 4.0~\GeV$
with an integrated luminosity of 397.6~fb$^{-1}$.
In that study, we compared the high-energy behavior of 
the cross section with the QCD-based calculations or 
models~\cite{bl, handbag}.
Signals for the $\chi_{c0}$ and $\chi_{c2}$ charmonium states 
were observed.
Here, we extend the c.m. energy lower limit down to 1.05~GeV
and investigate the intermediate-mass resonances
with higher statistics data.

We report
the first measurement of the differential 
cross section for $\gamma\gamma \to \ks \ks$ below 2.4~GeV.
Previously, only the event distributions were obtained
for this process~\cite{tasso1, pluto, cello, L3} 
and the integrated cross section was presented with 
limited statistics~\cite{tasso2}.
In analyzing the differential cross section, we
measure the phase difference between the $a_2(1320)$ and $f_2(1270)$ 
as well as the parameters (mass, width and
product of the two-photon partial decay width and
decay branching fraction to the $K \bar{K}$,
$\Gamma_{\gamma \gamma}\B(K \bar{K})$ )
of the $f_2'(1525)$ including the interference.
Resonance-like enhancements are investigated
in the region $W>1.6$~GeV.
We also provide some new information on possible glueball candidates
such as the $f_0(1710)$ and 
$f_J(2220)$~\cite{scalar, scalar2, scalar3, scalar4}.

We 
then 
update the measurements of the
parameters of the $\chi_{c0}$ and $\chi_{c2}$ states.
Possible contributions from the radially
excited states $\chi_{cJ}(2P)$ are investigated.
The $\chi_{c2}(2P)$ was discovered and confirmed in 
two-photon collisions~\cite{pdg2012}, and the $X(3915)$ found
in the $\gamma \gamma \to \omega J/\psi$ process has been 
identified recently as the $\chi_{c0}(2P)$ state~\cite{pdg2013}.
In addition, we also report searches for
 the $P$- and $CP$-violating decay 
$\eta_c \to \ks \ks$
and set a new upper limit for its branching fraction.
Finally, we compare the cross section dependence on $W$ and 
$|\cos \theta^*|$ for $W > 2.6$~GeV with QCD predictions.

This article is organized as follows.
First we describe the details of the data selection
(Sec.~\ref{sec:selec}),
background subtraction (Sec.~\ref{sec:backg}), 
efficiency determination (Sec.~\ref{sec:effic}) and derivation
of the differential cross section (Sec.~\ref{sec:cross}).
We then present results on resonance analysis (Sec.~\ref{sec:reson}),
update the properties of several charmonia (Sec.~\ref{sec:charm}), and 
model the cross-section behavior for $W>2.6~\GeV$ 
(Sec.~\ref{sec:qcd}).
Finally, we present a summary and draw conclusions 
(Sec.~\ref{sec:summary}).

\section{The experimental apparatus and
selection of signal candidates}
\label{sec:selec}
In this section, we describe the Belle detector, data
sample, triggers, Monte Carlo simulation program 
and selection of signal candidates.

\subsection{Experimental apparatus}
\label{sub:apara}
Data were collected with the Belle detector operated at the KEKB
asymmetric-energy $e^+ e^-$ collider~\cite{kekb, kekb2}.
A comprehensive description of the Belle detector is
given elsewhere~\cite{belle, belle2}.
In this paper we briefly discuss only those
detector components that are essential for the described measurement.
Charged tracks are reconstructed from hit information in 
the silicon vertex detector and the central drift chamber (CDC).
The CDC is used as the main device to trigger readout for the
events with charged particles.
A barrel-like arrangement of time-of-flight (TOF) counters
and trigger scintillation counters (TSC) are used to supplement the CDC 
trigger on charged particles and to measure their time of flight.
Particle identification (ID) is achieved by including information from
an array of aerogel threshold Cherenkov counters.
Photon detection and energy measurements are performed with a CsI(Tl) 
electromagnetic calorimeter (ECL).
All of the above detectors are located inside a superconducting 
solenoid coil that provides a uniform 1.5~T magnetic field.
The detector solenoid is oriented along the $z$ axis, pointing
in the direction opposite that of the positron beam. 
The $r \varphi$ plane is transverse to this axis.

\subsection{Data sample}
\label{sub:data}
This analysis is based on a data sample corresponding to an
integrated $e^+ e^-$ luminosity of 972~fb$^{-1}$. 
Data were collected at the energy of the $\Upsilon(4S)$
resonance ($\sqrt{s} = 10.58$~GeV) and 60~MeV below it 
(784~fb$^{-1}$),
at energies between 10.6~GeV and 11.1~GeV (151~fb$^{-1}$,
mainly near the $\Upsilon(5S)$ resonance at 10.88~GeV),
and at lower energies between 9.4~GeV and 10.3~GeV (38~fb$^{-1}$,
primarily near the $\Upsilon(2S)$ resonance at 10.02~GeV).
We analyze these data with a common algorithm for 
selecting $\ks$ pair candidates from a zero-tag two-photon process
because the process is independent of incident $e^{\pm}$ energies.

\subsection{Triggers and filtering}
\label{sub:trigg}
The analysis is based on data recorded with triggers
that are sensitive to low-transverse-momentum ($p_t$) pions 
from $\ks \to \pi^+ \pi^-$ decays. 
Signal low-$p_t$ pions have large curvatures in the CDC and deposit 
only a small amount of energy in the ECL;
as a result, the trigger efficiency for the signal pions
decreases steeply toward 
the threshold energy for $\ks \ks$ production.
To reduce the uncertainty in the trigger efficiency,
we select data events recorded inclusively with
triggers A, B and C as described below.
These triggers make use of full- (short-) length charged tracks 
in  the CDC volume that have $p_t > 0.3~\GeV/c$ 
($0.2~\GeV/c < p_t < 0.3~\GeV/c$)~\cite{cdctrig}.
Trigger A requires two or more full-length tracks in the CDC wire layers 
with an opening angle of roughly $135^{\circ}$ or larger
in the $r\varphi$ plane~\cite{a135}, and
at least two TOF/TSC-module hits~\cite{kichimi}
and energy deposit with more than 0.11 GeV in at least one ECL
trigger segment.
Trigger B requires two CDC tracks, of which
at least one track is a full-length one,
with the opening angle requirement of trigger A, 
as well as a low-energy threshold condition (LowE~\cite{cheon}) 
of 0.5~GeV for the ECL total energy.
By design, there is a large redundancy between triggers A and B.
Trigger C is a three-track trigger 
with TOF/TSC-module and ECL segment/energy requirements.
This trigger is sensitive to short and full tracks, 
but must have hits in the TOF and ECL.
Details of the efficiencies and correlations of the three triggers are
discussed in Sec.~\ref{sub:trgeff}.

To be recorded, a candidate event must pass the 
level-4 software trigger (L4, see Ref.~\cite{l4}), 
in which a fast track-finding program reconstructs one or more 
tracks with transverse momentum $p_t > 0.3$~GeV/$c$, 
each satisfying the requirements on the point of
closest approach of the track to the $z$ axis of 
$dr < 1$~cm and $|dz| < 4$~cm, where $dr$ and $dz$ are 
the distances between this point and the interaction point (IP) 
in the $r \varphi$ plane and along the $z$ direction, respectively.

\subsection{Monte Carlo simulation}
\label{sub1:mc}
The signal Monte Carlo (MC) events for $e^+e^- \to e^+e^- \ks \ks$ are
generated using the MC code TREPS~\cite{treps} 
at 81 fixed $W$ points between 1.0
and 4.1~GeV and isotropically in $|\cos \theta^*|$.
Variables with (without) the asterisk represent observables
in the c.m. (laboratory) reference frame.
As we cannot measure the $\gamma\gamma$ collision axis
directly, in the measurement we approximate it 
by the $e^+e^-$-collision axis in the $e^+e^-$ c.m. frame.

In our simulation,
we use the experimental setup and background files
for runs at $\sqrt{s}=10.58$~GeV.
To study the dependence of the analysis on run conditions 
and beam energy, 
we have generated additional signal MC events at 14 $W$ points
for each of the different run periods at $\sqrt{s}=10.58$~GeV,
and at 12 and 6 $W$ points with $\sqrt{s}= 10.88$~GeV and 10.02~GeV,
respectively.
We embed background hit patterns from random trigger data
into MC events, thus taking into account the efficiency 
dependence on run conditions.

In the signal MC generator, the $Q^2_{\rm max}$ parameter,
a maximum virtuality of the incident space-like
photons is set to 1.0~GeV$^2$. 
The form factor $\sigma_{\gamma\gamma}(0,Q^2)=\sigma_{\gamma\gamma}(0,0)/
(1+Q^2/W^2)^2$ is assumed.
This assumption does not affect the results
of our analysis, because we select events 
with $\sqrt{Q^2} \approx |\sum \vec{p}_t^*| < 0.1~\GeV/c$,
thus requiring $Q^2/W^2$ to be much smaller than 1,
where $|\sum \vec{p}_t^*|$ is the transverse momentum of the $\gamma \gamma$
system in the $e^+e^-$ c.m. frame.
Although the maximum $Q^2$ value determined from the requirement of the 
non-detection range of the scattered electron/positron is
about 2 GeV$^2$, 
the $|\sum \vec{p}_t^*|$ condition applied to data limits $Q^2$ 
more tightly to be less than $\sim 0.01~\GeV^2$.
The  $Q^2_{\rm max} = 1~\GeV^2$ used in the MC is larger than this
experimental limit, and in this case the choice of $Q^2_{\rm max}$ 
in the MC  does not affect the final $\gamma\gamma$-based cross section 
results; {\textit i.e.,} the $Q^2_{\rm max}$ value is included in the  
definitions of the luminosity function calculated by TREPS, as well as 
in the efficiency.
As a result, their effects are compensated in the cross section
derivation (see Eq.~(\ref{eqn:dsdc})).

A sample of 400,000 events is generated at each $W$ point per 
experimental setup.
These events are then processed through 
the detector and trigger simulations and 
reconstructed using the same algorithms as for the real data.
The decay of the $\ks$ meson is managed in the GEANT-based
detector simulation~\cite{geant}.

\subsection{Selection criteria}
\label{sub:selec}
We select $\ks \ks$ two-photon event candidates 
in which each $\ks$ decays to $\pi^+ \pi^-$
and neither scattered lepton is
detected, 
\textit{i.e.}, in the zero-tag mode.
Such candidates 
are required to contain exactly four charged tracks 
with small total transverse momentum in which
two pairs of oppositely charged tracks 
form $\ks$ candidates with vertices significantly away from the IP.

In order to reduce the background contribution from $e^+ e^-$ 
annihilation processes, the sum of the absolute momenta of the
four tracks must be less than 6~GeV/$c$
and the total energy of all ECL clusters must
be less than 6~GeV.

To reduce the systematic uncertainty arising from 
reconstruction efficiency,
we use only good-quality tracks
that have $p_t > 0.1$~GeV/$c$, $dr < 5$~cm and
$|dz| < 5$~cm.
The vector sum of the transverse momenta of the four tracks
$|\sum \vec{p}_t|$ must be less than 0.2~GeV/$c$,
using the azimuthal direction of the tracks at their closest approach
to the nominal IP
on its curved trajectory in the magnetic field. 
Each of the four tracks has to be identified as a pion 
from the particle-ID detectors with a $K/\pi$ likelihood ratio:
${\cal L}(K)/({\cal L}(K)+{\cal L}(\pi))<0.8$.
The pion identification efficiency is larger than 99\% for 
$p < 0.6$~GeV/$c$  and  95\% for $p = 0.8$~GeV/$c$.
To further reduce the annihilation contribution,
the invariant mass of the four tracks with the pion mass assignment 
is required to be less than 5~GeV/$c^2$.
To eliminate backgrounds that include $\pi^0$ mesons,
we require that there be no $\pi^0$ candidates with 
$p_t > 0.1$~GeV/$c$ and $\chi^2 < 4$ in the mass-constrained fit 
of the available two-photon combinations.

\begin{figure}
\centering
\includegraphics[width=5.5cm]{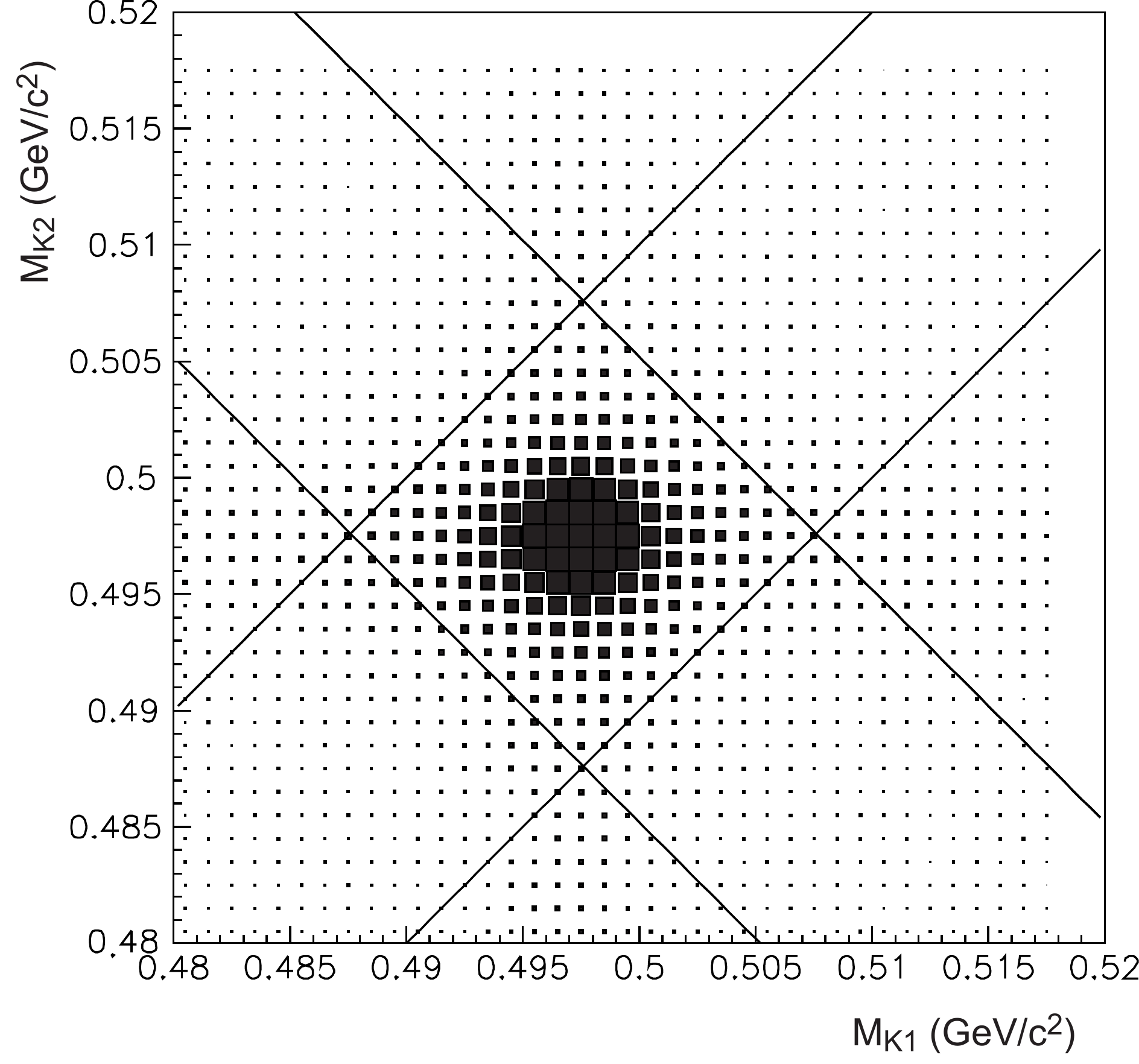}
\centering
\caption{Reconstructed masses of the two
$\ks$ candidates in data. 
The labels $K1$ and $K2$ are randomly assigned in each event.
The diamond region near the center indicates the 
signal region.
}
\label{fig:mass2d}
\end{figure}

Each pair of tracks forming a $\ks$ candidate must have a 
difference in $z$ coordinates at their point of closest 
approach in the $r \varphi$ plane, $|\Delta z|$, satisfying 
$|\Delta z| < (p_{K} + 1.6)$~cm, where $p_K$ is 
the $\ks$ momentum in GeV/$c$.
The momentum dependence here incorporates the effect of 
resolution in the vertex determination.
The reconstructed invariant mass of the two pions,
$M_{\pi \pi}$, should satisfy
$|M_{\pi \pi} - m_K |< 20~\MeV/c^2$, where 
$m_K$ is the nominal $\ks$ mass.
We require a unique assignment of the four pions as the decay 
products from the two $\ks$ by rejecting events that have
ambiguous combinations.
We further require that exactly two $\ks$ candidates that are 
reconstructed from non-overlapping combinations of two charged 
tracks are found in the event.
Figure~\ref{fig:mass2d} shows a two-dimensional plot 
of the two measured $\ks$ masses
where $K1$ and $K2$ are randomly assigned in each event. 

To further reduce the background contribution and to select
well-reconstructed events, we require 
the difference of the reconstructed masses of the two $\ks$  
to satisfy $|M_{K1} - M_{K2}| < 10$~MeV/$c^2$.
We define the average of the reconstructed masses of the two $\ks$
as $\langle M_K \rangle \equiv (M_{K1}+M_{K2})/2 $,
which must satisfy $|\langle M_K \rangle - m_K| < 5~\MeV/c^2$.
These selection criteria are depicted in Fig.~\ref{fig:mass2d}
with diagonal lines. Then, the decay position and momentum vector 
of each $\ks$ are determined by a kinematical fit.

The radial displacement of each $\ks$ vertex
from the nominal IP, $r_{V}$,
must satisfy the condition
$r_{Vi} > {\rm max}(0, W - 2) \times 0.1$~cm, where $W$ is in GeV.
This requirement does not apply to events with $W< 2$~GeV.

Backgrounds from the non-$\ks \ks$ two-photon four-charged-pion 
production process (the ``four-pion'' process) are strongly 
suppressed 
if we require the two $\ks$ vertices to be 
spatially separated, using combinations of two-dimensional 
($d_{Vr}$) and three-dimensional ($d_V$) distances. 
The signed distance between the two vertices in the 
$r \varphi$ plane, $d_{Vr}$,
defined according to
\begin{equation}
d_{Vr} = |\vec{r}_{V2}-\vec{r}_{V1}|\frac{(\vec{r}_{V2}-\vec{r}_{V1}) \cdot (\vec{p}_{t2}-\vec{p}_{t1})}{|(\vec{r}_{V2}-\vec{r}_{V1}) \cdot (\vec{p}_{t2}-\vec{p}_{t1})|} ,
\label{eqn:dvr}
\end{equation}
must satisfy $d_{Vr} > + 0.05$~cm, where
$\vec{r}_{Vi}$ and $\vec{p}_{ti}$ are two-dimensional vectors
projected onto the $r \varphi$ plane of the decay vertex and transverse
momentum, respectively, for each $\ks$.
The event must satisfy either $d_V > 0.7$~cm or $d_{Vr} > +0.3$~cm,
where $d_V$ is a distance between the
two vertices in the three-dimensional space.

\begin{figure}
\centering
\includegraphics[width=7cm]{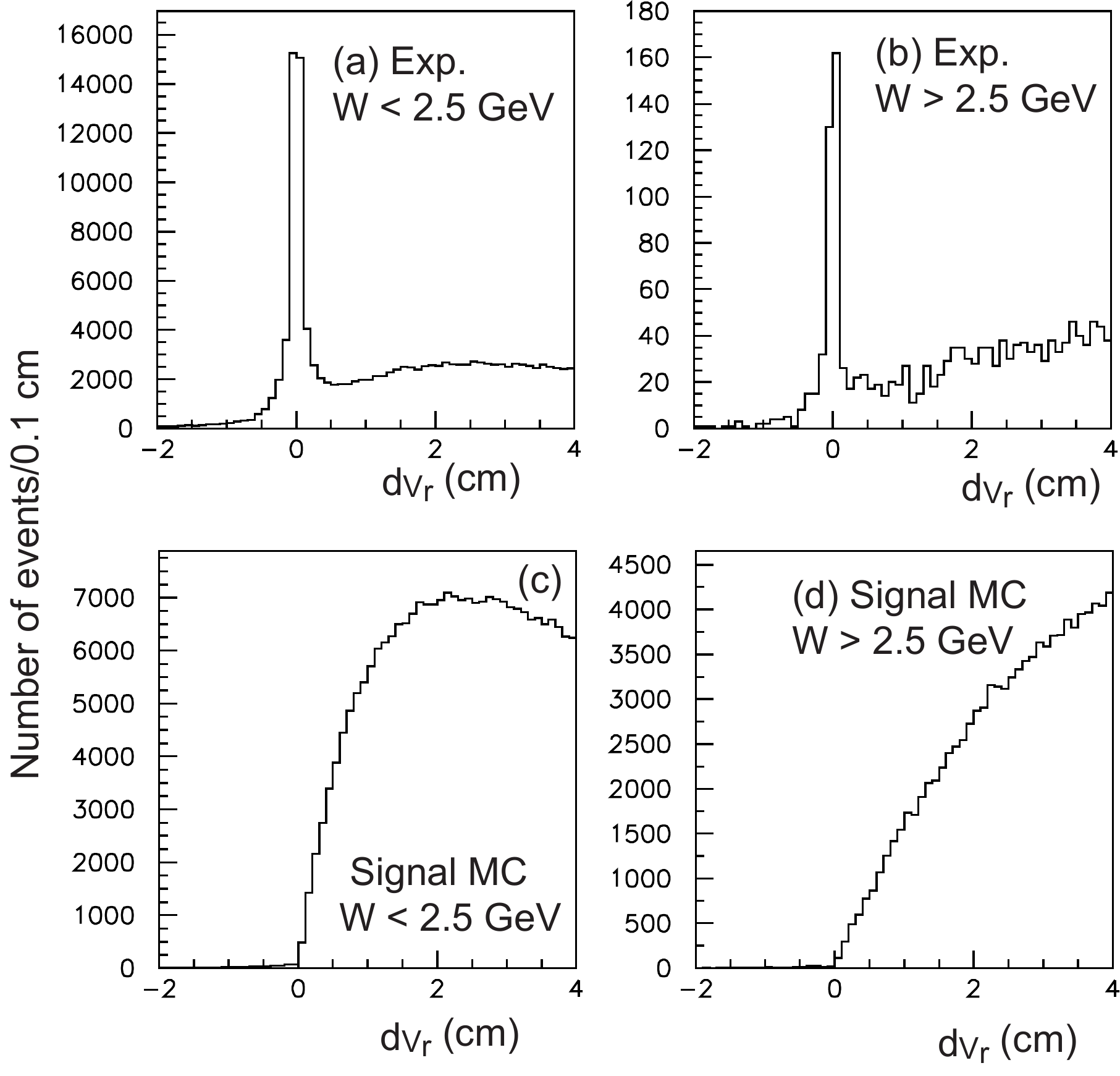}
\centering
\caption{Distribution of $d_{Vr}$ (the signed distance between 
the two vertices in the $r \varphi$ plane) for the data (a,b) 
and MC (c,d) samples in two $W$ regions.}
\label{fig:cutvr}
\end{figure}

\begin{figure}
\centering
\includegraphics[width=7cm]{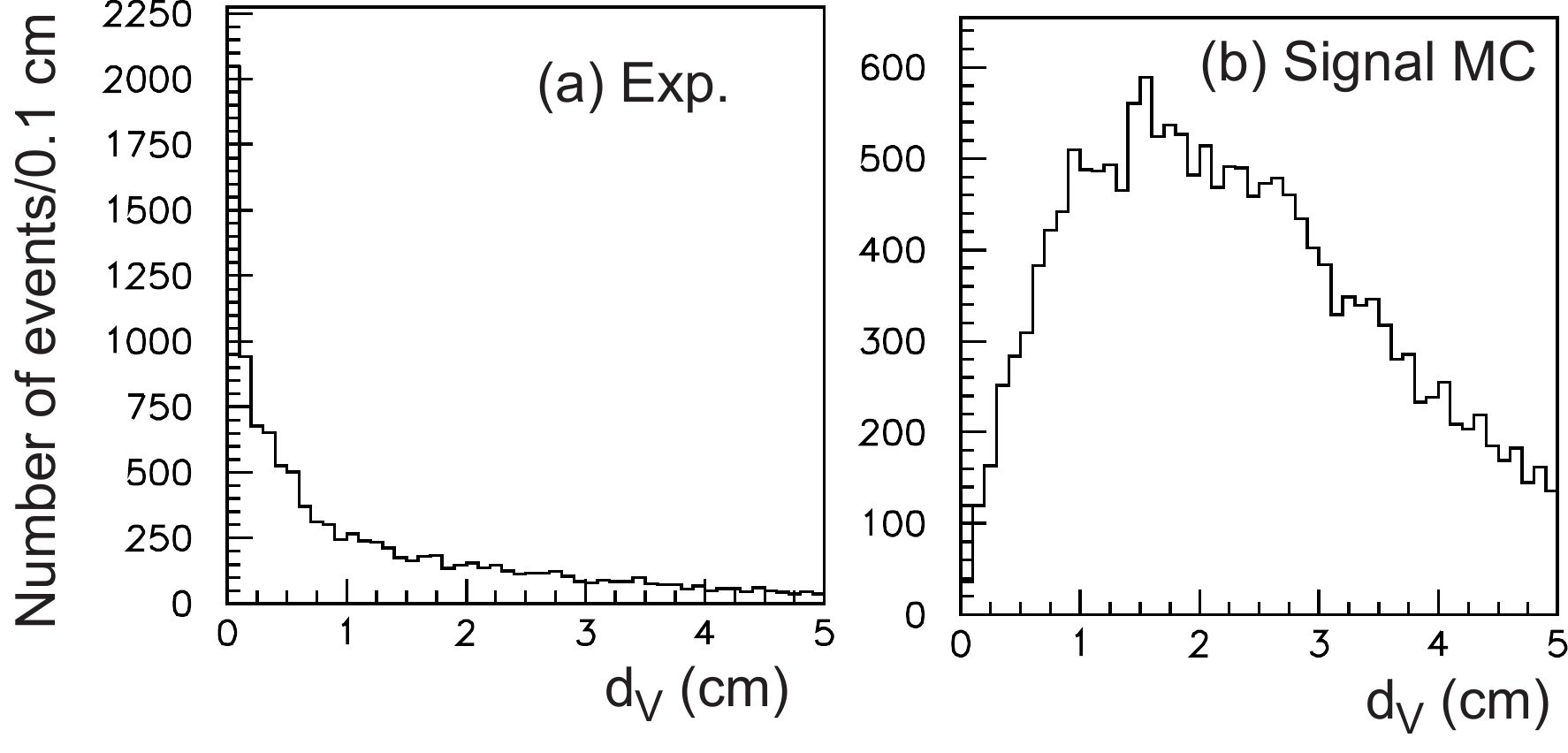}
\centering
\caption{Distribution of $d_V$ (the distance of the two vertices
in the three-dimensional space) for the data (a) and MC (b) samples.}
\label{fig:cutv}
\end{figure}

Figures~\ref{fig:cutvr} and \ref{fig:cutv} show the distributions
for these distances in the data (before the above selection criteria 
on them are applied) and signal MC samples. 
The peaks near zero in the data are due to the four-pion process
$\gamma \gamma \to \pi^+\pi^-\pi^+\pi^-$
whose cross section is larger than the signal one. 
This process is discussed in Secs.~\ref{sub:nonks} and \ref{sub:valid}. 
Note that events with $d_{Vr} < + 0.05$~cm or 
$d_V < 0.3$~cm are rejected by our selection criteria
and the relation $|d_{Vr}| \leq d_V$.
We further require the projection of the 
distance between the vertices in the $r \varphi$ plane 
onto the vector of the transverse momentum difference, 
$\delta_V$, defined by 
\begin{eqnarray}
\delta_V &=& \frac{|(\vec{r}_{V2}-\vec{r}_{V1})
\times (\vec{p}_{t2}-\vec{p}_{t1})|}{|\vec{p}_{t2}-\vec{p}_{t1}|}
\nonumber \\
 &=&  |d_{Vr}\sin \Delta \varphi| ,
\label{eqn:delv}
\end{eqnarray}
to satisfy $\delta_V < 0.7$~cm,
where $\Delta \varphi$ is the azimuthal-angle difference between the
vertex-position difference vector and the transverse-momentum 
difference vector.

To further eliminate events with significant photon activity,
we require the total energy deposit in the ECL to satisfy
$E_{\rm ECL} < E_{K1} + E_{K2} - 0.3$~GeV, where $E_{Ki}$ 
is the total energy of each $\ks$. 
This selection criterion is determined by a study
based on the signal MC in order not to lose any significant
efficiency even if a pion deposits energy
in the ECL after a nuclear interaction.

Finally, 
the $p_t$ balance of the $\ks$ pair in the $e^+e^-$ c.m. frame
is required to satisfy $|\sum \vec{p}_t^*| < 0.1~\GeV/c$.

\begin{figure}
\centering
\includegraphics[width=7cm]{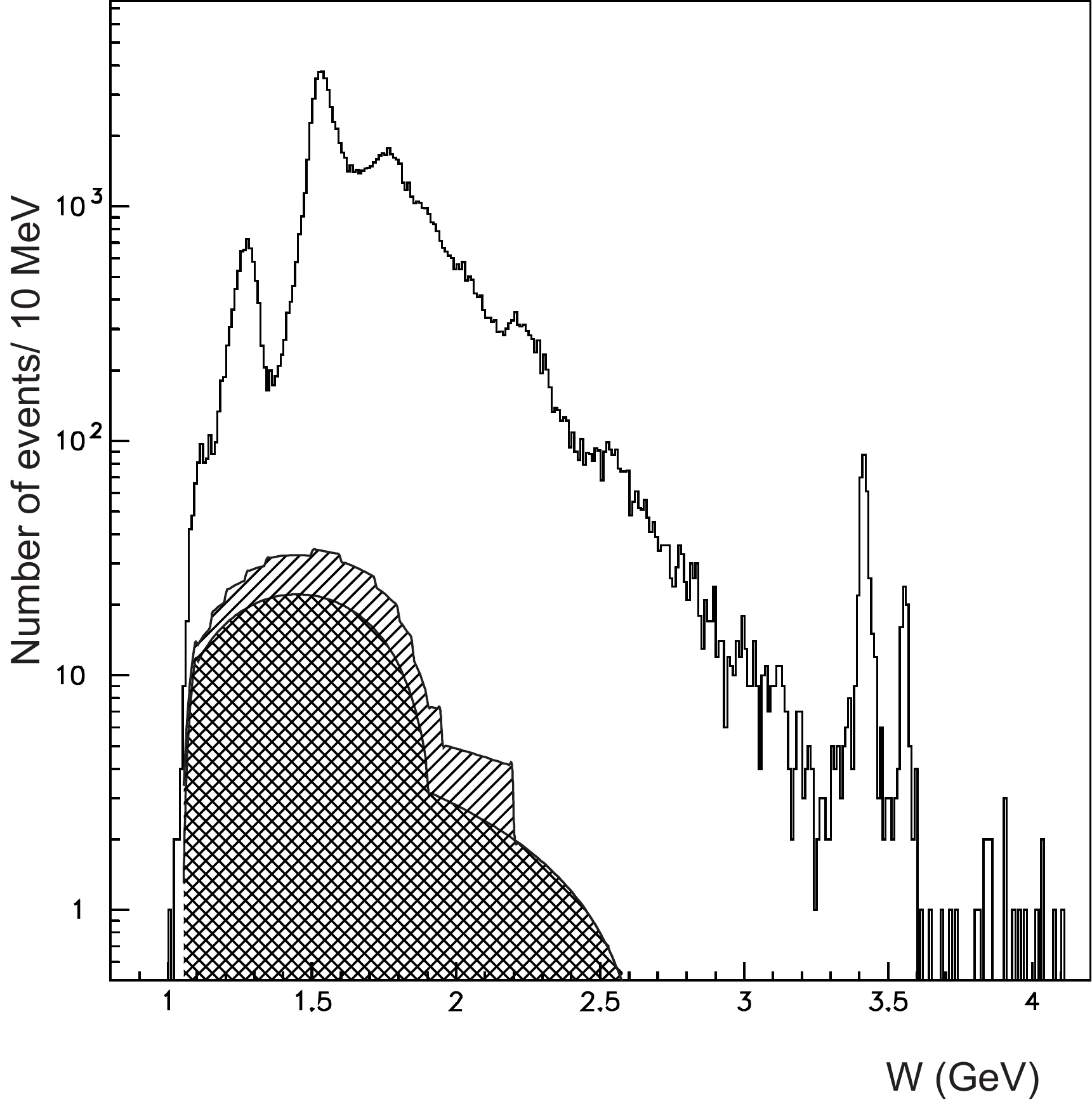}
\centering
\caption{Distribution of $W$ for candidate events (solid histogram),
as well as for the estimated 
non-exclusive background ($\ks \ks X$, cross hatched)
and non-$\ks \ks$ four-pion background (hatched, modeled as
a multi-step function).
The requirement $|\cos \theta^*|<0.8$ is applied.}
\label{fig:invm_sideband}
\end{figure}

We select candidates in the region $1.05~\GeV \le W \le 4.10~\GeV$
and $|\cos \theta^*| < 0.8$.
The $W$ distribution of the selected $\ks \ks$ 
candidate events is shown in Fig.~\ref{fig:invm_sideband}.

\section{Background subtraction}
\label{sec:backg}
We first consider non-exclusive background of
the type $\ks \ks X$, where $X$ is one or more particles.
Then we discuss four-track events: $\pi^+ \pi^- \pi^+ \pi^-$
and $\ks K^{\pm} \pi^{\mp}$.

\subsection{Non-exclusive background}
\label{sub:nonex}
The contamination  by the non-exclusive background 
process, $\ks \ks X$,
is estimated by fitting the $p_t$-balance ($|\sum \vec{p}_t^*|$)
distribution with a function in which both the signal and background
are considered in the region below 0.18~GeV/$c$. 
The region above 0.18~GeV/$c$ is not used in this estimate
because the $p_t$-balance requirement effectively suppresses
events in this region.
We approximate the signal distribution with a 
function that is determined empirically from a signal MC study:
\begin{equation}
f_s(x) = \frac{Ax}{x^\alpha + Bx + C} \; ,
\label{eqn:ptdist}
\end{equation}
where $x \equiv  |\sum \vec{p}_t^*|$,  
$\alpha = 1.56$ is determined from signal MC,
and the parameters $A$, $B$ and $C$ are floated in the fits
in each bin of $W$ and $|\cos \theta^*|$.

The background distribution is approximated with first- and  
second-order polynomials connected smoothly
at $x = 0.05$~GeV/$c$:
\begin{eqnarray}
f_b(x) &=& ax \ \ \ \  \hspace{14mm} (x<0.05~\GeV/c)\\
       &=& bx^2 + cx + d \ \ \ (x \geq 0.05~\GeV/c) .
\end{eqnarray}
We verify this approximation
in our analyses of the $\pi^0 \pi^0$ and $\eta \pi^0$
two-photon production where we observed a large amount
of non-exclusive background of the same 
type~\cite{pi0pi0,pi0pi02,etapi0}.
The fit is performed for data 
in two-dimensional ($W$, $|\cos \theta^*|$) bins
of width $\Delta W = 0.1~$GeV (0.2~GeV) for $W$ below (above)
2.0~GeV and $\Delta |\cos \theta^*| = 0.2$.

\begin{figure}
\centering
\includegraphics[width=8cm]{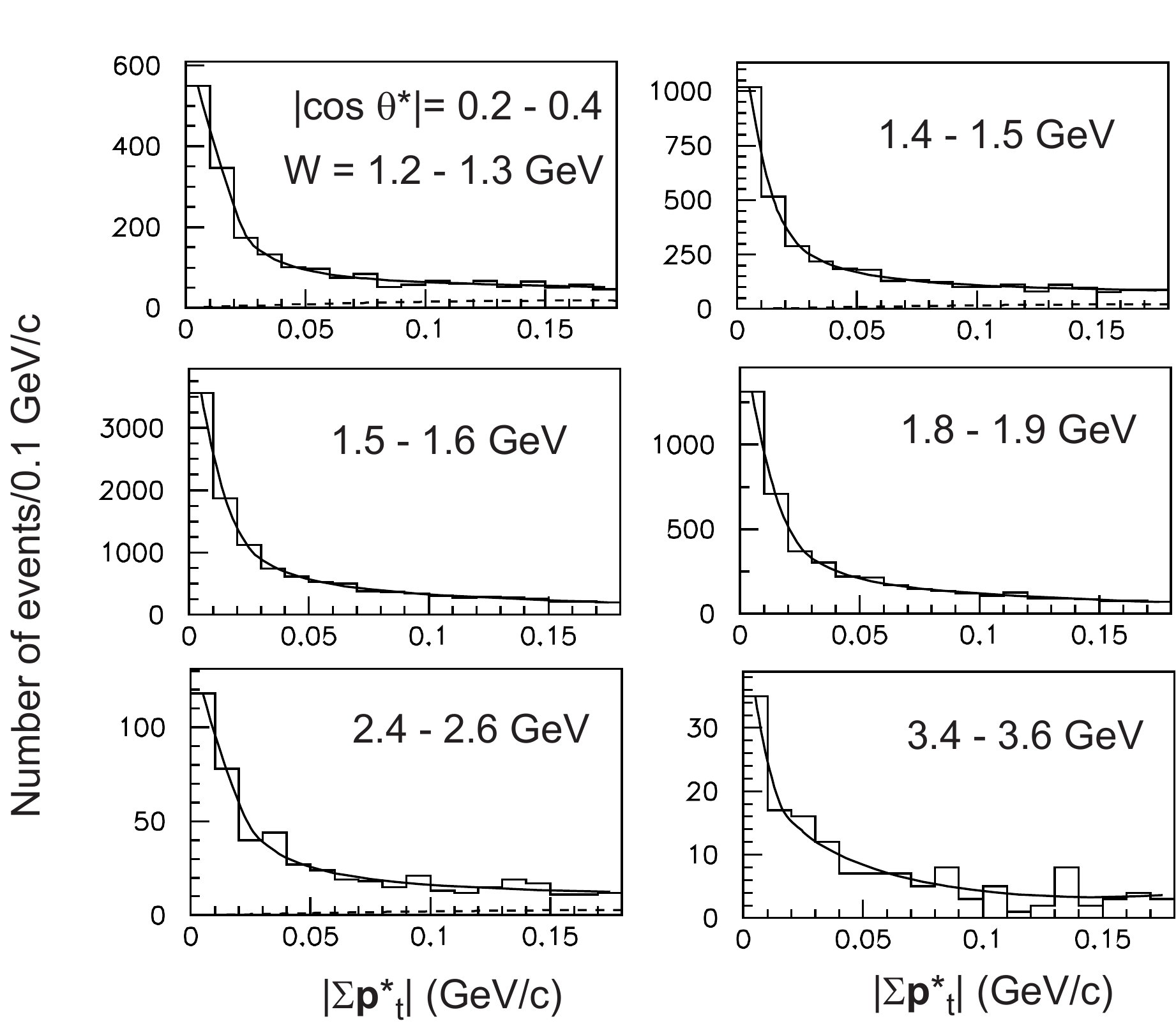}
\centering
\caption{The $|\sum \vec{p}_t^*|$ distributions 
for several regions of $W$ in data
for the angular region $0.2 < |\cos \theta^*| <0.4$.
The solid (dashed) curve shows the total (background) 
contribution obtained from the fit.
}
\label{fig:ptfit}
\end{figure}

The results of several such fits are shown for the 
$0.2 < |\cos \theta^*| <0.4$ region in Fig.~\ref{fig:ptfit}.
The background component is small in the signal region
where the data are well described by our parameterization.

To extract the signal yields from data, we subtract
the background contributions from our fits.
The ($W$, $|\cos \theta^*|$) dependence of the background
is approximated with a continuous function
that is quadratic in most of the $W$ range (connected to linear in a 
subset of this range) and linear in $|\cos \theta^*|$.
The background yields in each $W$ region,
integrated over the angular bins, are shown in 
Fig.~\ref{fig:invm_sideband}.

We estimate the systematic uncertainty associated with
this background and its subtraction
as half of the subtracted component. 
We add another 2\% error in quadrature to account for
the uncertainty in the background $p_t$ fit procedure.

\subsection{Non-$\ks\ks$ background -- four-pion process}
\label{sub:nonks}
\begin{figure}
\centering
\includegraphics[width=7cm]{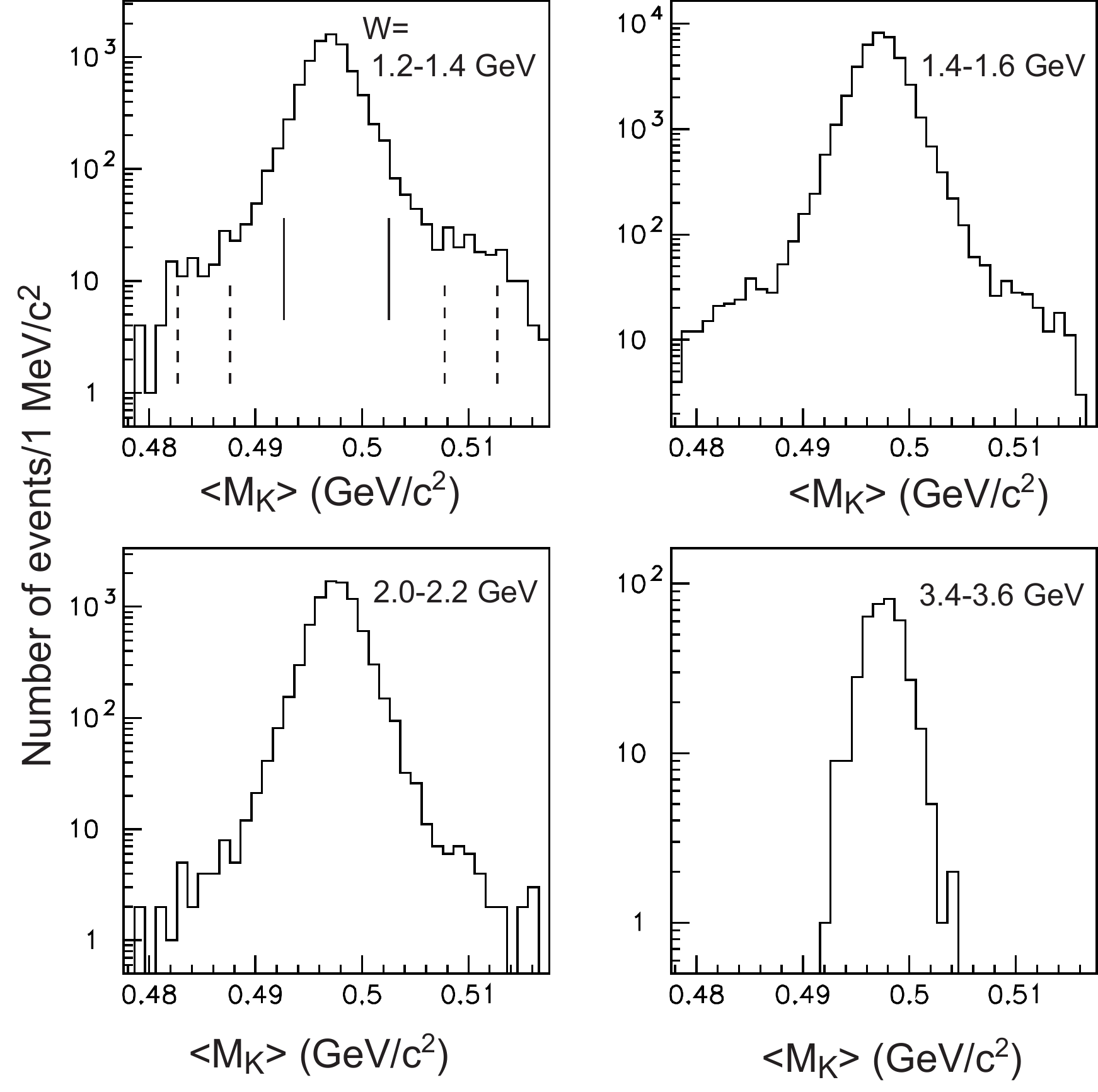}
\centering
\caption{ $\langle M_K \rangle$ data distributions
for four $W$ regions. 
The vertical solid lines and the pairs of dashed vertical 
lines indicate the signal region and two sideband regions used 
for background subtraction, respectively.
}
\label{fig:massmean}
\end{figure}

Background from the four-pion process is
estimated using the summed yield in the $\langle M_K \rangle$
sideband regions, 0.4826--0.4876~GeV/$c^2$ and
0.5076--0.5126~GeV/$c^2$; 
the sum of the widths 
is the same
as that for the signal region (0.4926--0.5026~GeV/$c^2$).
We show $\langle M_K \rangle$ distributions 
for data in some $W$ regions in Fig.~\ref{fig:massmean}.
The background contribution is appreciable in
the region $W<2.2$~GeV only;
as this background is always less than 1\% for $W>2.2$~GeV,
we incorporate the uncertainty in our estimate of this 
contribution in the systematic error but perform no
subtraction in this $W$ region.

We obtain the $W$ distribution of the 
$\langle M_K \rangle$-sideband yields
for the four separate $|\cos \theta^*|$ bins
with a bin width of 0.2.
To subtract the four-pion background,
we approximate the ($W$, $|\cos \theta^*|$) dependence of 
the background with a multi-step function for $W$ (as shown
in Fig.~\ref{fig:invm_sideband})
and a linear function for $|\cos \theta^*|$.

If there were an overlap in the two kinds of
backgrounds, \textit{i.e.}, if \textit{non-exclusive} 
four-pion events 
($\pi^+\pi^-\pi^+\pi^- X$) were to mimic the $\ks \ks X$ background,
these contributions would be doubly counted and over-subtracted.
We find no significantly large non-$\ks \ks (X)$ contribution in
the  $\langle M_K \rangle$ distribution for the
$p_t$-unbalanced events with 
$0.1~\GeV/c < |\sum \vec{p}^*_t| < 0.2~\GeV/c$, 
and therefore estimate the systematic uncertainty associated with
the background subtraction as a half of the subtracted component.
The possible effect of the overlap is included
in this systematic uncertainty.

\subsection{Non-$\ks\ks$ background -- $\ks K \pi$ process}
\label{sub:kskpi}
The $\ks K^\mp \pi^\pm$ two-photon production,
which has a cross section about ten times larger than that
of the signal, would contaminate the signal sample if the
charged kaon were misidentified as a pion.

According to 
our MC-based studies, the
probability that a generated $\ks K \pi$ event
is selected as a $\ks \ks$ signal candidate is smaller than
$\sim 10^{-4}$ 
for $W>2.0$~GeV.
This probability is so small
because of the requirement on the decay vertex distances 
$r_{V1}$ and $r_{V2}$ imposed to reject this background.

We use two data-based methods to estimate the remaining $\ks K \pi$ 
background:
from a study of the $r_V$ distributions near the IP
and using our previous measurement of the  $\ks K \pi$ production 
process~\cite{nkzwc}.

In the first method, we investigate the $r_V$ distribution after 
identifying one $\ks$ with a large $r_{Vj}$, $r_{Vj} > 1$~cm on the 
opposite side.
An excess of events near $r_V = 0$~cm is
observed in data for $W < 2.0$~GeV/$c$. 
This is due to the $\ks K \pi$ background process,
constituting between 0.1\% and 4\% of the sample
at larger $r_{V}$.
This component is observed primarily in the $W$ region 
below 1.5~GeV. 
The concentration of the background in the $W$ region may
be partially due to four-pion final processes, where
one pion track is misreconstructed, resulting in
a fake reconstructed vertex.
Since we do not separate the four-pion and $\ks K^{\mp} \pi^{\pm}$
backgrounds clearly in the low-$W$ region, we subtract
this background assuming 
the contribution to be $2\% \pm 2\%$ of the signal
in the $W$ region below 1.5~GeV. 
For $W>1.5~\GeV$, the excess in the $r_V$ distribution is small;
this is supported by a  study using the measurement of $\ks K \pi$
production.

In the second method, the observed yield from the 
process $\gamma \gamma \to \ks K \pi$ 
is an order of magnitude larger than that of the signal process
for $W>2.5$~GeV~\cite{nkzwc}, 
but this background is suppressed by a factor of $\sim 1000$ 
in the data sample after our selection criteria are applied.
Thus, it contributes less than 1\% to the signal sample.
We take this possible contamination into account
as a systematic uncertainty of 1\% for $W > 1.5$~GeV.

\section{Efficiency and efficiency corrections}
\label{sec:effic}
In this section, we describe efficiency estimates including 
the factors from the L4 filter,
triggers and $\ks \ks$ reconstruction.
Then we discuss corrections for beam energy dependence.

\subsection{The L4 efficiency}
\label{sub:l4eff}
\begin{figure}
\centering
\includegraphics[width=7cm]{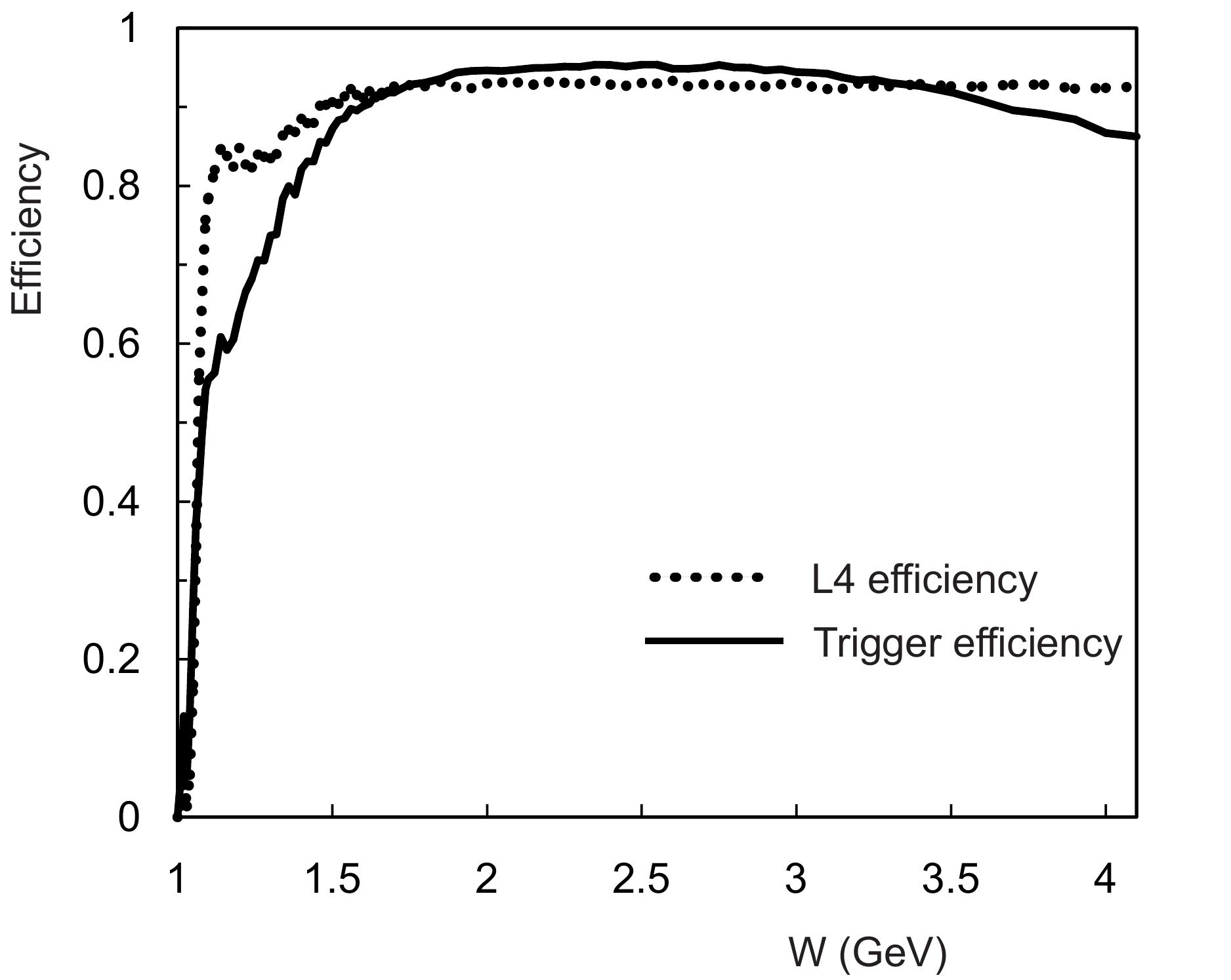}
\centering
\caption{ $W$ dependence of the L4 efficiency (dotted line)
and trigger efficiency (solid line)
estimated using the signal MC, where $\ks \ks$ were
generated isotropically
in the $\gamma \gamma$ c.m. frame at each $W$ point
for the $W$ region 1.05 -- 4.0~GeV.
The L4 efficiency (trigger efficiency) is defined for the
sample that passes through the trigger (L4) and
all the selection criteria.}
\label{fig:l4trgeff}
\end{figure}

Some loss of efficiency is introduced by the L4 software filter
that is designed to suppress beam-gas
and beam-wall events.
Figure~\ref{fig:l4trgeff} shows the dependence of the
L4 efficiency on $W$ for signal MC events 
that pass the trigger and all the
selection criteria for an assumed isotropic angular dependence.
The efficiency is significantly reduced for $W < 1.1$~GeV and 
is stable, in the range between 80\% and 94\%, for $W > 1.1$~GeV.
For very low $W$, the inefficiency 
is dominated by the low reconstruction efficiency for
tracks with small $p_t$; for high $W$, it is explained 
by tracks with large $dr$.

\subsection{Trigger efficiency}
\label{sub:trgeff}
\subsubsection{Tuning of the simulator for trigger B}
\label{sub:tuneloe}
We tune the energy threshold for the ECL trigger
(LowE), whose nominal value is 0.5~GeV, by comparing the 
efficiency curves of trigger B between the data and MC events 
in the four-pion process. 
With this tuning study in the trigger simulator (TSIM), 
the optimal value is determined to be 0.52~GeV.

In addition, 
we find a disagreement of about 20\% between data and MC for
the energy deposition in the ECL by a low-energy pion.
As it is impractical to make dedicated changes in the
trigger or detector simulation to describe the
detector response to low-energy charged pions
for this analysis,
we have effectively shifted the LowE
threshold by +110~MeV (to 0.63~GeV) to compensate for the
pion-energy deposition mismatch.

This shift could affect the efficiency of the selection criterion
based on $E_{\rm ECL}$.
We study this possible effect and conclude that
it is small, because of the loose criterion
on $E_{\rm ECL}$.
As our studies indicate that 
we could underestimate the efficiency by
$\sim 1$\% because of the ECL energy shift, 
we correct its value by this amount
and assign 1\% to the systematic uncertainty of this 
selection in the entire kinematic region.

\subsubsection{Estimation of the trigger efficiency}
\begin{figure*}
\centering
\includegraphics[width=11cm]{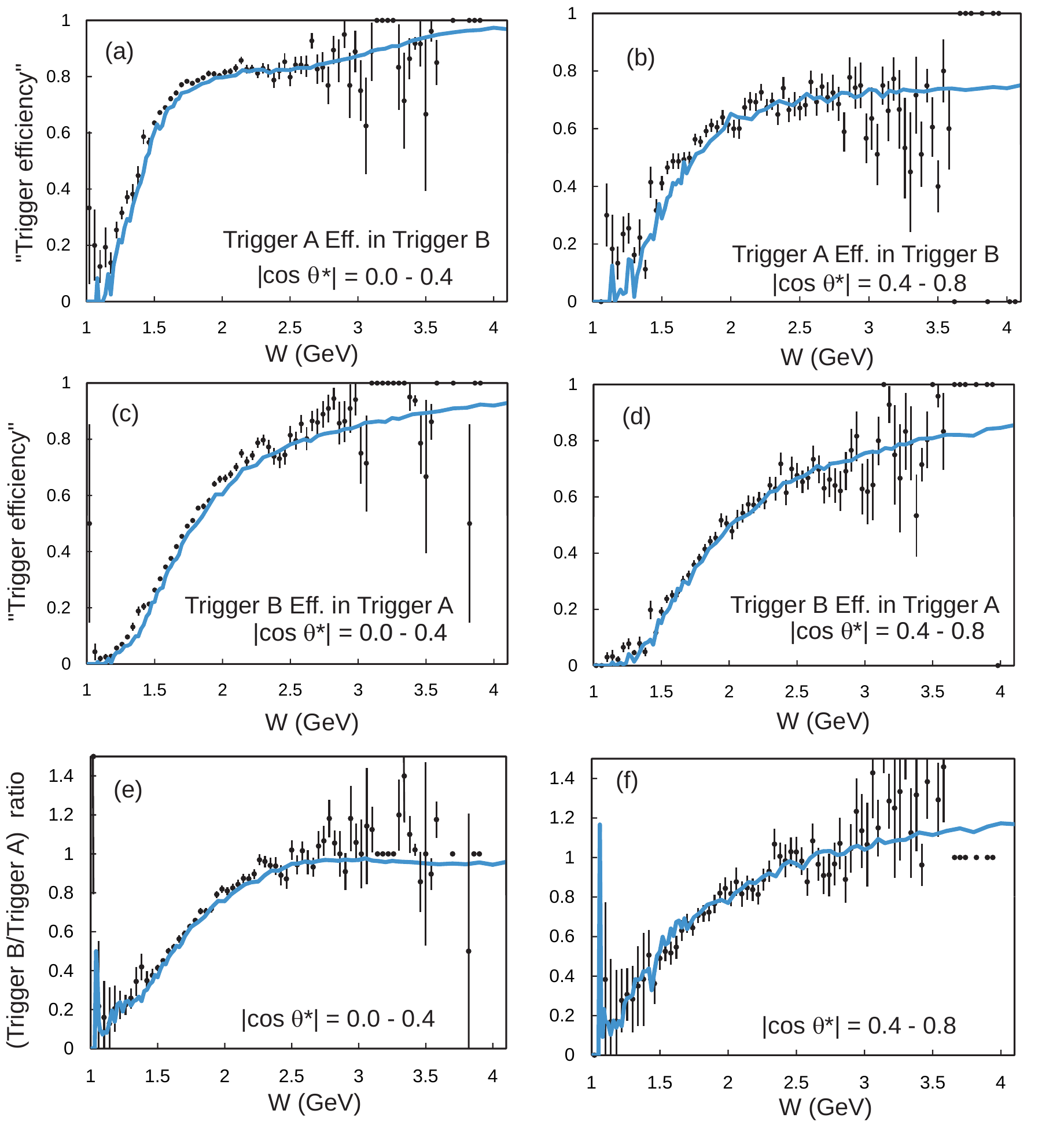}
\centering
\caption{ (a,b) The trigger-A efficiency 
and (c,d) the trigger-B efficiency (as defined in the text);
(e,f) the ratio of the number of selected events
from the two trigger samples, A and B.
Data (dots with error bars) and signal MC (curves) samples 
are subdivided into the two $|\cos \theta^*|$ angular bins 
as labeled on the plots.
}
\label{fig:trgrat}
\end{figure*}

Using TSIM, we estimate the trigger efficiency for
the combination of triggers A, B and C.
Its validation using data is non-trivial,
because we do not have mutually exclusive triggers
to precisely measure the trigger efficiency from 
the data alone. 
We find that the contribution of trigger C 
to the combined efficiency is very small (0.3\% -- 2.0\%,
depending on $W$ and $|\cos \theta^*|$),
so its contribution to the systematic error is negligible.
To estimate the systematic uncertainty 
of the combined trigger efficiency,
we study ``trigger-A efficiency'' $N({\rm A \cap B})/N({\rm B})$
and ``trigger-B efficiency'' $N({\rm A \cap B})/N({\rm A})$, 
where $N({\rm A \cap B})$ is the number of events 
recorded with both triggers, while $N({\rm B})$ 
($N({\rm A})$) is that recorded with trigger B (A).
These values represent the true trigger-A and -B efficiencies
if triggers A and B are uncorrelated.
Even though it is impossible to estimate the
trigger correlation from data, it is useful to compare data and MC.
We show the trigger-A and -B efficiencies in Fig.~\ref{fig:trgrat}(a-d)
for data and MC.
In Fig.~\ref{fig:trgrat}(e,f), the ratio $N({\rm B})/N({\rm A})$
is shown for data and MC.
The figures are shown separately for the two angular regions,
$|\cos \theta^*| = $ 0.0 -- 0.4 and 0.4 -- 0.8.

The difference in the angular distribution between the MC 
and data could cause an apparent deviation of the trigger
efficiencies and their ratios in the comparison: in MC, we implement 
a flat distribution while, in data, steep changes of the 
distribution are seen for the small angles (typically, in 
$0.5 < |\cos \theta^*| < 0.8$) in some energy regions. 
To reduce this artifact in the plot for the region 
$0.4 < |\cos \theta^*| < 0.8$ (Fig.~\ref{fig:trgrat}(b,d,f)),
we subdivide the region into 
two bins with the same width, 0.2, and take an average for
the two bins, for the trigger-A and -B efficiencies and 
the ratio. The trigger efficiencies estimated by the 
data and the MC simulation 
agree within 0.05 except for a low-statistics region.
The assumption of flat angular distributions in MC 
introduces no bias in the efficiency calculation
for cross section derivation because the efficiency is 
estimated on a bin-by-bin basis 
with a further narrow bin width, 0.05, in $|\cos \theta^*|$, 
whose resolution is much finer than 
the bin width, as described in Sec.~\ref{sub:resol}.

In Fig.~\ref{fig:l4trgeff}, we show the TSIM trigger efficiency
as a function of $W$
for isotropically simulated MC events 
that satisfy the L4 and all other 
selection criteria in our analysis.
The trigger efficiency rises steeply
from 3\% near $W=1.05$~GeV to 90\% near $W=1.6$~GeV.

The systematic uncertainty of the trigger efficiency
is estimated using the differences in
the trigger-A and -B efficiencies and ratios 
between data and MC, taking into account the
correlation between the triggers A and B as
estimated from MC. 
It is evaluated to be 5\%--7\%, with a weak $W$ dependence.

\subsubsection{Validation of the trigger efficiency}
\label{sub:valid}
\begin{figure}
\centering
\includegraphics[width=7cm]{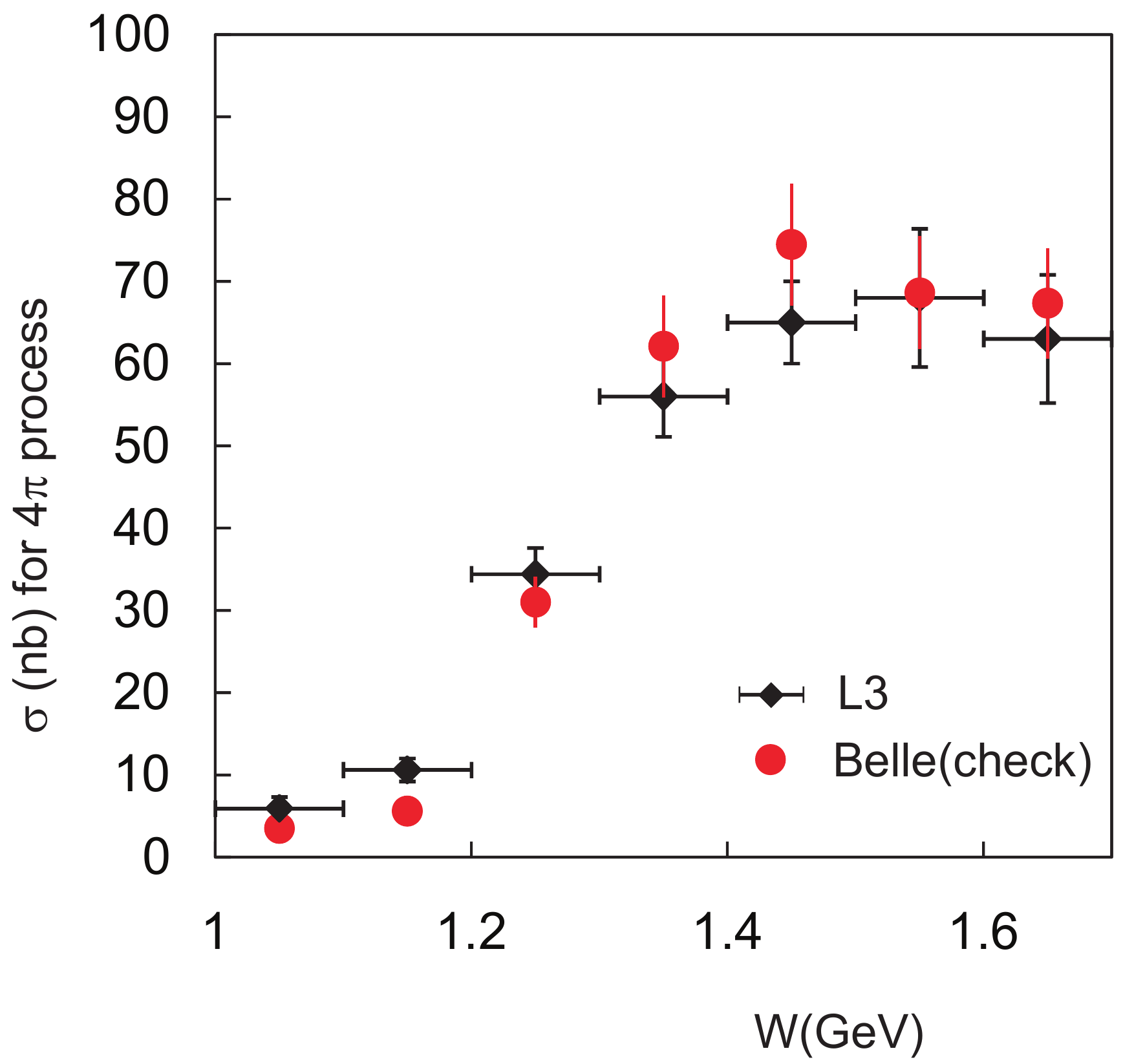}
\centering
\caption{The measured cross section 
for the $\gamma \gamma \to \pi^+\pi^-\pi^+\pi^-$ process
including $\rho^0\rho^0$ 
from Belle (closed circles) and L3 
(diamonds)~\cite{l34pi}. 
The error bars include both statistical and systematic
uncertainties, with a uniform 10\% estimate used for Belle.
These distributions are used solely for trigger-efficiency validation.
}
\label{fig:fourpi}
\end{figure}

We compare our data
with the results from the L3 experiment for the cross 
section of the $\gamma \gamma \to 4\pi$ process~\cite{l34pi}, 
where the $\pi^+\pi^-\pi^+\pi^-$ final-state 
includes $\rho^0\rho^0$ production but not
$\ks \ks$ production.
Ideally we would prefer to compare our results directly
with $\ks \ks$ data obtained in previous experiments;
however, no such high-statistics data are available.
Figure~\ref{fig:fourpi} shows a comparison between
Belle and L3 for the cross section of the four-pion process
(excluding $\ks \ks$) at seven $W$ points (the $W$ bin widths 
being different between Belle and L3). 
The Belle selection for the four-pion process
has a $p_t$ balance cut at 50~MeV/$c$ and non-exclusive
backgrounds are subtracted using the $p_t$ distribution.

The relative systematic error of the Belle result is
estimated to be 10\%, while
the statistical error is much smaller than that of L3.
The Belle result is consistent with the L3 results, but
no accurate comparison at a level better than 10\% is possible.
We assume the L3-determined fractions of the $\rho^0 \rho^0$ 
components with spin 0 and 2.
Note that the efficiency of the four-pion final state depends
on this assumption.

\subsection{Reconstruction efficiency for the $\ks$ pair}
\label{sub:kspair}
The systematic error associated with
the selection efficiency of the $\ks$ pairs is
estimated by varying the selection criteria in the signal MC.
When we do not find two $\ks$ candidates with our nominal criteria,
we loosen the |$\Delta z$| criterion to
$|\Delta z| < 10$~cm, remove
the requirements on $\ks$ vertices and
loosen the requirement on $\langle M_K \rangle$
to $|\langle M_K \rangle - m_K |< 10$~ MeV/$c^2$,
keeping all other criteria at their nominal values.
These changes increase both signal efficiency and 
backgrounds, and we evaluate them with the same methods. 
The increase of the efficiency is 3\%--10\% (10\%--20\%) 
for $W>1.15$~GeV ($W<1.15$~GeV).

After the background subtraction, we use the differences 
in the fractional increase of the efficiency
between the original and the loose cuts as its systematic uncertainty.
It is difficult to evaluate backgrounds below $W<1.3$~GeV because 
the contamination is larger than the efficiency change and
the two different types of non-$\ks \ks$ backgrounds are
not well separated. 
As the systematic uncertainty is not expected to strongly depend 
on $W$, we assign 
3\% for $W < 2.6$~GeV and 5\% for $W>2.6$~GeV as the uncertainty 
in the efficiency reconstruction for the $\ks$ pairs. 

\begin{figure}
\centering
\includegraphics[width=7cm]{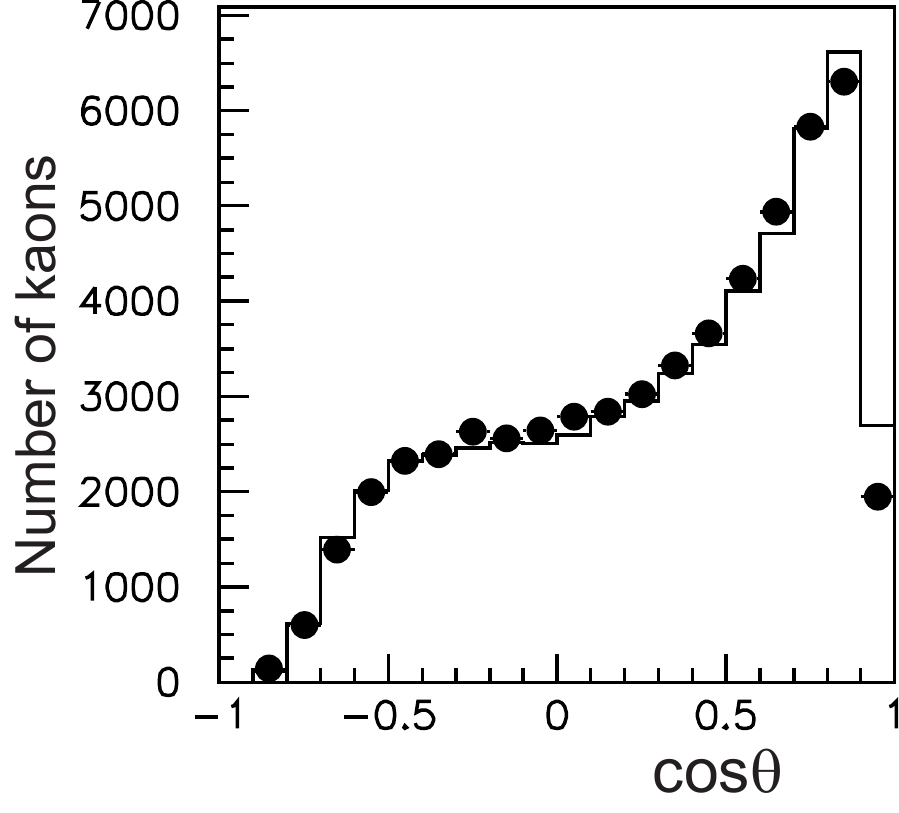}
\centering
\caption{
Distribution of $\cos \theta$ of $\ks$ in the $\ks \ks$
candidate events at $W =$ 1.7 -- 1.9~GeV and
$|\cos \theta^*|< 0.8$ (two entries per event) for the data
(dots) and MC (histogram).  
MC distribution is normalized to have the same number of kaons as
observed in data.
}
\label{fig:revfig1}
\end{figure}

Figure~\ref{fig:revfig1} shows the
distribution of $\cos \theta$ (cosine of the laboratory angle) 
of $\ks$ for the signal candidates
at $W =$1.7 -- 1.9~GeV for the data and MC.
Good agreement between the data and MC is obtained 
except for the forward-most bin ($\cos \theta > 0.9$). 
The discrepancy there is due to the inadequate efficiency estimation, 
but its effect (about 3\%) is within the systematic 
uncertainty from tracking, $K^0_S$ reconstruction and 
trigger efficiencies (see Sec.~\ref{sec:cross}).

\begin{figure}
\centering
\includegraphics[width=7cm]{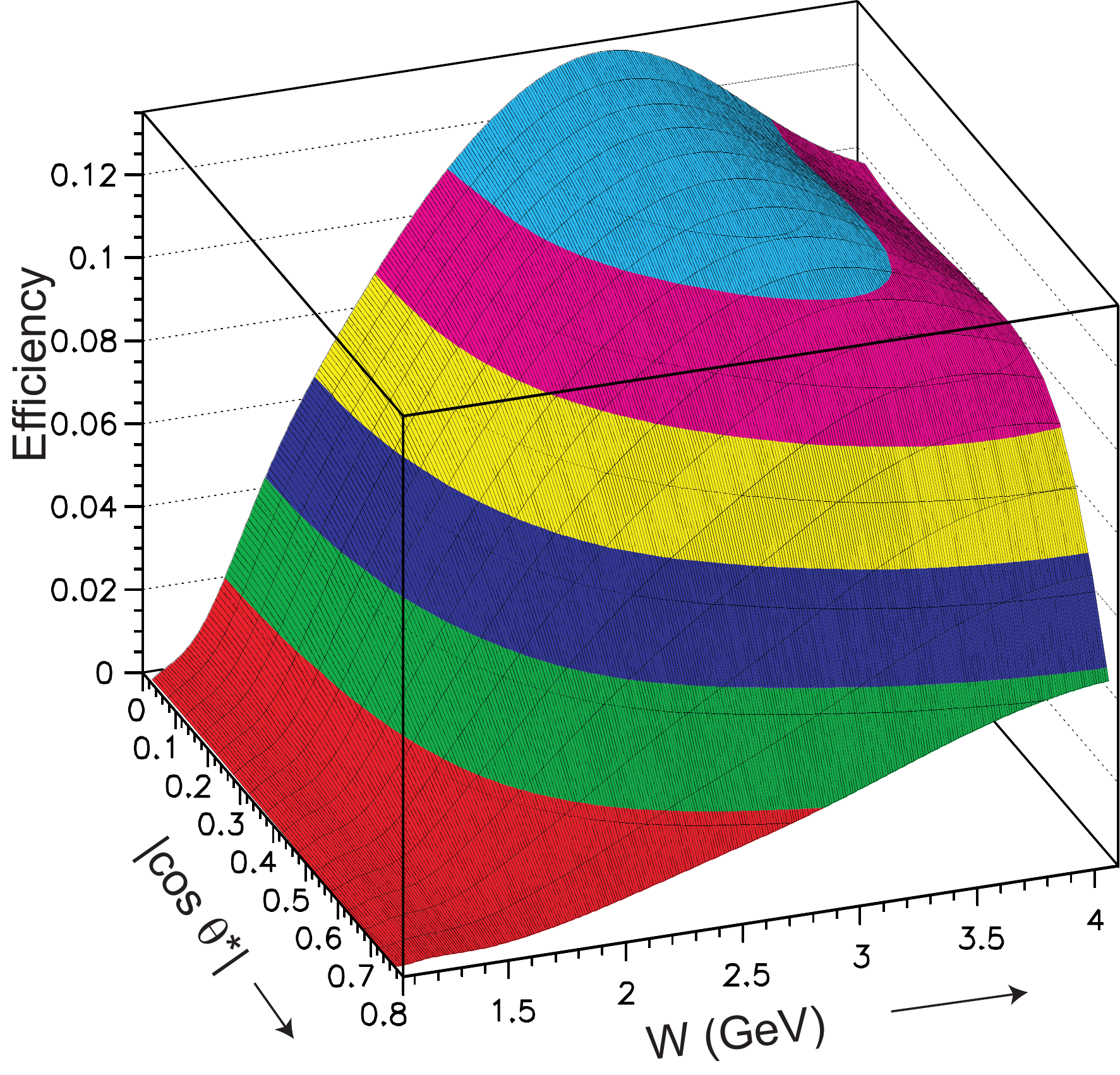}
\centering
\caption{The overall efficiency vs.
$W$ and $|\cos \theta^*|$.
}
\label{fig:effsur}
\end{figure}

\subsection{Beam energy dependence}
\label{sub:beame}
The beam-energy dependence of the luminosity function 
and the efficiency is studied at the three
energy points: $\Upsilon(4S)$ (10.58~GeV), 
$\Upsilon(5S)$ (10.88~GeV) and $\Upsilon(2S)$ (10.02~GeV),
with the signal MC samples generated at each energy.
We compare the luminosity function and
efficiency at several $W$ points among the three beam energies.
We use 10.58~GeV as the reference energy point and apply a correction
proportional to the integrated luminosity to each sample
at the other energies.

The luminosity function has a beam-energy dependence
with a factor depending on $W$;
for $W$ in (1.1~GeV, 2.0~GeV, 4.0~GeV),
the factor is ($-5\%$, $-6\%$, $-10\%$)
for 10.02~GeV and ($+2\%$, $+3\%$, $+5\%$) for 10.88~GeV.
Meanwhile, the efficiency depends on the beam energy:
$+3\%$ at 10.02~GeV and $-1\%$ at 10.88~GeV, which is 
opposite to the trend in the luminosity function.
It is also weakly dependent on $W$.

The overall effect of the beam-energy differences is negligible
when we apply the values of the efficiency and luminosity
function for 10.58~GeV to all the data, and it is estimated
to be at most 0.4\% at any $W$.
We do not correct for this effect and do not assign 
any systematic error.

\subsection{Invariant-mass and angular resolution}
\label{sub:resol}
We estimate a $\ks \ks$ mass resolution (\textit{i.e.}, 
a $W$ resolution) 
of $\sigma_W/W = 0.2\%$ for the entire $W$ region, with a 
small $W$ dependence, according to a signal MC study. 
As this is much smaller than the
bin width (at worst, $\sigma_M < 4$~MeV 
near $W=1.9$~GeV, where the bin width is 10~MeV), 
we do not apply unfolding. 
The estimated systematic shift
due to bin migrations associated with resolution 
is less than 1\% and is absorbed in the systematics.

The resolution for the c.m. angle measurement in each event
is typically $\sigma_{|\cos  \theta^*|}=0.0025$, which is 
much smaller than the bin width of 0.1.

\section{Differential cross section}
\label{sec:cross}
The differential cross section $d\sigma/d|\cos \theta^*|$ is
derived after the subtraction of the backgrounds and the application
of the corrections to the yields and efficiencies:
\begin{eqnarray}
\frac{d \sigma}{d |\cos \theta^*|} &=& 
\frac{1}{\int {\cal L}dt\ L_{\gamma\gamma} \ \Delta W\ 
\Delta |\cos \theta^*|} \times \nonumber \\
&& 
\frac{N - N_{\rm bkg}}{\epsilon \ {\cal B}(\ks \to \pi^+\pi^-)^2} \; ,
\label{eqn:dsdc}
\end{eqnarray}
where $N$ ($N_{\rm bkg}$) is the number of candidate (background) events,
$\int {\cal L}dt$ is the total integrated luminosity and
$L_{\gamma \gamma}$ is the two-photon luminosity function, calculated
as a function of $W$. 
$\Delta W$ and $\Delta |\cos \theta^*|$
are the bin widths, and $\epsilon$ is the efficiency that includes
all trigger/selection effects.
The $W$ and $|\cos \theta^*|$ dependence of the overall efficiency is
shown in Fig.~\ref{fig:effsur}.
The efficiency is smaller than 0.14 everywhere in the measurement range.
A major cause of the overall efficiency loss is associated with a
Lorentz boost of the two-photon system which results in
at least one $\ks$ falling outside of the detector's
acceptance typically more than half of the time.
Note that this efficiency loss strongly depends on 
$W$ and $|\cos \theta^*|$. 

We extract the differential cross section in the range
$|\cos \theta^*| <0.8$ and 1.1~GeV$< W < 3.3$~GeV, with
a $W$ bin width of 10~MeV for $W = 1.1 - 1.9$~GeV, 20~MeV 
for 1.9 -- 2.4~GeV, 40~MeV for 2.4 -- 2.6~GeV, and 100~MeV 
for 2.6 -- 3.3~GeV. 
In this extraction, we first evaluate the
differential cross section for finer bin widths, $\Delta W = 10$~MeV
and $\Delta |\cos \theta^*| = 0.05$ over the entire region, 
using the efficiency for the central point of each bin.
The values for these fine bins are then combined via a weighted average
into the coarser bins,
with a weight calculated from the statistical errors.

In the range $W = 1.05 - 1.10$~GeV, we extract only
the cross section integrated over $|\cos \theta^*| <0.6$,
assuming a flat angular dependence of the differential
cross section because of limited statistics and the limited
coverage in the forward angles in the vicinity of 
$|\cos \theta^*| \sim 0.6$.

In the range $W = 3.3 - 3.6$~GeV, we do not
extract the $\gamma \gamma \to \ks \ks$
cross section where the contributions
from the $\chi_{c0}$ and $\chi_{c2}$ resonances dominate
the yield; we cannot subtract leakages from
these narrow states reliably over the entire region.

In the range $W = 3.6 - 4.0$~GeV, we find some contribution
from the signal process. 
It is possible to extract the integrated cross section for 
$|\cos \theta^*| <0.8$ in this $W$ region;
however, we do not present differential cross section
due to small statistics.
There could be a contribution from high-mass charmonium resonance(s)
($\chi_{cJ}(2P)$ for example) at $W=3.80 - 3.95$~GeV,
as we find some events at large angles in this
$W$ range; these events are included in the total cross section
(see Fig.~\ref{fig:kskschic}).

At $W = 4.0 - 4.1$~GeV, we find only a small number of signal events 
that give a peak near $|\sum \vec{p}^*_t| = 0$, 
consistent with a large background contamination.
No cross section measurement is therefore performed in
the $W$ region above 4.0~GeV.

\begin{figure*}
\centering
\includegraphics[width=12cm]{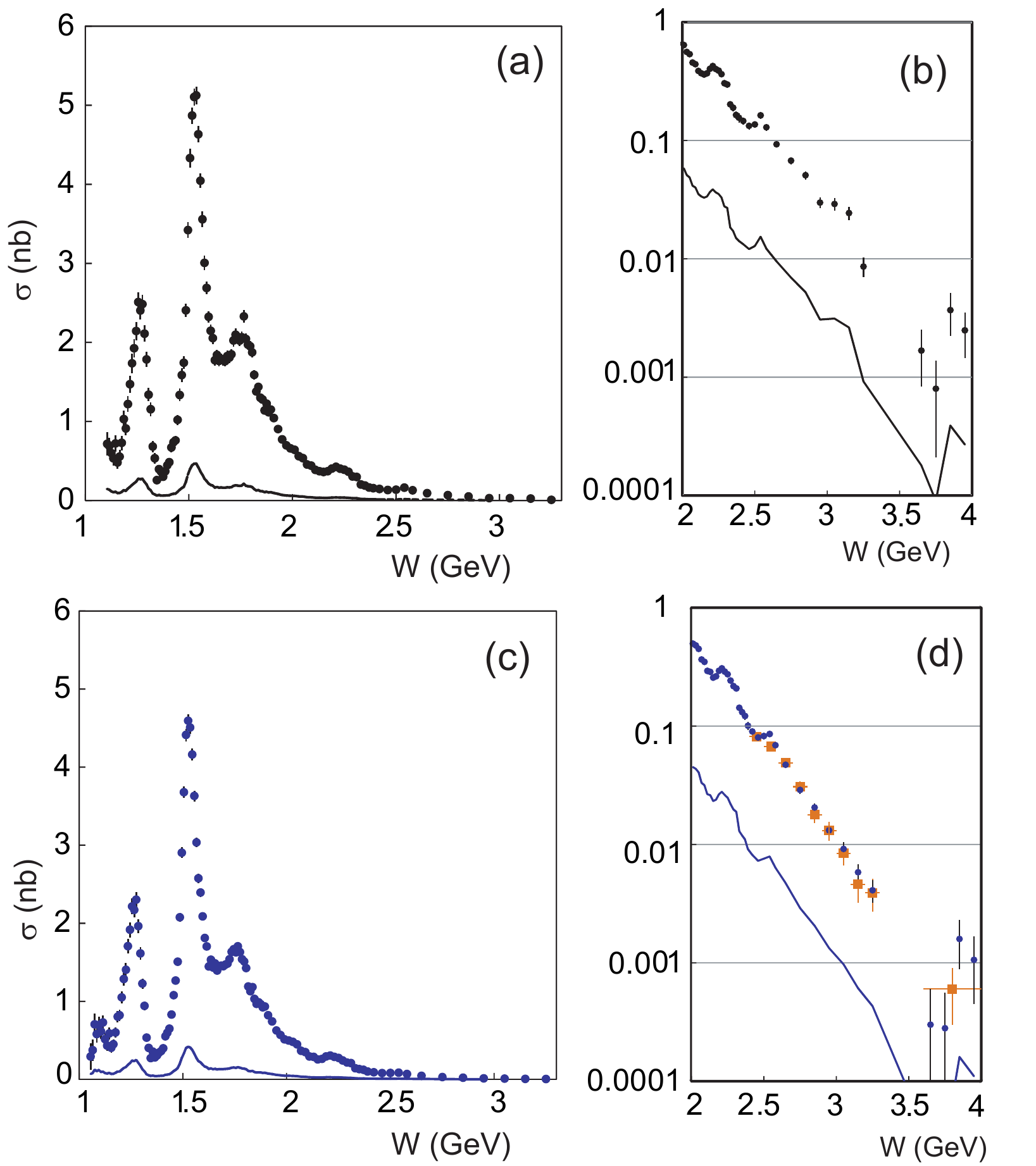}
\centering
\caption{
The $W$ dependence of the $\gamma \gamma \to \ks \ks$ cross section 
after integrating over the angle
up to (a,b) $|\cos \theta^*| < 0.8$ (black points) and
(c,d) $|\cos \theta^*| < 0.6$ (blue points).
The orange square markers in (d) are from our previous 
publication~\cite{chen} for $|\cos \theta^*| < 0.6$. 
The solid curves are the systematic uncertainties.
}
\label{fig:ksksres}
\end{figure*}

Figure~\ref{fig:ksksres} shows the cross section integrated
over $|\cos \theta^*|$. 
The integration is performed by summing
the differential cross section
for $|\cos \theta^*|<0.8$ or $|\cos \theta^*|<0.6$. 
The error bars are statistical only. 
The curves in the figure show the total systematic errors.

The systematic error includes contributions from the uncertainties in
tracking efficiency (2\% for 4 tracks), beam-background effects (1\%) 
estimated from the stability of yield ratios between the data and MC 
across all run periods, pion identification (2\% for four pions),
non-exclusive and four-pion backgrounds (described in 
Sec.~\ref{sub:nonex} and B),
geometrical coverage and fit uncertainty (4\% in total), $\ks K \pi$
background subtraction (Sec.~\ref{sub:kskpi}), $\ks$-pair 
reconstruction (Sec.~\ref{sub:kspair}), trigger efficiency 
(Sec.~\ref{sub:trgeff}), 
and the $E_{\rm ECL}$ cut (Sec.~\ref{sub:trgeff}).
We assign the uncertainty for the L4 efficiency
to be about 10\% of the inefficiency in different $W$ regions.
The systematic error associated with the uncertainty
in the integrated luminosity and luminosity function
includes the form-factor effect of space-like photons.
Summing in quadrature, the total systematic uncertainty
evaluated is typically 10\%.
The systematic uncertainties are summarized 
in Table~\ref{tab:sumsys}.

\begin{center}
\begin{table*}
\caption{Summary of systematic uncertainties (\%) in the cross section
integrated over the angle in a single $W$ bin. 
When a range is shown, the uncertainty varies between the 
values with decreasing $W$.
}
\label{tab:sumsys}
\begin{tabular}{lc} \hline \hline
Source & Uncertainty (\%) \\ \hline
Tracking efficiency (for 4 tracks) & 2  \\
Beam background effect & 1 \\
Pion identification (for 4 tracks) & 2  \\
Non-exclusive and four-pion backgrounds &  2 -- 19 \\
Geometrical coverage and fit uncertainty & 4  \\ 
$\ks K \pi$ background subtraction & 1 -- 2 \\
$\ks$-pair reconstruction & 5 -- 3 \\
Trigger efficiency & 5 -- 7 \\
$E_{\rm ECL}$ cut & 1 \\
Integrated luminosity and luminosity function & 5 -- 4  \\
L4 efficiency & 1 -- 10 \\ \hline
Total & 9 -- 25, typically 10 \\
\hline \hline
\end{tabular}
\end{table*}
\end{center}

\section{Study of resonances}
\label{sec:reson}
\begin{figure}
 \centering
   {\epsfig{file=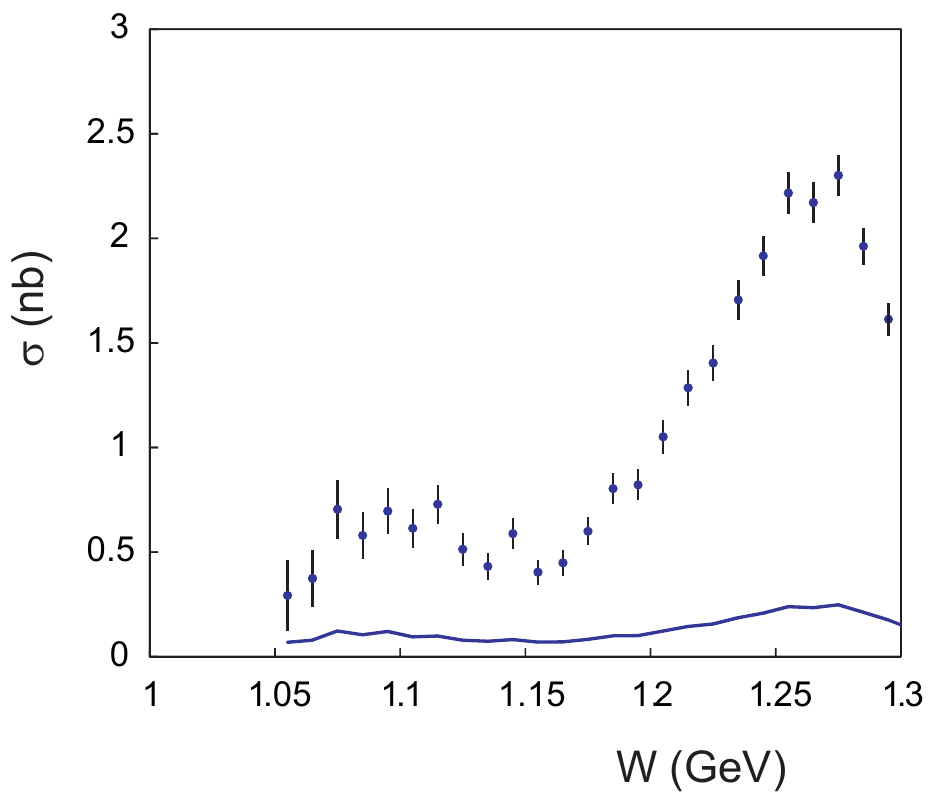,width=65mm}}
 \caption{A close-up view of
the measured integrated cross section ($|\cos \theta^*| \leq 0.6$)
near the threshold
for the process $\gamma \gamma \to \ks \ks$.
The solid curve is the systematic uncertainty.}
\label{fig:totthr}
\end{figure}

Figure~\ref{fig:ksksres}(a) shows the measured
integrated cross section ($|\cos \theta^*| \leq 0.8$),
where prominent peaks are observed near
1.3, 1.5 and 1.8~GeV.
Enhancements are also observed around 2.3 and 2.6~GeV.
A close-up view of the integrated cross section 
($|\cos \theta^*| \leq 0.6$) near the threshold is shown in 
Fig.~\ref{fig:totthr}, where the cross section is 
small ($<1$~nb), in agreement with theoretical
predictions~\cite{achasov, achasov2}.

In this section we describe the extraction of partial wave
information from our data by fitting the
differential cross section using suitable
parameterizations to estimate the mass, total width and
$\Gamma_{\gamma \gamma}\B(K \bar{K})$ of the $f_2'(1525)$,
to derive the phase difference between the $f_2(1270)$ and $a_2(1320)$
and to identify the nature and obtain the parameters
of the resonances near 1.8, 2.3 and 2.6~GeV.

\subsection{Partial wave amplitudes}
\label{sub:pwa}
In the $\gamma \gamma \to \ks \ks$ channel, only the partial waves of
even angular momenta contribute.
Furthermore, in the energy region $W \simlt 3~\GeV$, the $J > 4$ partial 
waves may be ignored,
so only S, D and G waves are considered in the fit.
The differential cross section can then be expressed as
\begin{eqnarray}
\frac{d \sigma (\gamma \gamma \to \ks \ks)}{d \Omega} 
 &=& \left| S \: Y^0_0 + D_0 \: Y^0_2  + G_0 \: Y^0_4 \right|^2 \nonumber \\
&& + \left| D_2 \: Y^2_2  + G_2 \: Y^2_4 \right|^2 \; ,
\label{eqn:diff}
\end{eqnarray}
where $S$ is the S-wave amplitude,
$D_0$ and $G_0$ ($D_2$ and $G_2$) denote the helicity-0 (2) components
of the D and G wave~\cite{pw}, respectively, and $Y^m_J$ are 
the spherical harmonics.
The angular dependence of the cross section is governed by the 
spherical harmonics while the energy dependence is determined by
the partial waves.

Since the spherical harmonics are not independent of each other,
a unique decomposition of partial waves cannot be determined using 
the measured differential cross section.
To overcome this problem, we rewrite Eq.~(\ref{eqn:diff}) as
\begin{eqnarray}
\frac{d \sigma (\gamma \gamma \to \ks \ks)}{4 \pi d |\cos \theta^*|} 
 &=& \hat{S}^2 \: |Y^0_0|^2  + \hat{D}_0^2 \: |Y^0_2|^2 
 + \hat{D}_2^2  \: |Y^2_2|^2 \, \nonumber \\ 
&& + \hat{G}_0^2  \: |Y^0_4|^2  \, + \hat{G}_2^2  \: |Y^2_4|^2  \, .
\label{eqn:diff2}
\end{eqnarray}
The ``hat-amplitudes'' $\hat{S}^2$, $\hat{D}_0^2$, $\hat{D}_2^2$, 
$\hat{G}_0^2$ and $\hat{G}_2^2$ can be negative because of
interference terms, and 
correspond to $|S|^2, \; |D_0|^2, \; |D_2|^2, \; |G_0|^2$
and $|G_2|^2$, respectively, when interference 
terms are ignored~\cite{pi0pi0}.

As the absolute squares of the spherical harmonics are independent of each other,
we can fit the differential cross section in each $W$ bin to obtain 
$\hat{S}^2$, $\hat{D}_0^2$, $\hat{D}_2^2$, $\hat{G}_0^2$
and $\hat{G}_2^2$.
The fit with the value of $J \le 4$ is named the ``$SDG$ fit.''
At low energy, we expect that the contribution from
$J=4$ is negligible,
so we also perform a separate fit by setting 
$\hat{G}_0^2 = \hat{G}_2^2 =0$, which is named
the ``$SD$ fit.'' 

The differential cross section is fitted according to 
Eq.~(\ref{eqn:diff2}), 
where statistical errors only are taken into account.
The differential cross section for $|\cos \theta^*| \leq 0.8$ is
extracted for $1.1~\GeV \leq W \leq 3.3~\GeV$.
In the $SDG$ fit, two consecutive data points of $\Delta W = 0.01$~GeV
are merged, resulting in bins of width 0.02~GeV.

\begin{figure*}
 \centering
   {\epsfig{file=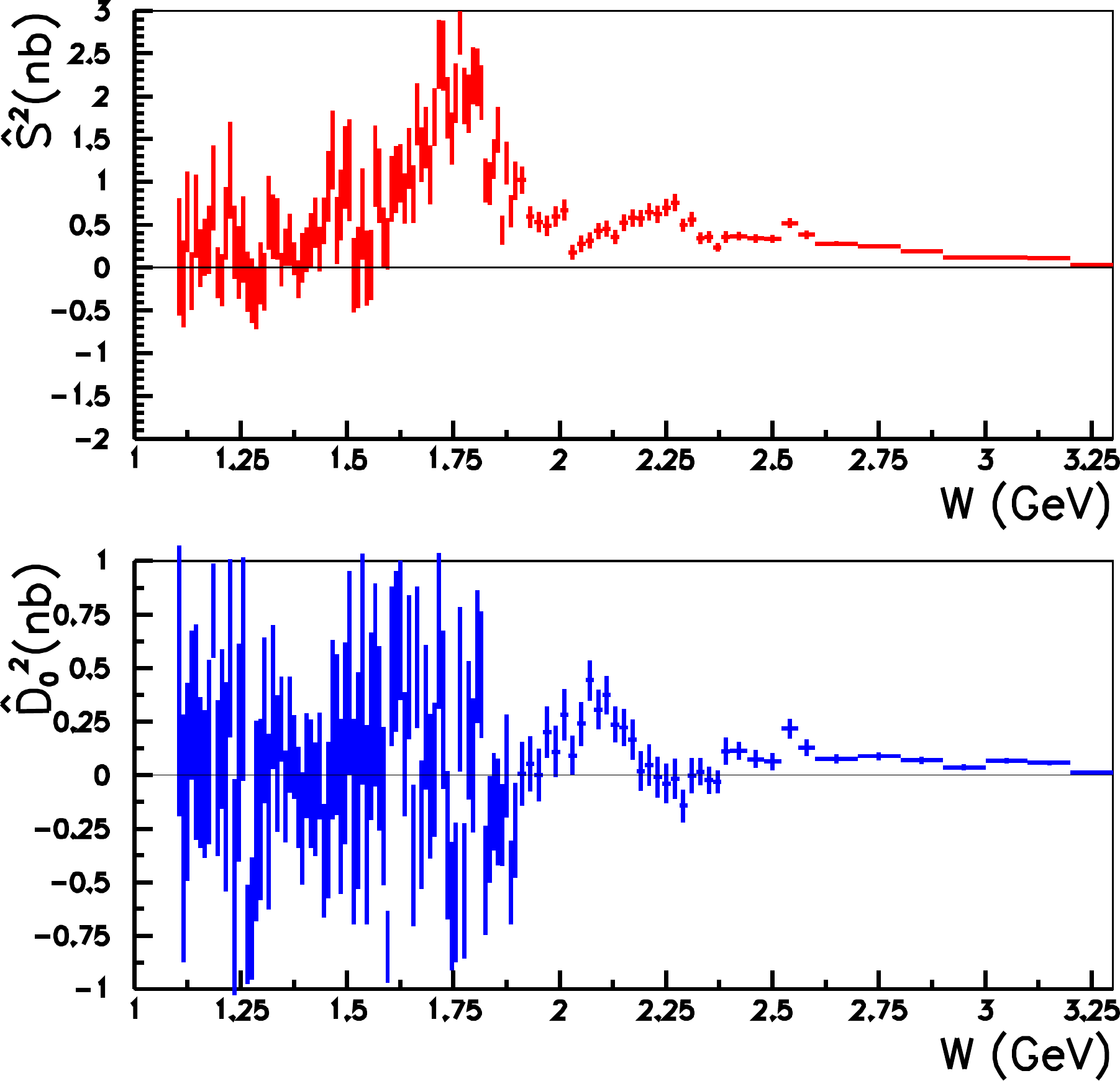,width=65mm}}
   {\epsfig{file=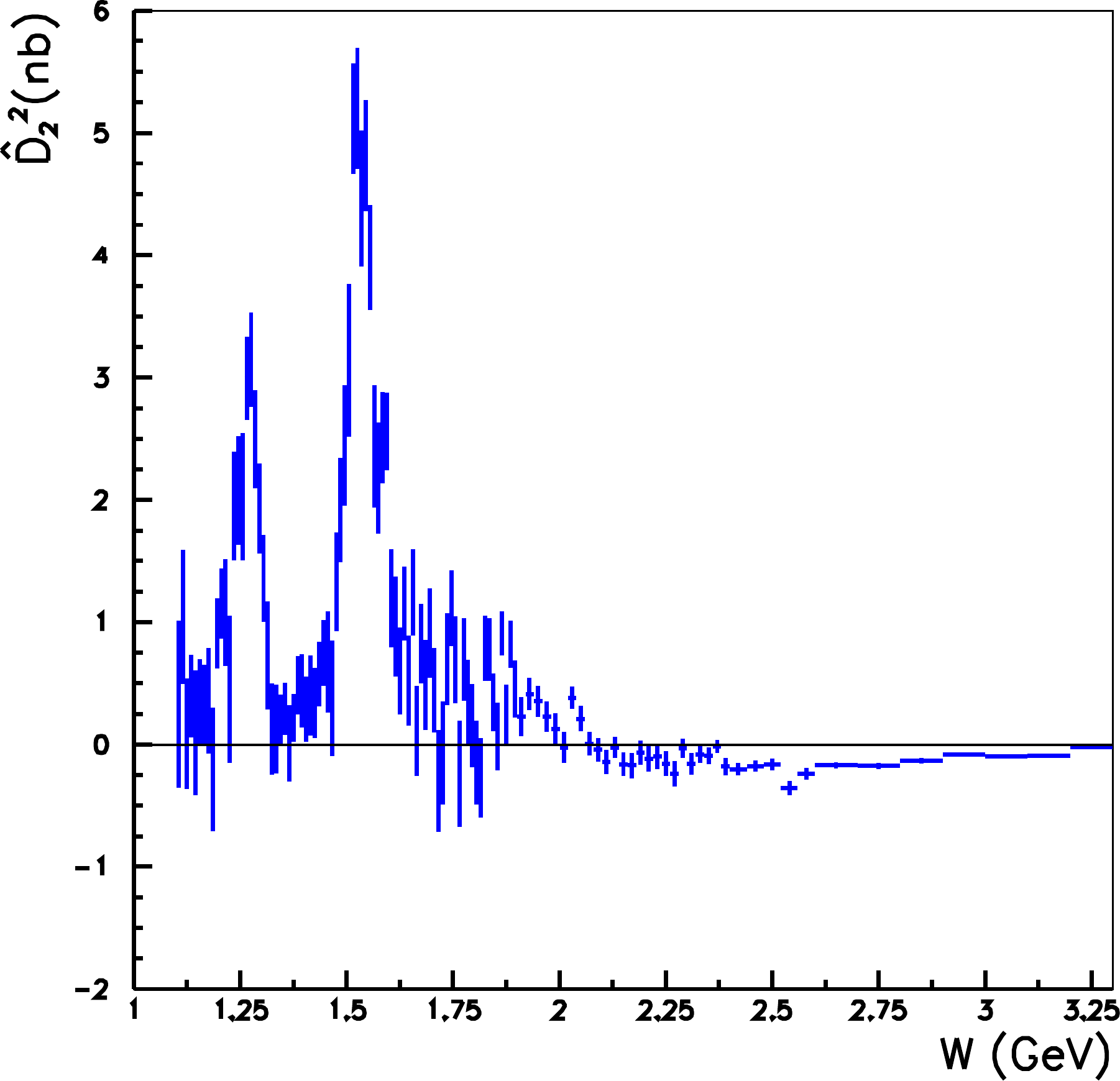,width=65mm}}
 \caption{Amplitudes $\hat{S}^2$ (left top),
 $\hat{D}_0^2$ (left bottom) and $\hat{D}_2^2$ (right)
obtained from the $SD$ fit. 
The error bars represent the statistical uncertainties
when no correlations among the fit parameters are included.
}
\label{fig:dsd02}
\end{figure*}

\begin{figure*}
 \centering
   {\epsfig{file=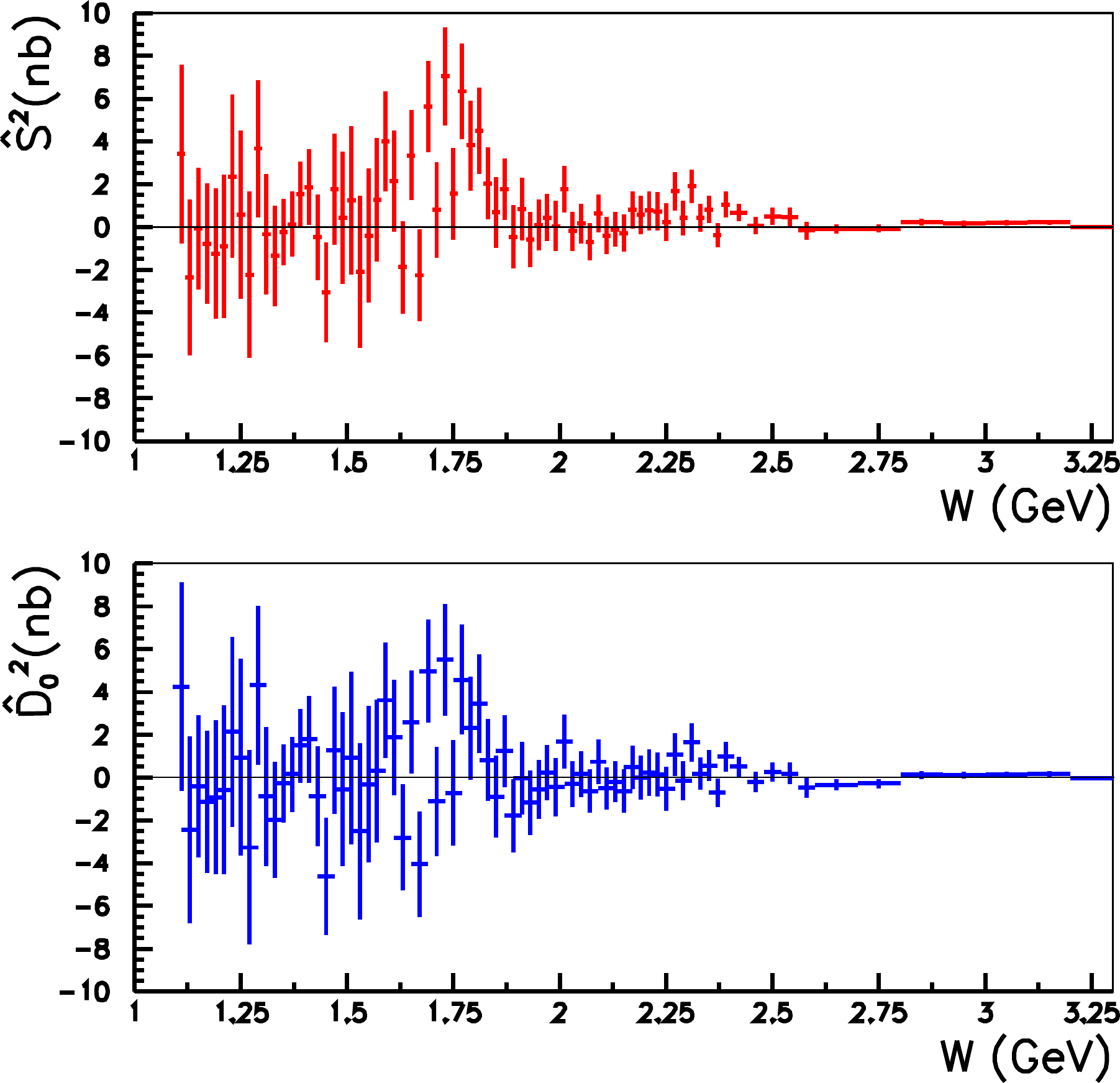,width=65mm}}
   {\epsfig{file=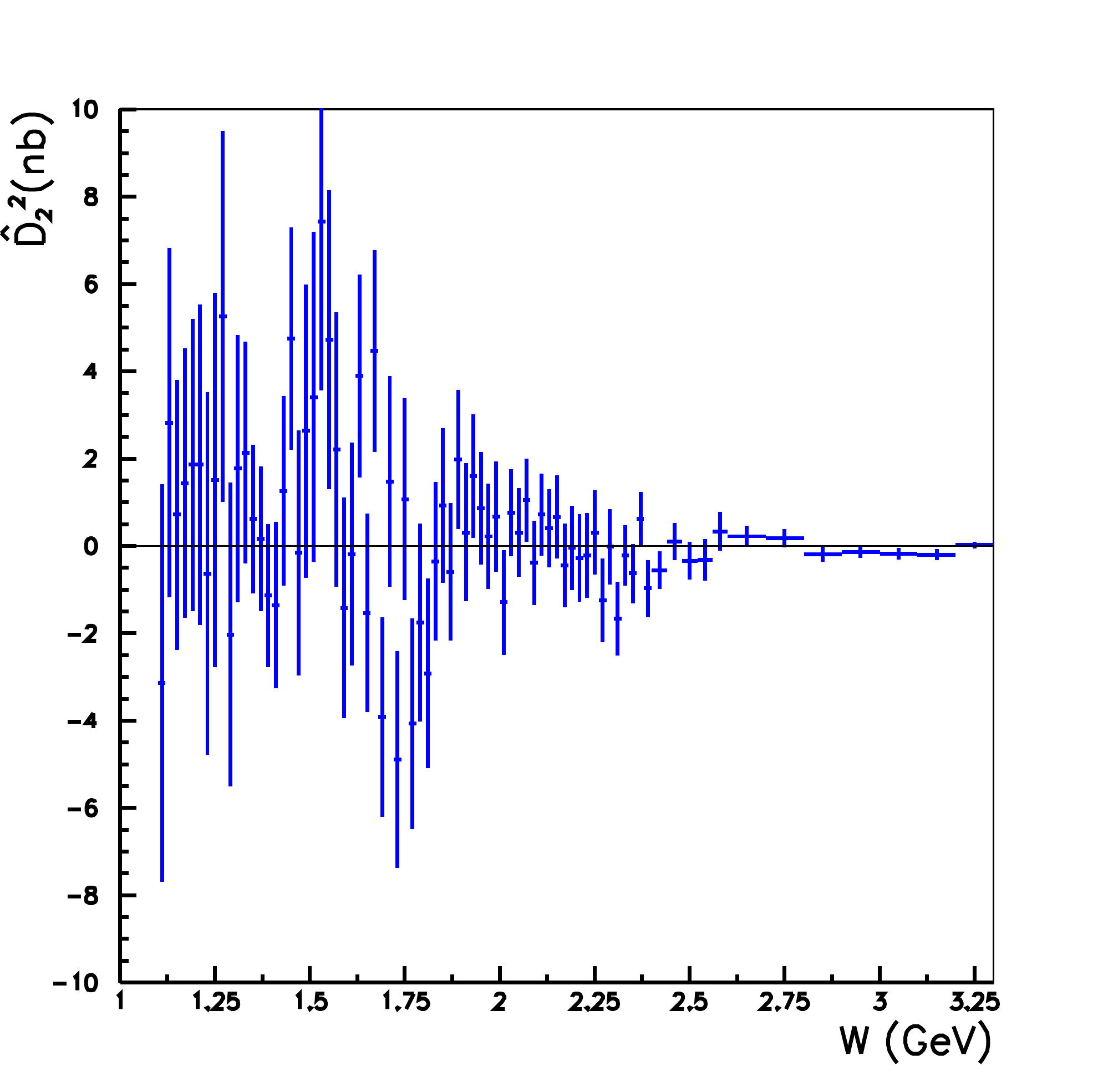,width=65mm}}
 \caption{Amplitudes $\hat{S}^2$ (left top),
 $\hat{D}_0^2$ (left bottom) and $\hat{D}_2^2$ (right)
obtained from the $SDG$ fit. 
The error bars represent the statistical uncertainties
when no correlations among the fit parameters are included.
}
\label{fig:gdsd02}
\end{figure*}

\begin{figure*}
 \centering
   {\epsfig{file=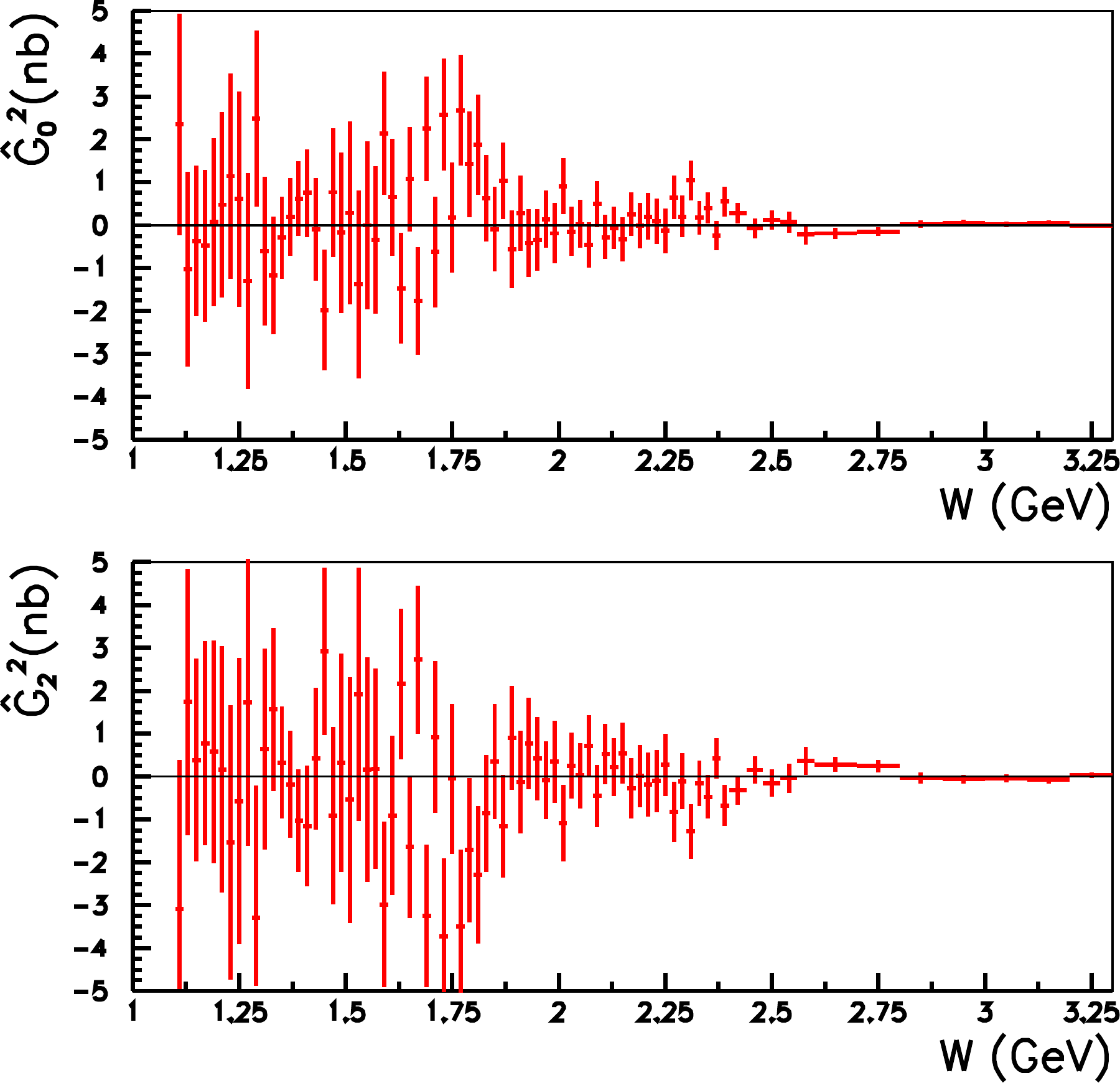,width=65mm}}
   {\epsfig{file=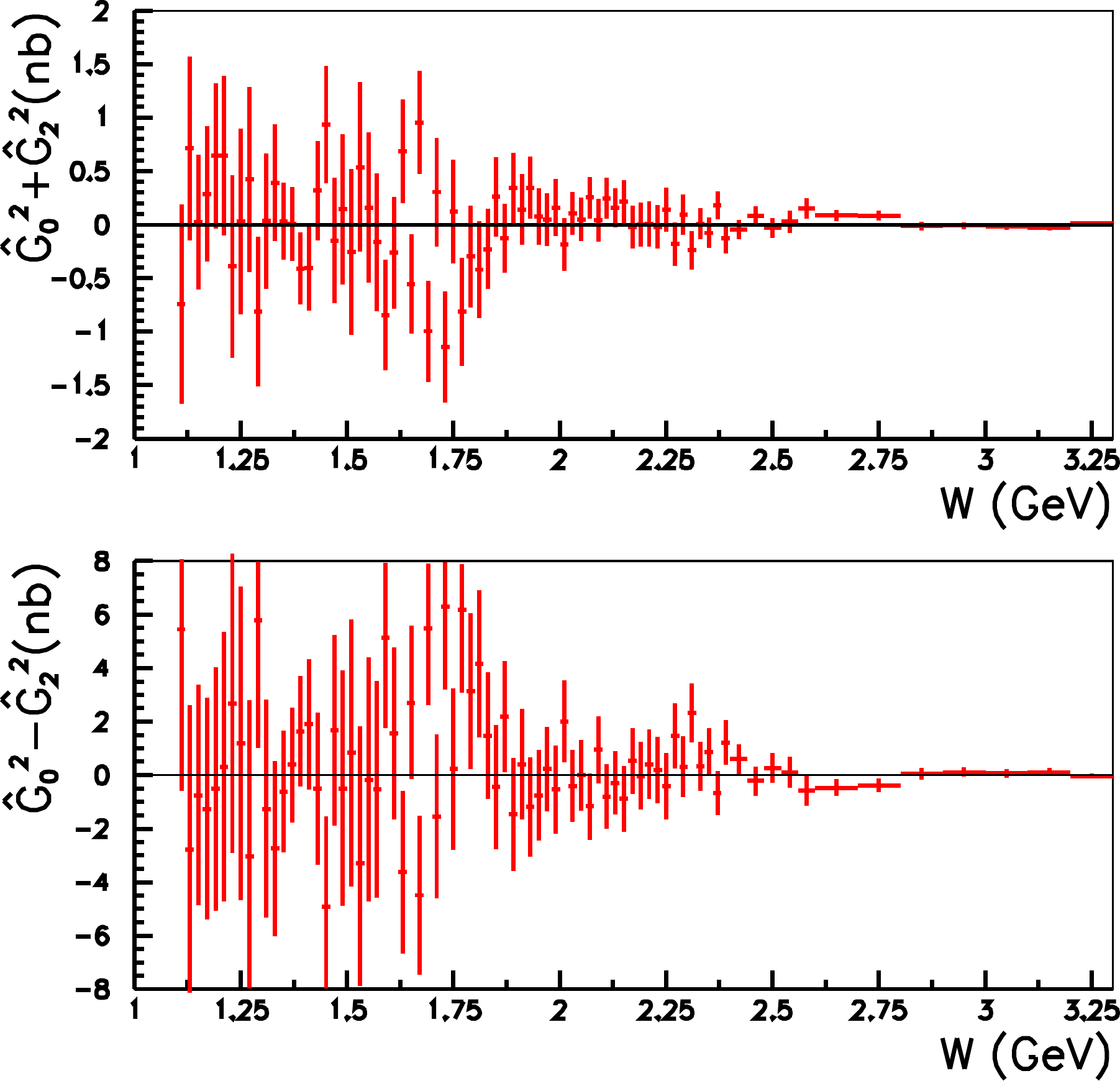,width=65mm}}
 \caption{Amplitudes $\hat{G}_0^2$ (left top),
 $\hat{G}_2^2$ (left bottom) and
$\hat{G}_0^2 \pm \hat{G}_2^2$ (right)
obtained from the $SDG$ fit.
The error bars represent the statistical uncertainties
when no correlations among the fit parameters are included.
}
\label{fig:gdg}
\end{figure*}

\begin{figure}
 \centering
   {\epsfig{file=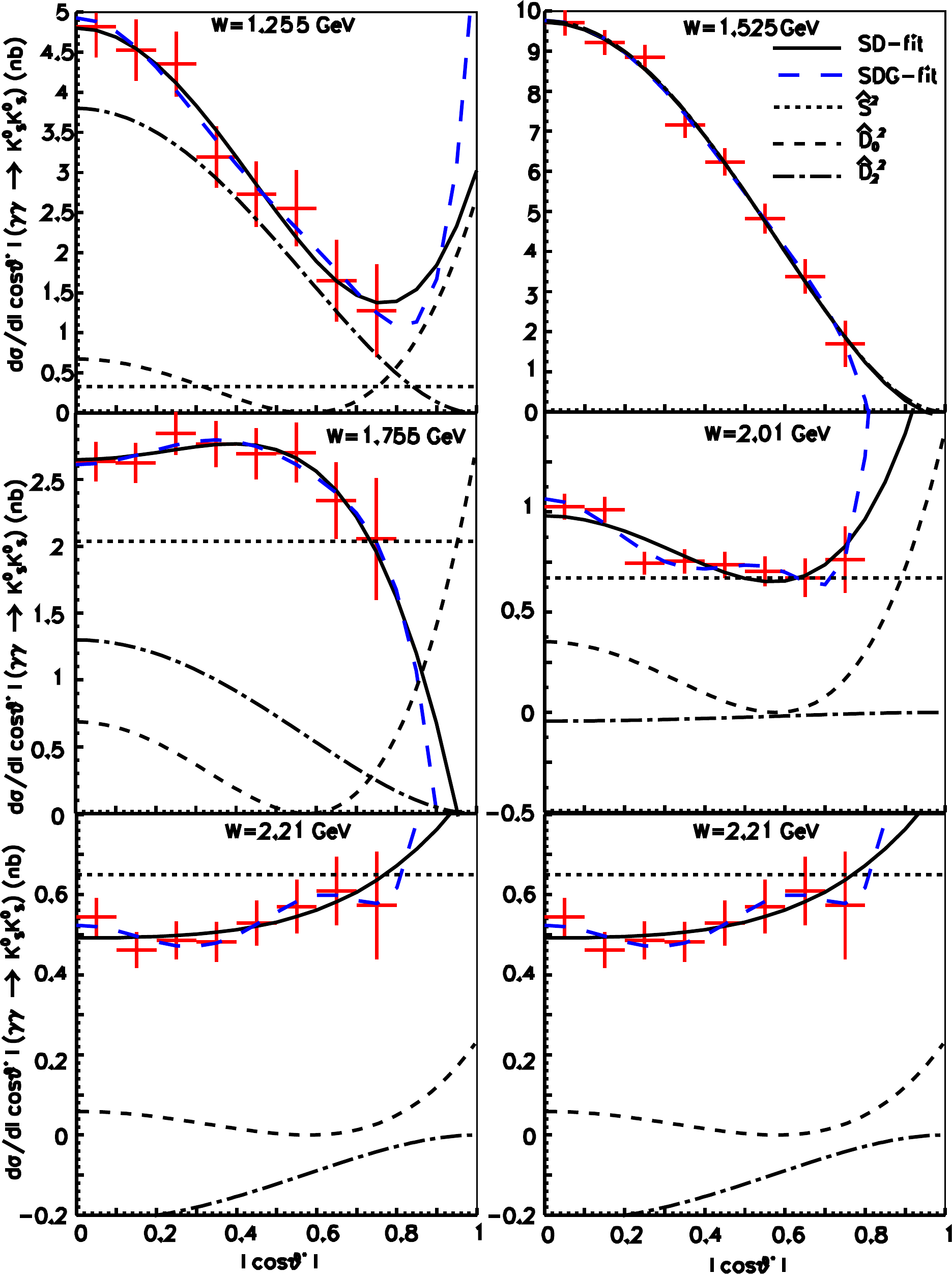,width=65mm}}
 \caption{
Results of the $SD$- (solid line) and 
$SDG$- (long-dashed line) fits of the differential cross section
in selected $W$ bins. 
The number in each panel denotes the $W$ bin.
Also shown are $\hat{S}^2$ (dotted line), $\hat{D}_0^2$ (dashed line)
and $\hat{D}_2^2$ (dot-dashed line) obtained from the $SD$ fit.
}
\label{fig:dhsel}
\end{figure}

The obtained spectra of $\hat{S}^2$, $\hat{D}_0^2$ and $\hat{D}_2^2$ 
for the $SD$ fit are shown in Fig.~\ref{fig:dsd02}.
Figures \ref{fig:gdsd02} and \ref{fig:gdg}
show the hat-amplitudes for the $SDG$ fit.
$\hat{G}_0^2 \pm \hat{G}_2^2$ are also plotted in Fig.~\ref{fig:gdg},
since the angular dependences of $|Y^0_4|^2$ and $|Y^2_4|^2$ are 
similar for $|\cos \theta^*| \simlt 0.6$.
In the $SDG$ fit, the structures in $\hat{D}_2^2$ are less visible 
and the G waves appear to be small for $W \leq 3.3$~GeV.
Figure~\ref{fig:dhsel} shows the differential cross section
for selected $W$ bins with
the fitted $\hat{S}$, $\hat{D}_0$ and $\hat{D}_2$ waves.

Although the derived hat-amplitudes $\hat{S}^2$, $\hat{D}_0^2$,
$\hat{D}_2^2$, $\hat{G}_0^2$ and $\hat{G}_2^2$ in fact
contain interference terms such as $\Re (S^* D_0)$,
they do provide useful information about partial wave amplitudes.
Two prominent peaks are observed in the $\hat{D}_2^2$ spectrum:
the peak near 1.3~GeV is due to the interference between
the $f_2(1270)$ and $a_2(1320)$ and
the second peak is due to the $f_2'(1525)$.
No other notable structures are observed in
Fig.~\ref{fig:dsd02} (right).
In the $\hat{S}^2$ spectrum shown in Fig.~\ref{fig:dsd02} (left top),
three peaks around 1.8, 2.3 and 2.6~GeV are observed.
The lowest may be due to the $f_0(1710)$
(\textit{not} a tensor meson, as discussed in past 
experiments~\cite{tasso1, L3, kabe}). 
This might be an $a_0$ meson, though no such mesons 
have been observed previously in this mass region~\cite{pdg2012}.
$\hat{D}_0^2$ is rather small and featureless except around 2.1
and 2.6~GeV, and hence the D$_0$ wave may also be small but non-zero:
there appears to be an interference term between S and D$_0$.

We use our assumptions for the partial wave amplitudes and fit
the data to extract the parameters of the resonances.
Note that we do not fit the obtained spectra of hat-amplitudes
$\hat{S}^2$, $\hat{D}_0^2$ and 
$\hat{D}_2^2$, but rather fit the differential cross section directly
using Eq.~(\ref{eqn:diff}).
In our analysis, we fit the energy region $W \le 3.0~\GeV$,
with separate fits for $W \leq 2$~GeV and $W > 2$~GeV.

\subsection{Fitting the region $W \le 2.0~\GeV$}
\label{sub:below}
In this section, we describe our fits in the $W \leq 2.0$~GeV region.
Motivated by the spectra of $\hat{D}_2^2$ and  $\hat{S}^2$,
we include the $f_2(1270)$, $a_2(1320)$ and $f_2'(1525)$ 
in the D$_2$ wave
and test the hypothesis of a scalar meson (coined the $f_0(1710)$)
in the S wave.
In this analysis, we measure the relative phase probing the destructive 
interference between the $f_2(1270)$ and $a_2(1320)$
and determine relevant parameters of the $f_2'(1525)$, in particular,
$\Gamma_{\gamma \gamma}\B(K \bar{K})$.

\subsubsection{Parameterization of amplitudes}
\label{sub:param}
Based on the above observation, the amplitudes
$S$, $D_0$ and $D_2$ are parameterized as follows:
\begin{eqnarray}
S &=& A_{f_0(1710)} e^{i \phi_{f0}}  + B_S  , \nonumber \\
D_0 &=& B_{D0}e^{i \phi_{D0}} , \nonumber \\
D_2 &=& A_{f_2(1270)} + A_{a_2(1320)} e^{i \phi_{a2}}
+ A_{f_2'(1525)} e^{i \phi_{f2p}} \nonumber \\
&& + B_{D2}e^{i \phi_{D2}} ,
\label{eqn:param}
\end{eqnarray}
where $A_{f_0(1710)}$, $A_{f_2(1270)}$,  $A_{a_2(1320)}$
and $A_{f_2'(1525)}$ are the amplitudes describing the resonances;
$B_S$, $B_{D0}$ and $B_{D2}$ are the non-resonant background
amplitudes for the S, D$_0$ and D$_2$ waves; and
$\phi_{f0}$, $\phi_{a2}$, $\phi_{f2p}$, $\phi_{D0}$ and $\phi_{D2}$
are the phases of the resonances and background amplitudes.
The phases are defined relative to $B_S$ ($f_2(1270)$) for
helicity-0 (2) amplitudes.
Using this convention, the relative phase between the $f_2(1270)$ 
and $a_2(1320)$ is given by $\phi_{a2}$.
We also study the case in which the $f_0(1710)$ is replaced by a
tensor meson (labeled the $f_2(1710)$ here, although
the $f_2(1810)$ is listed in PDG~\cite{pdg2012}) in $D_2$.
To investigate if our approximation could describe the data well
without this resonant contribution, we also perform a fit
assuming no resonance at 1.8~GeV.

We assume the background amplitudes to be quadratic in $W$ 
multiplied by the threshold factor $\beta^{2 l +1}$ for all waves:
\begin{eqnarray}
B_S &=&  \beta (a_S W'^2  + b_S W' + c_S) 
\;, \nonumber \\
B_{D0} &=& \beta^5 (a_{D0} W'^2  + b_{D0} W' + c_{D0}) \;, \nonumber \\
B_{D2} &=& \beta^5 (a_{D2} W'^2  + b_{D2} W' + c_{D2}) \;, 
\label{eqn:para2}
\end{eqnarray}
where $W'=W-2m_{\ks}$,
$\beta=\sqrt{1-4m_{\ks}^2/W^2}$ is the velocity of the $K^0_S$
divided by the speed of light, $m_{\ks}$ is the mass of the $\ks$,
and $l$ is 0 (2) for $S$ ($D_0$ and $D_2$).

We use the parameterization of the $f_2(1270)$ and  $f_2'(1525)$ 
given in Ref.~\cite{pi0pi0} and that of the $a_2(1320)$ in 
Ref.~\cite{etapi0}.
We note that ${\B}(R \rightarrow \ks \ks)/
{\B}(R \rightarrow K \bar{K}) = 1/4$ for any $f_J$ or $a_J$
resonance $R$.

The amplitude $A_R(W)$ for each spin-$J$ resonance $R$ of mass $m_R$ 
is parameterized using the relativistic Breit-Wigner formula
\begin{eqnarray}
A_R^J(W) &=& \sqrt{\frac{8 \pi (2J+1) m_R}{W}} 
\nonumber \\
&& \times \frac{\sqrt{ \Gamma_{\rm tot}(W)
\Gamma_{\gamma \gamma}(W) \B(K  \bar{K}})}
{m_R^2 - W^2 - i m_R \Gamma_{\rm tot}(W)} \; .
\label{eqn:arj}
\end{eqnarray}
Hereinafter, we consider the case $J=2$.
The energy-dependent total width $\Gamma_{\rm tot}(W)$ is given by
\begin{equation}
\Gamma_{\rm tot}(W) = \sum_{X_1 X_2} \Gamma_{X_1 X_2} (W) \; ,
\label{eqn:gamma}
\end{equation}
where $X_1, \; X_2$ is $\pi$, $K$, $\eta$, $\gamma$, etc.
For $J=2$,
the partial width $\Gamma_{X_1 X_2}(W)$ is parameterized as~\cite{blat}:
\begin{eqnarray}
\Gamma_{X_1 X_2} (W) &=& \Gamma_R \B(R \rightarrow X_1 X_2) 
\left( \frac{q_X(W^2)}{q_X(m_R^2)} \right)^5 \nonumber \\
&& \times
\frac{D_2\left( q_X(W^2) r_R \right)}{D_2 \left( q_X(m_R^2) r_R 
\right)} \;,
\label{eqn:gamx}
\end{eqnarray}
where $\Gamma_R$ is the total width at the resonance mass,
\begin{eqnarray}
q_X(W^2) &=& \frac{1}{2W} \left[
\left( W^2 - (m_{X_1} + m_{X_2})^2 \right) \right. \nonumber \\
&& \times \left. \left( W^2 - (m_{X_1} - m_{X_2})^2  \right)
\right]^{\frac{1}{2}}
, \nonumber \\ 
D_2(x) &=& \frac{1}{9 + 3 x^2 +x^4} \; ,
\label{eqn:qx}
\end{eqnarray}
and $r_R$ is an effective interaction radius that varies from 
1~$(\GeV/c)^{-1}$ to 7~$(\GeV/c)^{-1}$ in different hadronic 
reactions~\cite{grayer, grayer2, grayer3}.
For the three-body and other decay modes,
\begin{equation}
\Gamma_{\rm other} (W) = \Gamma_R \B(R \rightarrow {\rm other})
\frac{W^2}{m_R^2}
\label{eqn:gam3}
\end{equation}
is used instead of Eq.~(\ref{eqn:gamx}).
This formalism is used for the $f_2(1270)$, $a_2(1320)$ and 
$f_2'(1525)$.
All parameters of the $f_2(1270)$ and $a_2(1320)$ 
are fixed at the PDG values~\cite{pdg2012}
except for $r_R$, which is fixed at the value determined in 
Refs.~\cite{mori1, mori2}, as summarized in Tables~\ref{tab:f2fit}
and \ref{tab:a2para}.

\begin{center}
\begin{table}
\caption{A summary of the parameters assumed in our fits.}
\label{tab:f2fit}
\begin{tabular}{ccccc} \hline \hline
Parameter & $f_2\lr{1270}$ & $f_2'\lr{1525}$  & Unit 
& Reference \\ \hline
 Mass & $1275.1 \pm 1.2$ & $1525 \pm 5$ 
& $\MeV/c^2$ & \cite{pdg2012}\\
$\Gamma_{\rm tot}$ & $185.1 ^{+2.9}_{-2.4}$ 
& $73 ^{+6}_{-5}$ & MeV & \cite{pdg2012}\\
$\B(\pi \pi)$ & $84.8^{+2.4}_{-1.2}$ 
& $0.82 \pm 0.15$ & \% & \cite{pdg2012} \\
$\B(K \bar{K})$ & $4.6 \pm 0.4$ 
& $88.7 \pm 2.2$ & \%  &\cite{pdg2012} \\
$\B(\eta \eta)$ & $ 0.40 \pm 0.08$
& $10.4 \pm 2.2$ & \%  & \cite{pdg2012} \\
$\B(\gamma \gamma)$ & $16.4 \pm 1.9$
& $1.11 \pm 0.14$ & $10^{-6}$ & \cite{pdg2012}\\
$r_R$  & $3.62 \pm 0.03$ & $3.62 \pm 0.03$ 
& $(\GeV/c)^{-1}$ & \cite{mori1, mori2} \\
\hline \hline
\end{tabular}
\end{table}
\end{center}

\begin{center}
\begin{table}
\caption{Parameters of the $a_2(1320)$~\cite{pdg2012}.}
\label{tab:a2para}
\begin{tabular}{ccc} \hline \hline
Parameter & Value & Unit  \\ \hline
 Mass & $1318.3 \pm 0.6$ & $\MeV/c^2$ \\
$\Gamma_{\rm tot}$ & $107 \pm 5$ & MeV \\
$\B(\rho \pi)$ & $70.1 \pm 2.7$ & \% \\
$\B(\eta \pi)$ & $14.5 \pm 1.2$ & \% \\
$\B(\omega \pi \pi)$ & $10.6 \pm 3.2$ & \% \\
$\B(K \bar{K})$ & $4.9 \pm 0.8$ & \% \\
$\B(\gamma \gamma)$ & $9.4 \pm 0.7$ & $10^{-6}$ \\
\hline \hline
\end{tabular}
\end{table}
\end{center}

Finally, the parameterization of the $f_0(M)$ meson for 
$M=1710~\MeV/c^2$ is taken to be:
\begin{equation}
f_0(M) = \sqrt{\frac{8 \pi m_{f_0}} {W}}
\frac{ \sqrt{ \Gamma_{f_0} 
\Gamma_{\gamma \gamma} \B(K \bar{K})_{f_0}}} 
{m^2_{f_0} -W^2 -i m_{f_0} \Gamma_{f_0}}
 \; ,
\label{eqn:f0y}
\end{equation}
where $\Gamma_{\gamma \gamma}\B(K \bar{K})_{f_0}$
is the product of the two-photon width and the branching
fraction to $K \bar{K}$ for the $f_0(M)$ meson.
Its PDG~\cite{pdg2012} parameters are summarized in 
Table~\ref{tab:param}, together with the parameters (when known)
of the $f_2(1810)$ and $a_2(1700)$.

\begin{center}
\begin{table}
\caption{Parameters (when known) of the $f_0(1710)$, $a_2(1700)$ 
and $f_2(1810)$~\cite{pdg2012}.}
\label{tab:param}
\begin{tabular}{lccc} \hline \hline
Parameter  & $f_0(1710)$  & $a_2(1700)$ & $f_2(1810)$  \\ \hline
Mass (MeV/$c^2$ ) & $1720 \pm 6$ & $1732 \pm 16$
& $1815 \pm 12$ \\
$\Gamma_{\rm tot}$ (MeV) & $135 \pm 8$ & $194 \pm 40$ 
& $197 \pm 22$ \\
$f_J/a_J \to K \bar{K}$ & seen & seen & unknown \\
$f_J/a_J \to \gamma \gamma$ & unknown & unknown & seen \\
\hline\hline
\end{tabular}
\end{table}
\end{center}

\subsubsection{Fit in the region $1.2~\GeV \le W \le 2.0~\GeV$}
We perform a fit for the region $1.2~\GeV \le W \le 2.0~\GeV$ 
by floating the mass, width, 
$\Gamma_{\gamma \gamma} \B(K \bar{K})$ 
and the relative phase of both the $f_2'(1525)$ and 
$f_J(1710)$ ($J=0$ or $J=2$).
Also floated are the relative phase of the $a_2(1320)$ and 
the parameters ($a, \; b$ and $c$ and the phases for $D_0$ and $D_2$) 
of the background amplitudes.
To remove arbitrary sign uncertainties, the coefficients 
$c_S$, $c_0$ and $c_2$ are chosen to be positive.

Twenty parameters describing the assumed amplitudes are obtained
by fitting the 
differential cross sections.
To search for the global minimum goodness of fit $\chi^2_{\rm min}$
to identify possible multiple solutions, about 1000 sets of randomly 
generated initial parameters 
are employed for fits performed using MINUIT~\cite{minuit}.
A fit is accepted as a satisfactory solution
if its $\chi^2$-value is within $\chi^2_{\rm min} + 10$ 
(corresponding to $3.2 \sigma$).

If the $f_0(1710)$ hypothesis is
assumed to explain the peak at $W \sim 1.8~\GeV$,
four solutions are obtained with $\chi^2_{\rm min}/ndf = 677.2/580$,
where $ndf$ is the number of degrees of freedom.
These solutions are distinguished by the 
$\Gamma_{\gamma \gamma} \B(K \bar{K})$ value, which ranges from 
6.3 to 216~eV for the $f_0(1710)$,
and from 83 to 104~eV for the $f_2'(1525)$,
as listed in Table~\ref{tab:fjall}.

When the $f_2(1710)$ hypothesis is used,
the two obtained solutions have lower quality
with $\chi^2_{\rm min}/ndf = 755.6/580$.
Their fitted values are also listed in Table~\ref{tab:fjall}.
As the $f_0(1710)$ solutions have lower $\chi^2_{\rm min}/ndf$,
they are favored over the $f_2(1710)$.

We conclude that the region $1.2~\GeV \le W \le 2.0~\GeV$ 
is too wide to fit in extracting the desired parameters at once.
We therefore fit individual parameters one at a time, keeping
in mind the limitations of this approach.

\begin{center}
\begin{table*}
\caption{Solutions of the $f_J(W \leq 2~\GeV)$ fit.}
\label{tab:fjall}
\begin{tabular}{l|cccc|cc} \hline \hline
Fit & \multicolumn{4}{c|}{$f_0(W \leq 2~\GeV)$ fit}
& \multicolumn{2}{c}{$f_2(W \leq 2~\GeV)$ fit} \\
Sol. & 1 & 2 & 3 & 4 & 1 & 2 \\ \hline
$\chi^2$ ($ndf=580$)
 &  677.2 &  682.3 &  685.4 &  686.7 &  755.6 &  759.6 \\ \hline
$\phi_{a2}$ (deg.)
 &  178.3 &  184.7 &  183.8 &  178.7 &  183.2 &  180.3 \\
Mass$(f_2')$ (MeV$/c^2$)
 & 1527.9 & 1527.2 & 1527.7 & 1526.1 & 1527.9 & 1527.5 \\
$\Gamma_{\rm tot}(f_2')$ (MeV) 
 &  85.5 &  86.3 &  85.8 &  81.5 &  85.5 &  83.5 \\
$\Gamma_{\gamma \gamma}\B(K \bar{K})_{f_2'}$ (eV)
 &  82.8 & 103.8 &  85.8 &  90.0 &  89.0 & 127.1 \\
$\phi_{f2p}$ (deg.)
 &  277  &  250  &  242  & 211  &  251 &  288 \\
\hline
Mass$(f_0)$ (MeV$/c^2$)
 & 1781  & 1780  & 1783  & 1761  & 1793  & 1782  \\
$\Gamma_{\rm tot}(f_0)$ (MeV) 
 &   99  &  110  &   96  &  119  &   93  &  104  \\
$\Gamma_{\gamma \gamma}\B(K \bar{K})_{f_0}$ (eV)
 & 216 &  6.3  &  189 &   10.3 &  89.0 & 127 \\
$\phi_{f0}$ (deg.)
 & 264  &  125  &  265  &   90  &  251  &  288  \\
\hline\hline
\end{tabular}
\end{table*}
\end{center}

\subsubsection{The ``$f_2'(1525)$ fit''}
Based on the above observation,
we first obtain the $f_2'(1525)$ parameters 
by fitting the c.m. energy range $1.15~\GeV \leq W \leq 1.65~\GeV$
and ignoring the contribution of the $f_J(1710)$.
The differential cross section is fit with the
parameterized amplitudes by floating the $f_2'(1525)$ parameters
as well as the relative phase
between the $a_2(1320)$ and $f_2(1270)$.
Hereinafter, this fit is referred to as the ``$f_2'(1525)$ fit.''
The background amplitudes are approximated with linear functions
because the fitting range is rather narrow.
There are thirteen parameters to extract from this fit.

\begin{figure*}
 \centering
  {\epsfig{file=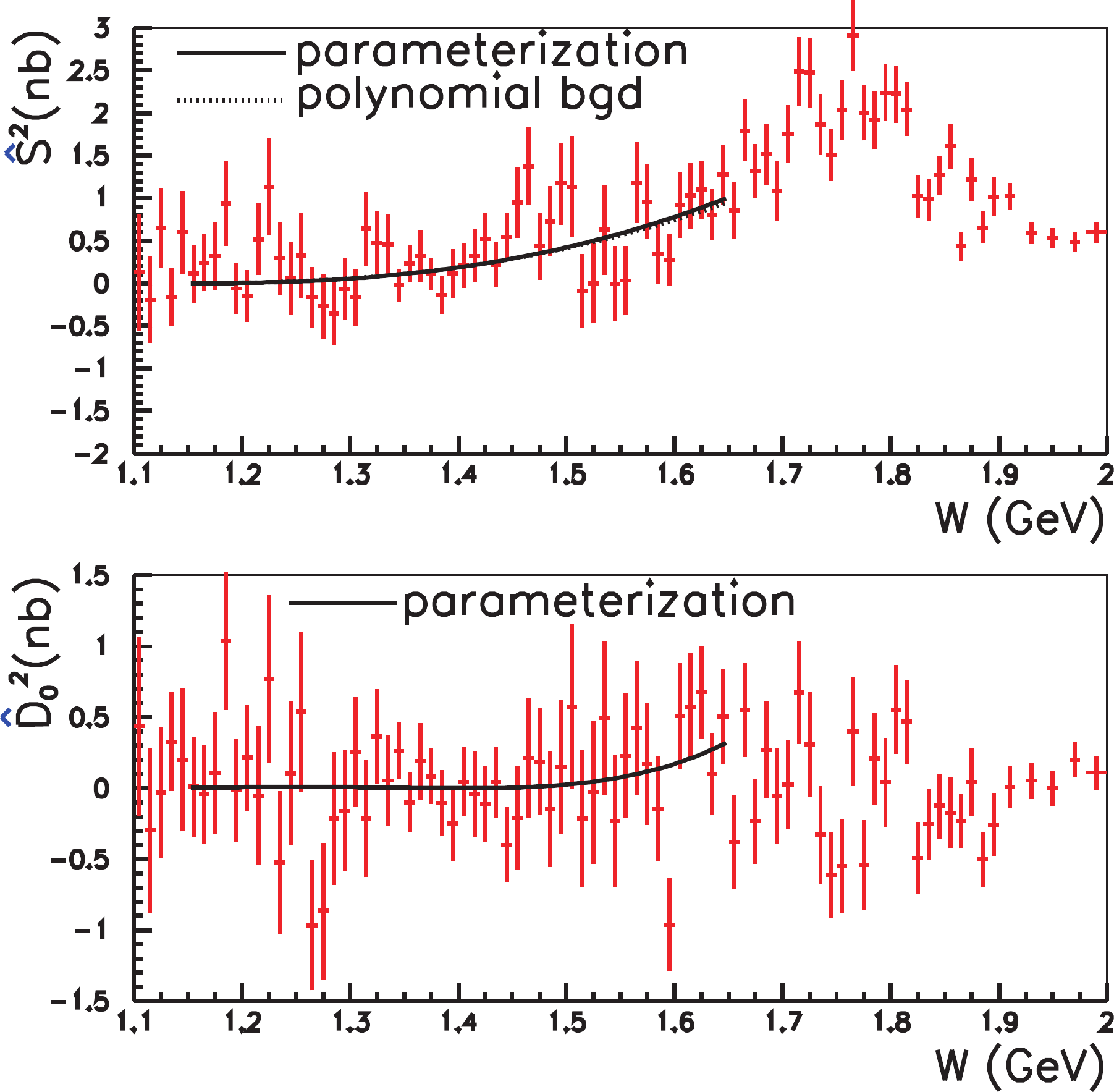,width=50mm}}
  {\epsfig{file=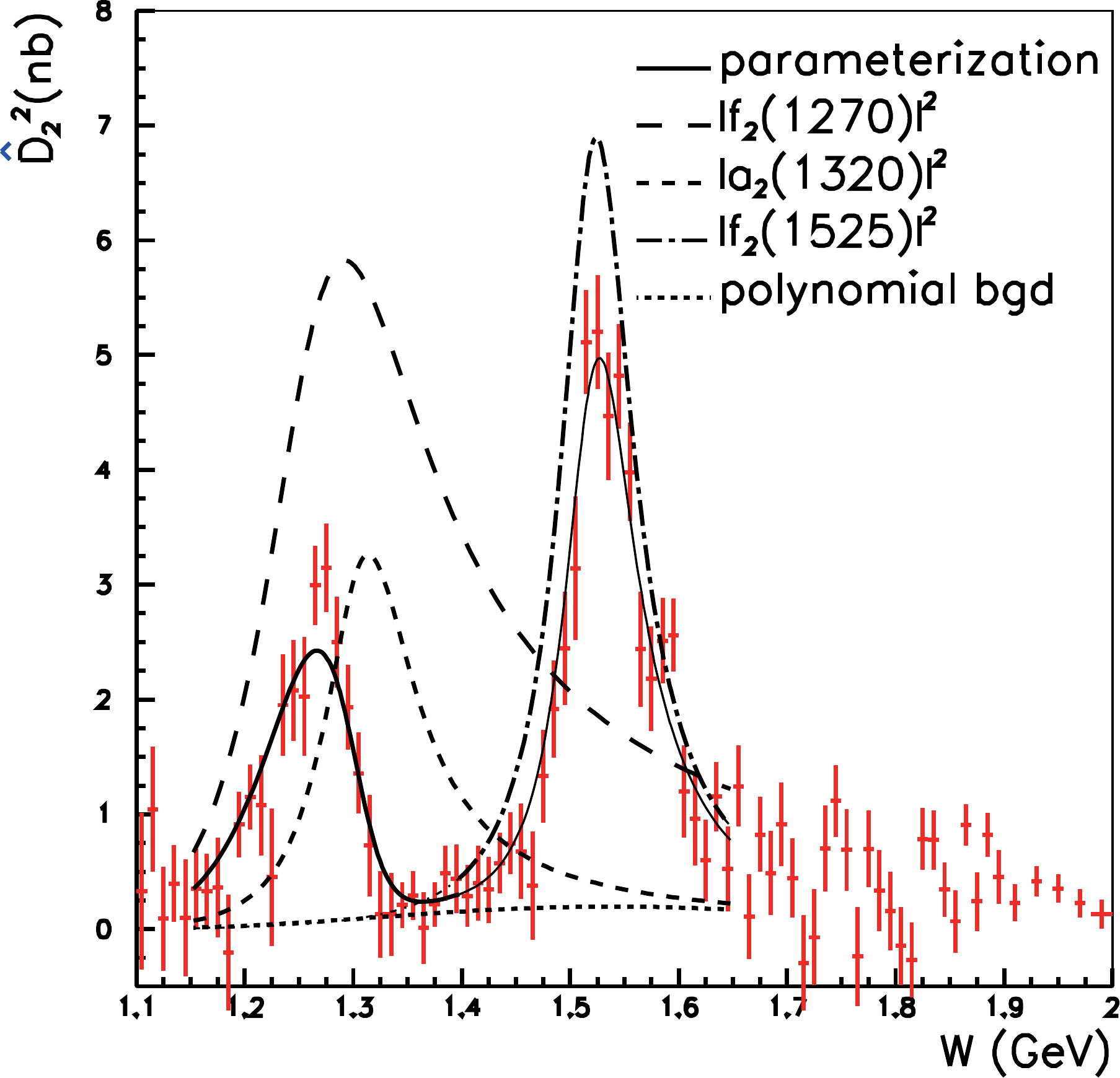,width=50mm}}
  {\epsfig{file=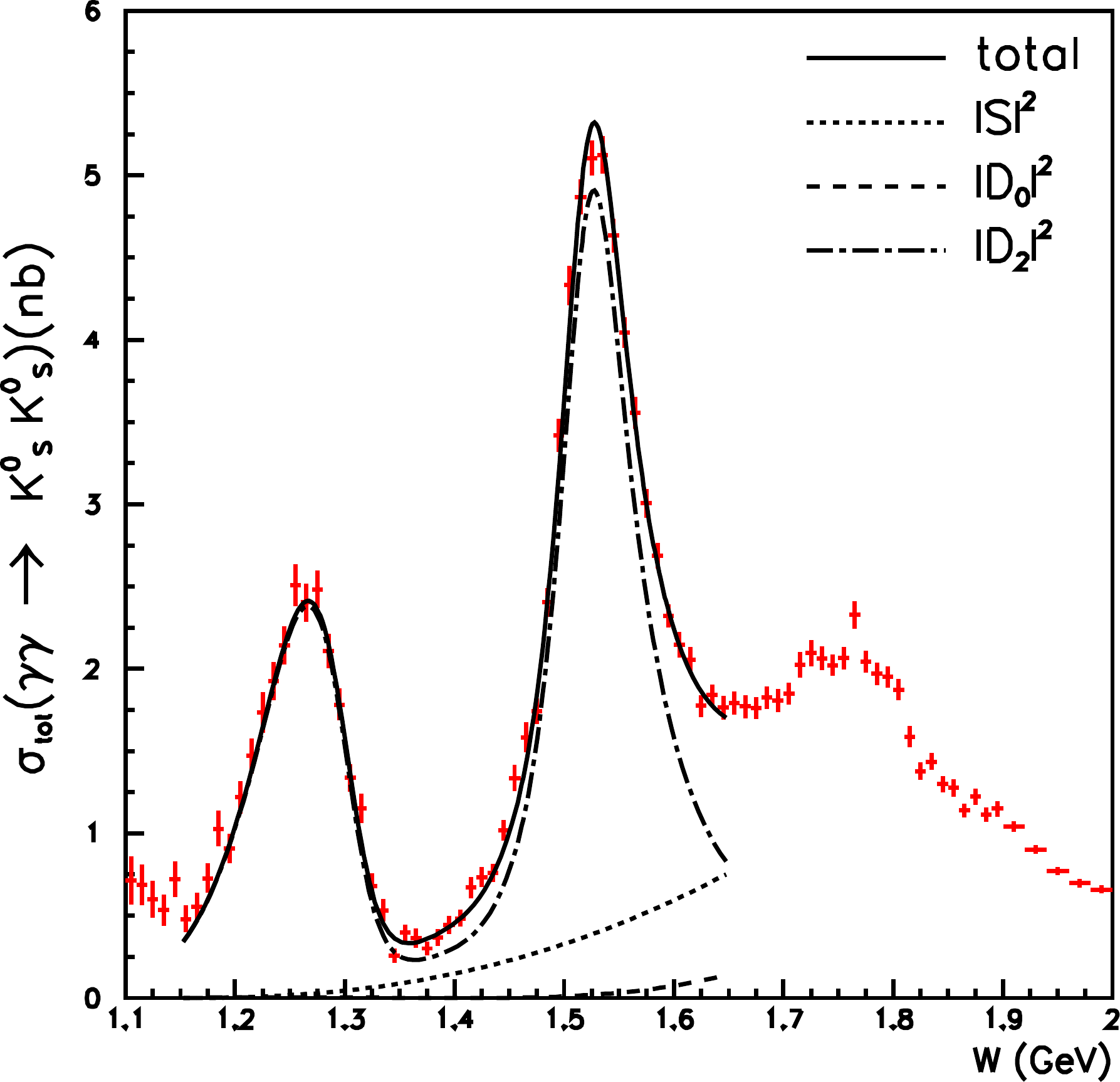,width=50mm}}
 \caption{The solution H of the $f_2'(1525)$ fit
(solid line) superimposed on the spectrum of 
$\hat{S}^2$ (left top), $\hat{D}_0^2$ (left bottom),
$\hat{D}_2^2$ (middle) and
on the integrated cross section (for $|\cos \theta^*| \leq 0.8$) (right).
In the $\hat{D}_2^2$ spectrum, the fitted results of the 
$f_2(1270)$ (long-dashed line),
$a_2(1320)$ (dashed line)
and
$f_2'(1525)$ (dot-dashed line) are also shown together with
the fitted non-resonant background $|B_{D2}|^2$ (dotted line).
In the integrated cross section, the fitted results of 
$|S|^2$ (dotted line), $|D_0|^2$ (dashed line) and
$|D_2|^2$ (dot-dashed line) are also shown.
}
\label{fig:p1sd02t}
\end{figure*}

\begin{figure*}
 \centering
  {\epsfig{file=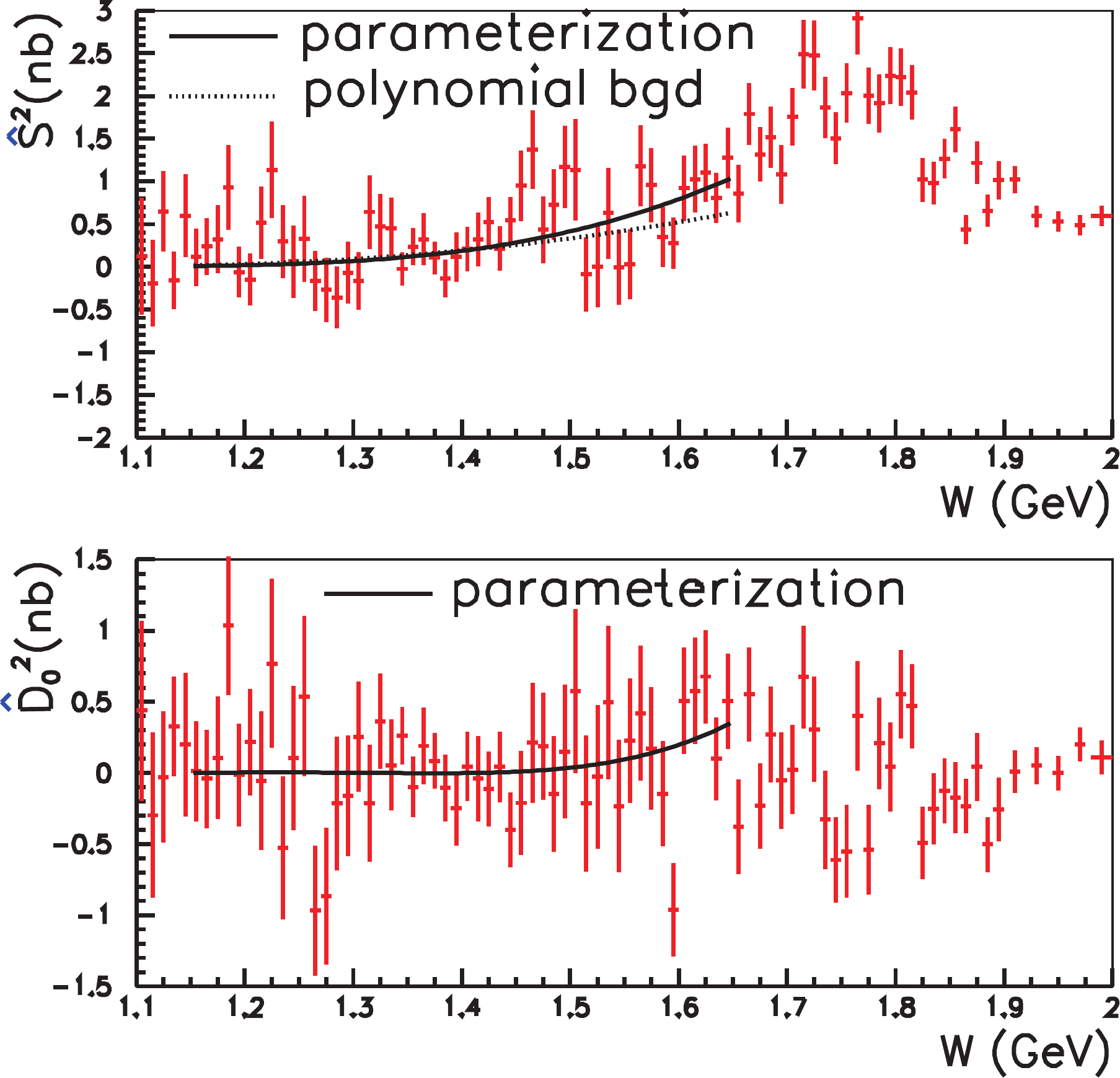,width=50mm}}
  {\epsfig{file=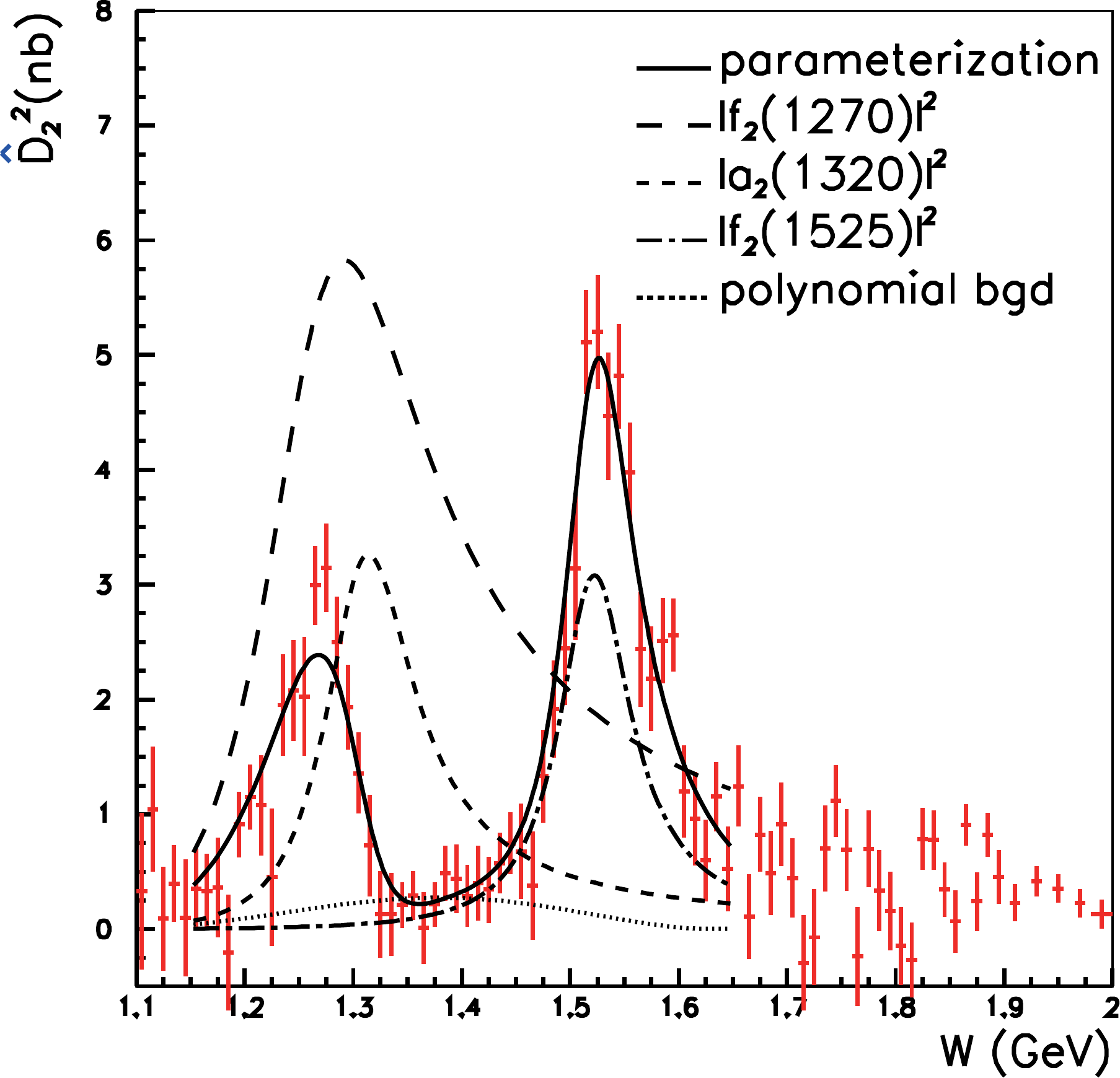,width=50mm}}
  {\epsfig{file=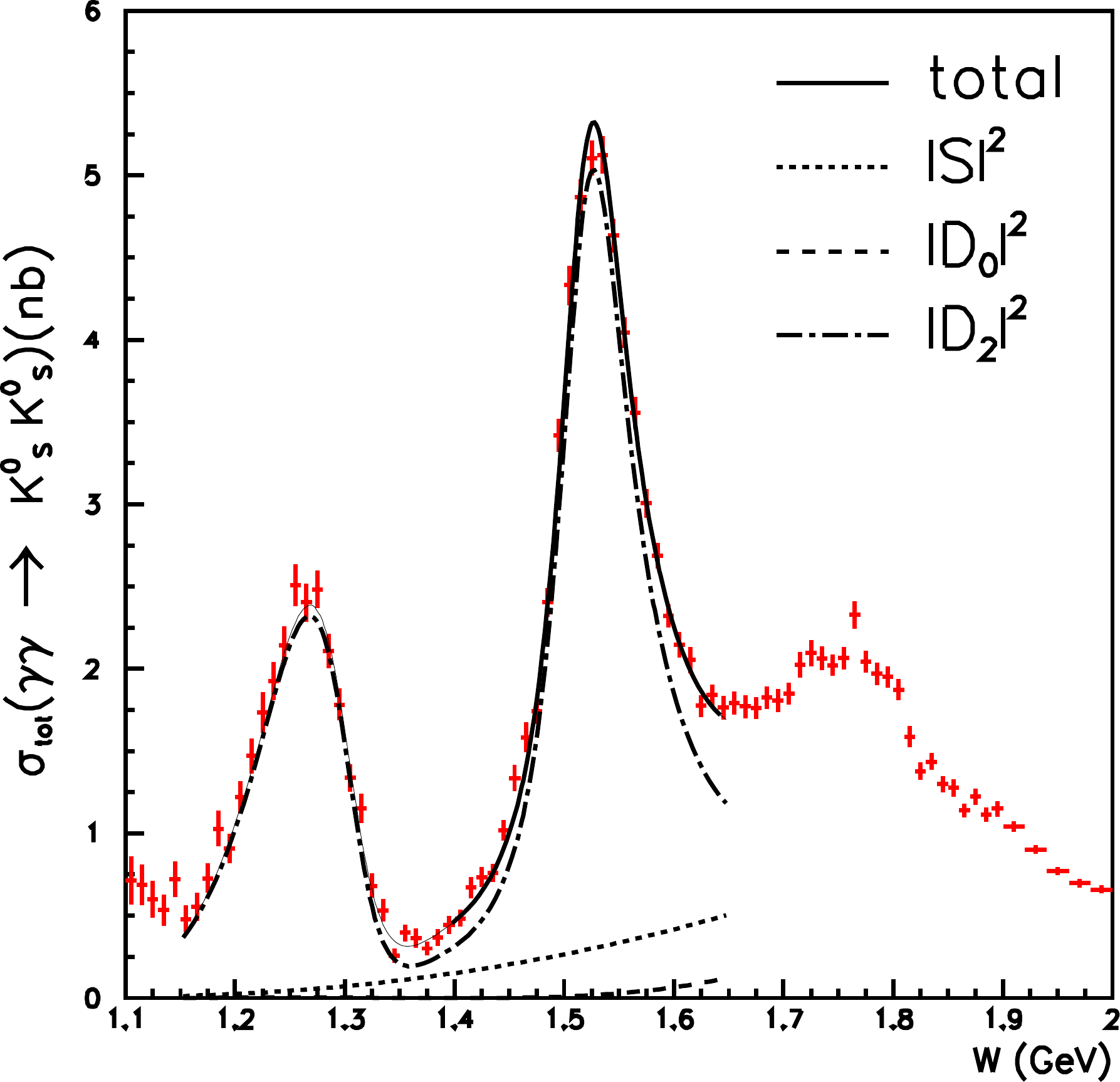,width=50mm}}
 \caption{The solution L of the $f_2'(1525)$ fit
(solid line)  superimposed on the spectrum of 
$\hat{S}^2$ (left top), $\hat{D}_0^2$ (left bottom),
$\hat{D}_2^2$ (middle), and
on the integrated cross section (for $|\cos \theta^*| \leq 0.8$).
See the caption of Fig.~\ref{fig:p1sd02t} for the line 
convention (also shown in the legends).
}
\label{fig:p2sd02t}
\end{figure*}

Two solutions are obtained, both with $\chi^2/ndf = 0.97$.
The main difference between the two solutions is the values of
$\Gamma_{\gamma \gamma}\B(K \bar{K})$ for the $f_2'(1525)$:
113 and 48~eV, with the two solutions referred to as 
H (high) and L (low), respectively.
They correspond to destructive and constructive interference
between the $f_2'(1525)$ and non-resonant $D_2$ background.
The fitted results are shown in Figs.~\ref{fig:p1sd02t}
and \ref{fig:p2sd02t} for the H and L solutions, respectively.
The fitted values are listed and compared
with those of PDG~\cite{pdg2012} in Table~\ref{tab:f2pfit}.
The quoted errors are MINOS statistical errors,
determined by evaluating the parameter values that give
$\chi^2_{\rm min} + 1$ for each variable being studied.
In the fit, all other parameters are floated.
In both solutions, the interference between the $f_2(1270)$
and $a_2(1320)$ is indeed destructive as predicted~\cite{lipkin},
\textit{i.e.}, the measured $\phi_{a2}$ is close to $180^{\circ}$.

We stress that the previous measurements of  
$\Gamma_{\gamma \gamma}\B(K \bar{K})$ for the $f_2'(1525)$~\cite{L3}
assumed no interference.
In order to check the consistency with past experiments,
an incoherent fit is also performed, 
where we replace 
$|D_2 Y_2^2|^2$ with $|(D_2 - A_{f_2'(1525)}e^{i \phi_{f2p}}) Y_2^2|^2 
+ |A_{f_2'(1525)} Y_2^2|^2$ in Eq.~(\ref{eqn:diff}).
The obtained value of $\Gamma_{\gamma \gamma}\B(K \bar{K})$ 
is $79.1 \pm 1.4$~eV, which is consistent with $76 \pm 6 \pm 11$~eV
reported by L3~\cite{L3}, 
$110^{+30}_{-20} \pm 20$~eV by CELLO~\cite{cello},
$100^{+40+30}_{-30-20}$~eV by PLUTO~\cite{pluto}
and $110 \pm 20 \pm 40$~eV by TASSO~\cite{tasso1}.
The results of our fits are also shown in Table~\ref{tab:f2pfit}.

\begin{center}
\begin{table*}
\caption{Parameters obtained from the $f_2'(1525)$ fit and incoherent fit.
For the H and L solutions, the first error is statistical and the second
systematic (itemized in Table~\ref{tab:sysf2p}).
The parameters where the H and L solutions are combined
are also shown (explained in Sec.~\ref{sub:finloww}).} 
\label{tab:f2pfit}
\begin{tabular}{l|ccc|c|c} \hline \hline
Parameter  & Solution H & Solution L & H, L combined & Incoherent fit 
& PDG~\cite{pdg2012} \\ \hline
$\chi^2/ndf$ & 375.09/387 & 375.22/387 & -- & 406.6/388 & -- \\
\hline
&&&&&\\[-10pt]
$\phi_{a_2(1320)} $ (deg.)
& $178.1 ^{+1.7+6.7}_{-1.3-12.5}$ 
& $172.6 ^{+1.3+6.7}_{-1.0-3.1}$ 
& $172.6^{+6.0+12.2}_{-0.7-7.0}$ 
& $173.6^{+1.3}_{-1.4}$& -- \\
Mass$(f_2'(1525))$ (MeV/$c^2$) 
& $1526.1^{+0.9+2.9}_{-1.0-2.8}$ 
& $1524.3^{+1.0+1.6}_{-0.9-1.1}$ 
& $1525.3^{+1.2+3.7}_{-1.4-2.1}$ 
& $1530.7 \pm 0.4$ & $1525 \pm 5$ \\
$\Gamma_{\rm tot}(f_2'(1525))$ (MeV) 
& $83.4^{+1.9+2.0}_{-1.7-3.4}$ 
& $81.8^{+2.3+4.4}_{-2.0-0.9}$ 
& $82.9^{+2.1+3.3}_{-2.2-2.0}$ 
& $82.7 \pm 1.4$ & $73^{+6}_{-5}$ \\
$\Gamma_{\gamma \gamma}\B(K \bar{K})(f_2'(1525))$ (eV) 
& $113^{+25+43}_{-28-77}$ 
& $48 \pm 4 {}^{+33}_{-10}$
& $48^{+67}_{-8}{}^{+108}_{-12}$
& $79.1 \pm 1.4$ & $72 \pm 7$\\
\hline\hline
\end{tabular}
\end{table*}
\end{center}

The following sources of systematic uncertainties for the fitted
parameters are considered: dependence on the fit region, 
normalization errors of the differential cross section,
assumptions on the background amplitudes and
assumed parameters of the $f_2(1270)$ and $a_2(1320)$.
In each study, a fit is performed that allows all the parameters to float;
the differences of the fitted parameters from the nominal values
are quoted as systematic uncertainties for both solutions, H and L.

Two fitting regions shifted by $\pm 0.05$~GeV (10\% of the $W$-range),
are used to estimate the systematics associated with the fitting range.
The systematic uncertainties associated with
the normalization are separated into two groups:
one from the overall normalization ($\pm 4.0$\%)
and the other from the distortion of the spectra in both
$|\cos \theta^*|$ and $W$.
To estimate the uncertainty associated with
the overall normalization, fits are performed
by shifting the cross sections coherently by $\pm 4$\%.
For point-by-point normalization, fits are performed 
by shifting the cross section by
$\pm |d \sigma / d \Omega| \times 
\sigma_{\epsilon(W, |\cos \theta^*|)}$, 
where $\sigma_{\epsilon}$ is the relative uncertainty of the efficiency
(referred to as Efficiency in Table~\ref{tab:sysf2p}).
For the spectral distortion studies, 
the differential cross sections are shifted by 
$ \pm 0.1 \times |d \sigma / d \Omega| \times (|\cos \theta^*| - 0.4))$
(referred to as $|\cos \theta^*|$ bias)
and $ \pm 0.08 \times |d \sigma / d \Omega| \times (W - 1.4~\GeV)$
(referred to as $W$ bias).
We use the absolute value of $d \sigma / d \Omega$
because some of the central values for measured differential 
cross sections are negative due to background subtraction.

For studies of background (BG) amplitudes, each background wave 
is approximated by a constant or a parabola. 
The value of $r_R$ is changed by $\pm 0.03~(\GeV/c)^{-1}$
according to Refs.~\cite{mori1, mori2}.
Finally, the parameters of the $f_2(1270)$ and $a_2(1320)$ 
are changed one by one by their uncertainties shown in PDG~\cite{pdg2012}.

The total systematic uncertainties are calculated by adding the individual 
uncertainties in quadrature.
The resulting systematic uncertainties are summarized in 
Table~\ref{tab:sysf2p}.
In some of our studies, 
the value of $\Gamma_{\gamma \gamma} \B(K \bar{K})$ for the $f_2'(1525)$
fluctuates between the H and L solutions.
The obtained results for the relative phase between the $a_2(1320)$ and
$f_2(1270)$ and parameters of the $f_2'(1525)$ are given
in Table~\ref{tab:f2pfit}.

\begin{table*}[h]
\caption{Systematic uncertainties for the $f_2'(1525)$ fit.
The left (right) number in each row for each observable indicates
a positive (negative) deviation from the nominal values.}
\label{tab:sysf2p}
\begin{center}
\begin{tabular}{l|rr|rr|rr|rr|rr|rr|rr|rr} \hline \hline 
& \multicolumn{8}{c|}{Solution H} & \multicolumn{8}{c}{Solution L} 
 \\ \cline{2-17}
& & & \multicolumn{6}{c|}{$f_2'(1525)$} & 
& & \multicolumn{6}{c}{$f_2'(1525)$} \\ \cline{4-9} \cline{12-17} 
Source & \multicolumn{2}{c|}{$\phi_{a2}$} 
& \multicolumn{2}{c|}{Mass} & 
\multicolumn{2}{c|}{$\Gamma_{\rm tot}$}
& \multicolumn{2}{c|}{$\Gamma_{\gamma \gamma} \B(K \bar{K})$}
& \multicolumn{2}{c|}{$\phi_{a2}$} 
& \multicolumn{2}{c|}{Mass} & 
\multicolumn{2}{c|}{$\Gamma_{\rm tot}$}
& \multicolumn{2}{c}{$\Gamma_{\gamma \gamma} \B(K \bar{K})$} \\
& \multicolumn{2}{c|}{(deg.)} & \multicolumn{2}{c|}{(MeV/$c^2$)}
 & \multicolumn{2}{c|}{~~~(MeV)~~~} & \multicolumn{2}{c|}{(eV)}
& \multicolumn{2}{c|}{(deg.)} & \multicolumn{2}{c|}{(MeV/$c^2$)}
 & \multicolumn{2}{c|}{~~~(MeV)~~~} & \multicolumn{2}{c}{(eV)}
\\ \hline 

$W$-range & 6.1 & $-0.3$ & 2.9 & 0.0 & 1.5 & 0.0 & 32 & $ 0$ & 1.7 & $-1.1$ & 0.7 & $-0.3$ & 3.2 & 0.0 & 0 & $-2$ \\
$W$ bias & 0.0 & $-3.0$ & 0.0 & $-0.2$ & 0.0 & $-0.1$ & 2 & $ 0$ & 0.3 & $-0.3$ & 0.1 & 0.0 & 0.2 & 0.0 & 0 & $ 0$ \\
Efficiency & 2.9 & $-2.8$ & 0.0 & $-0.2$ & 0.1 & $-0.2$ & 0 & $-4$ & 2.4 & $-1.3$ & 0.1 & $-0.1$ & 0.9 & 0.0 & 2 & $-6$ \\
Overall normalization & 1.0 & $-1.0$ & 0.0 & $-0.1$ & 0.1 & $-0.1$ & 1 & $-1$ & 0.9 & $-0.9$ & 0.1 & $-0.1$ & 0.1 & 0.0 & 2 & $-3$ \\
$|\cos \theta^*|$ bias & 0.1 & $-0.1$ & 0.1 & $-0.2$ & 0.7 & $-0.8$ & 1 & $-4$ & 0.3 & $-0.2$ & 0.2 & $-0.1$ & 0.8 & $-0.6$ & 1 & $-1$ \\ \hline
$B_S$ & 0.8 & $-1.8$ & 0.0 & $-1.2$ & 1.1 & $-1.3$ & 28 & $-29$ & 1.5 & $-1.2$ & 0.0 & $-0.4$ & 1.9 & 0.0 & 0 & $-3$ \\
$B_{D0}$ & 0.0 & $-3.1$ & 0.0 & $-0.8$ & 0.2 & $-1.9$ & 0 & $-25$ & 0.0 & $-0.2$ & 0.2 & 0.0 & 0.0 & $-0.6$ & 0 & $ 0$ \\
$B_{D2}$ & 0.0 & $-2.8$ & 0.0 & $-1.5$ & 0.0 & $-2.1$ & 0 & $-21$ & 5.0 & 0.0 & 1.3 & $-0.9$ & 0.6 & 0.0 & 32 & $-1$ \\ \hline
Mass($f_2(1270)$) & 0.0 & $-4.0$ & 0.0 & $-0.3$ & 0.0 & $-0.2$ & 2 & $-3$ & 1.1 & $-1.1$ & 0.1 & 0.0 & 0.2 & 0.0 & 1 & $-1$ \\
$\Gamma_{\rm tot}(f_2(1270))$ & 0.0 & $-2.9$ & 0.0 & $-0.1$ & 0.1 & 0.0 & 5 & $ 0$ & 0.2 & $-0.2$ & 0.2 & $-0.1$ & 0.2 & 0.0 & 1 & $-1$ \\
$\B(\gamma \gamma)(f_2(1270))$ & 0.0 & $-4.7$ & 0.1 & $-0.5$ & 0.3 & $-0.4$ & 0 & $-1$ & 1.8 & $-1.6$ & 0.3 & $-0.2$ & 1.3 & 0.0 & 3 & $-4$ \\
$r_R$ & 0.0 & $-3.0$ & 0.0 & $-0.2$ & 0.0 & $-0.2$ & 1 & $-1$ & 0.1 & $-0.1$ & 0.1 & 0.0 & 0.0 & 0.0 & 0 & $ 0$ \\ \hline
Mass($a_2(1320)$) & 0.0 & $-3.8$ & 0.0 & $-0.2$ & 0.0 & $-0.3$ & 0 & $-4$ & 0.6 & $-0.8$ & 0.1 & 0.0 & 0.2 & $-0.1$ & 0 & $ 0$ \\
$\Gamma_{\rm tot}(a_2(1320))$ & 0.0 & $-5.9$ & 0.0 & $-1.6$ & 0.0 & $-0.8$ & 3 & $-63$ & 0.2 & $-0.1$ & 0.3 & $-0.2$ & 0.4 & $-0.1$ & 0 & $-1$ \\
$\B(\gamma\gamma)(a_2(1320)$ & 0.0 & $-3.5$ & 0.0 & $-0.3$ & 0.0 & 0.0 & 6 & $-2$ & 0.7 & $-0.2$ & 0.1 & 0.0 & 1.1 & $-0.1$ & 1 & $-3$ \\ \hline
Total & ~~6.9 & $-12.5$ & ~~2.9 & $-2.7$ & ~~2.0 & $-3.4$ & ~~43 & ~$-77$ & ~6.5 & ~$-3.1$ & ~~1.6 & ~$-1.1$ & ~~4.3 & ~$-0.9$ & ~~~33 & ~~$-10$ \\
\hline \hline 
\end{tabular}
\end{center}
\end{table*}

\subsubsection{A fit including the $f_J(1710)$}
We fit the region $1.2~\GeV \leq W \leq 2.0~\GeV$ by
fixing the parameters of the $f_2'(1525)$ and $\phi_{a2}$ 
to either the H or L solution, and by including the contribution of the
$f_0(1710)$ (coined the ``$f_0(1710)$ fit'').
The background amplitude is assumed to be a second-order polynomial,
whose parameters are floated in the fit.

\begin{figure*}
 \centering
  {\epsfig{file=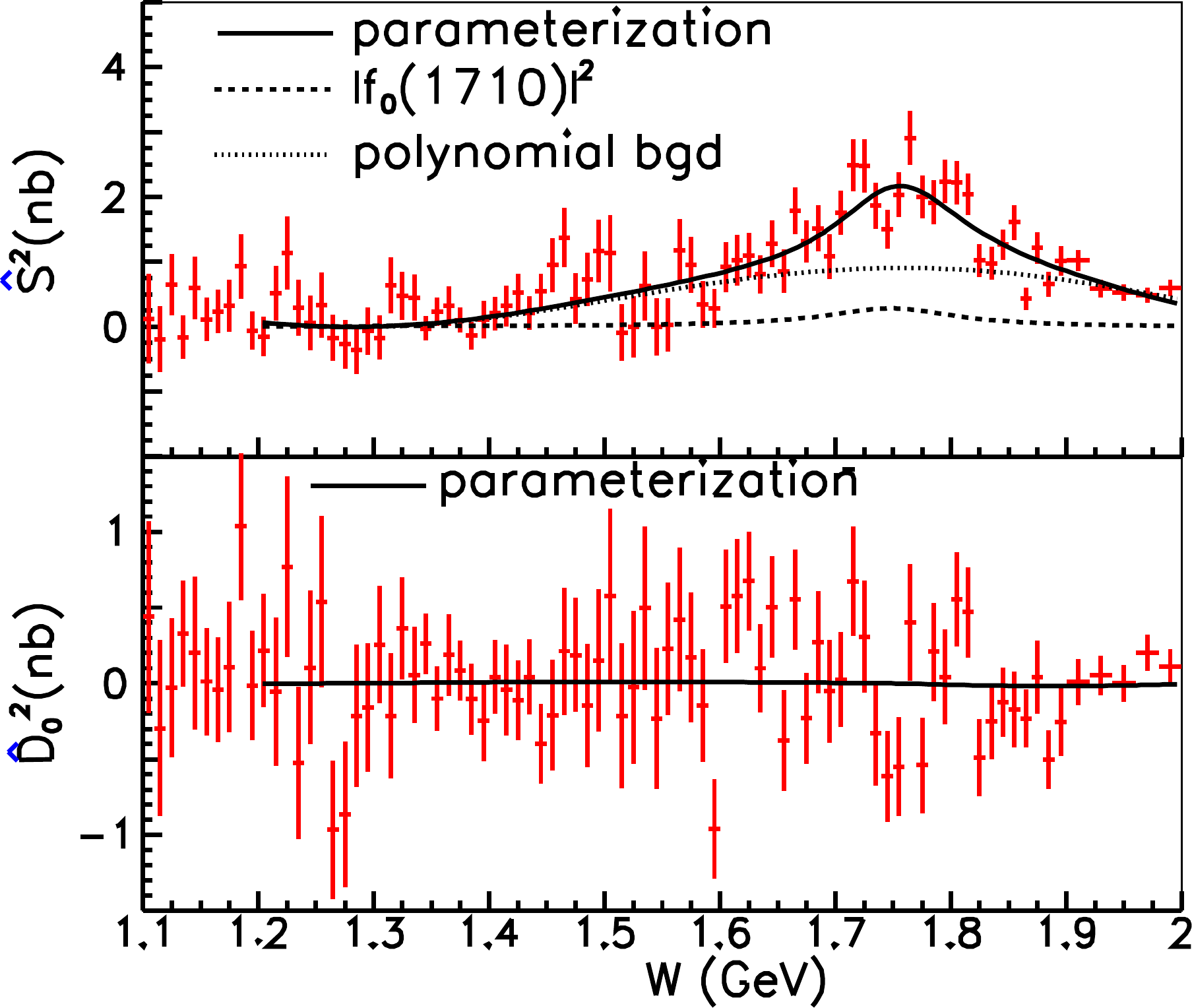,width=50mm}}
  {\epsfig{file=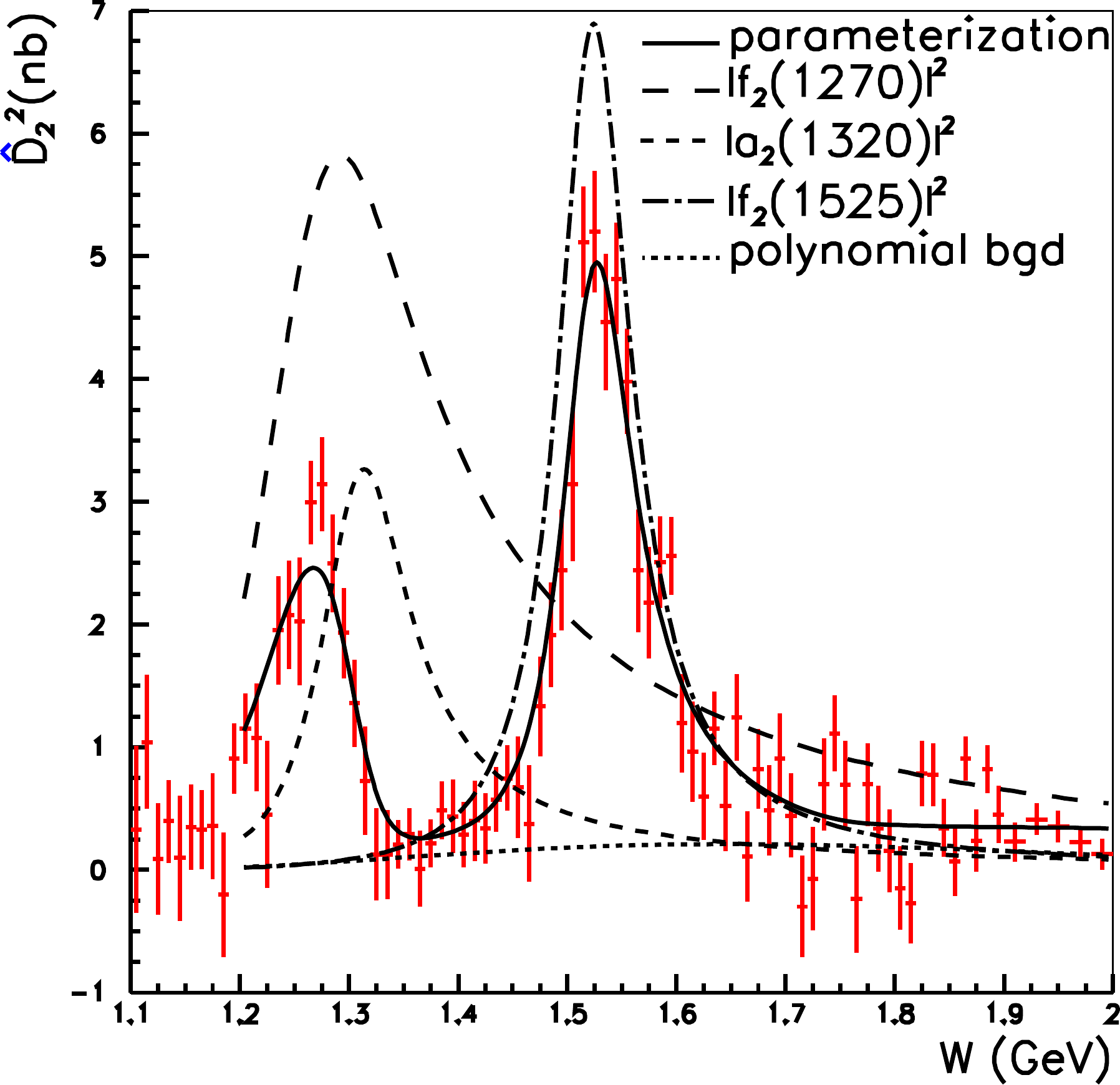,width=50mm}} 
  {\epsfig{file=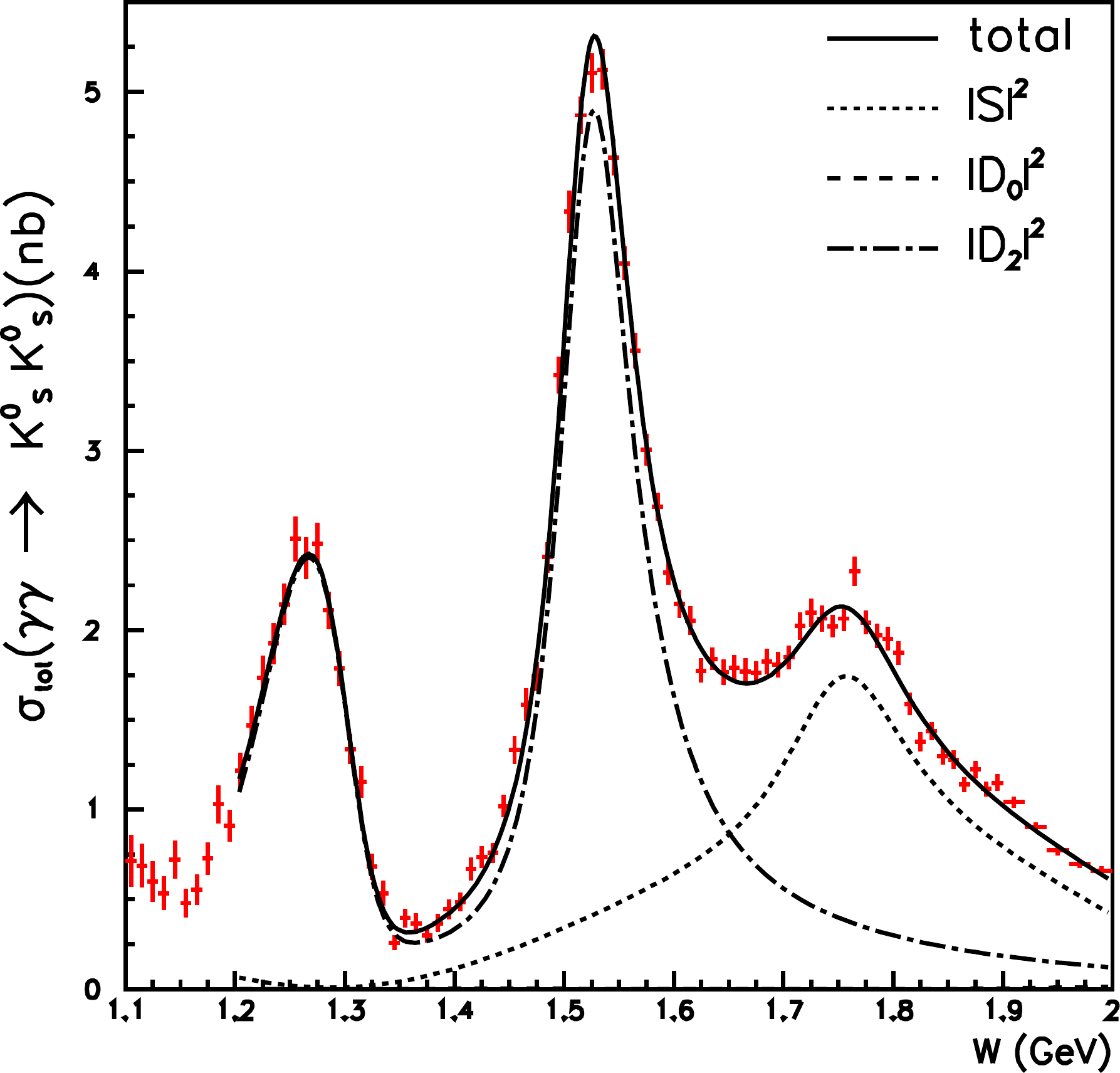,width=50mm}}
\caption{The fit-H of the $f_0(1710)$ fit
(solid line) superimposed on the spectrum of 
$\hat{S}^2$ (left top), $\hat{D}_0^2$ (left bottom),
$\hat{D}_2^2$ (middle) and
on the integrated cross section (for $|\cos \theta^*| \leq 0.8$) (right).
Shown in the $\hat{S}^2$ ($\hat{D}_2^2$) spectrum are the fitted results
of the $f_0(1710)$ (dashed line) and non-resonant background 
$|B_S|^2$ (dotted line)
($f_2(1270)$ (long-dashed line),
$a_2(1320)$ (dashed line),
$f_2'(1525)$ (dot-dashed line) and $|B_{D2}|^2$ (dotted line)).
In the integrated cross section, the fitted results of 
$|S|^2$ (dotted line), $|D_0|^2$ 
(dashed line, not visible) and
$|D_2|^2$ (dot-dashed line) are also shown.
}
\label{fig:f0Hsd02t}
\end{figure*}

\begin{figure*}
 \centering
  {\epsfig{file=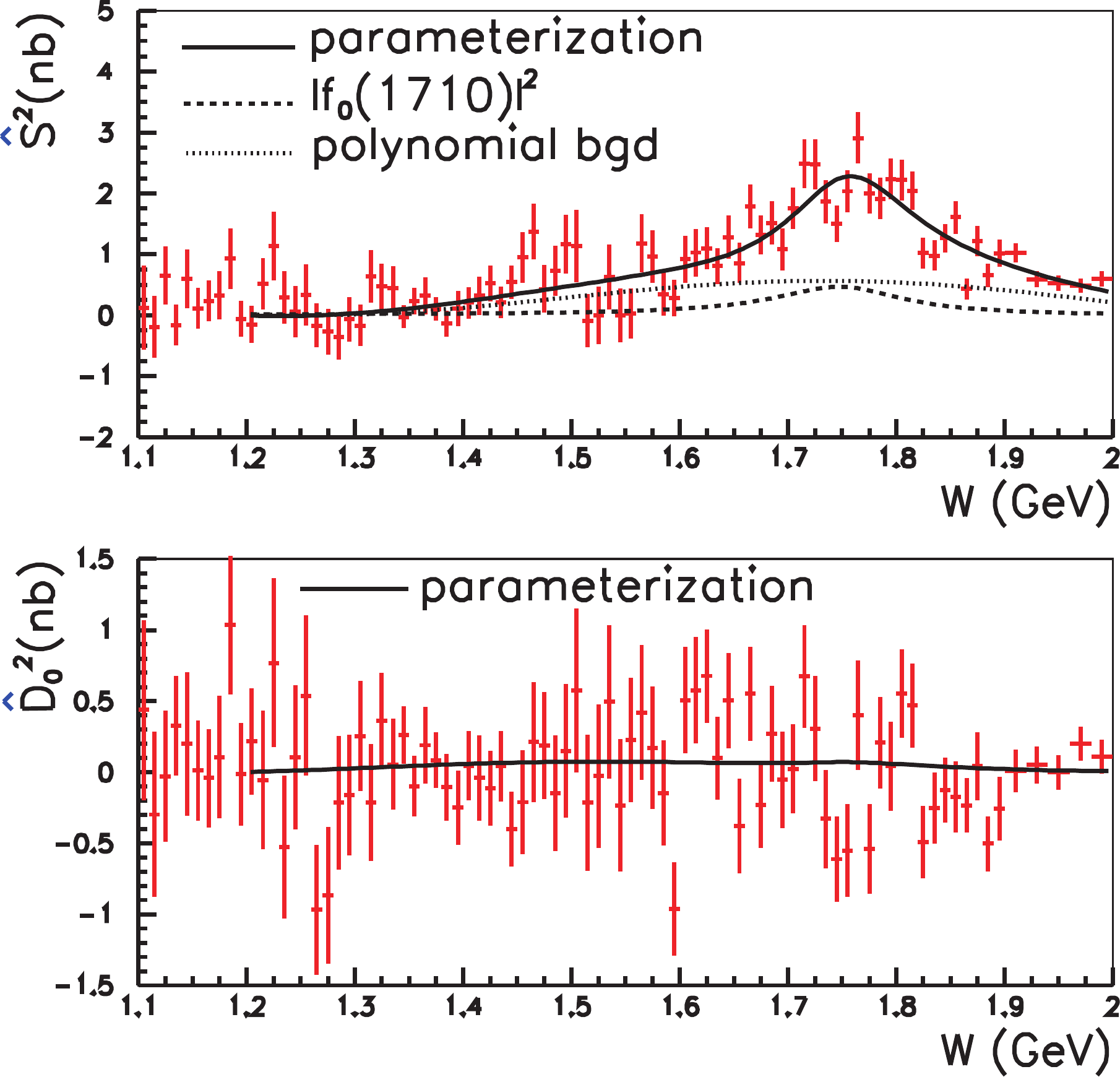,width=50mm}}
  {\epsfig{file=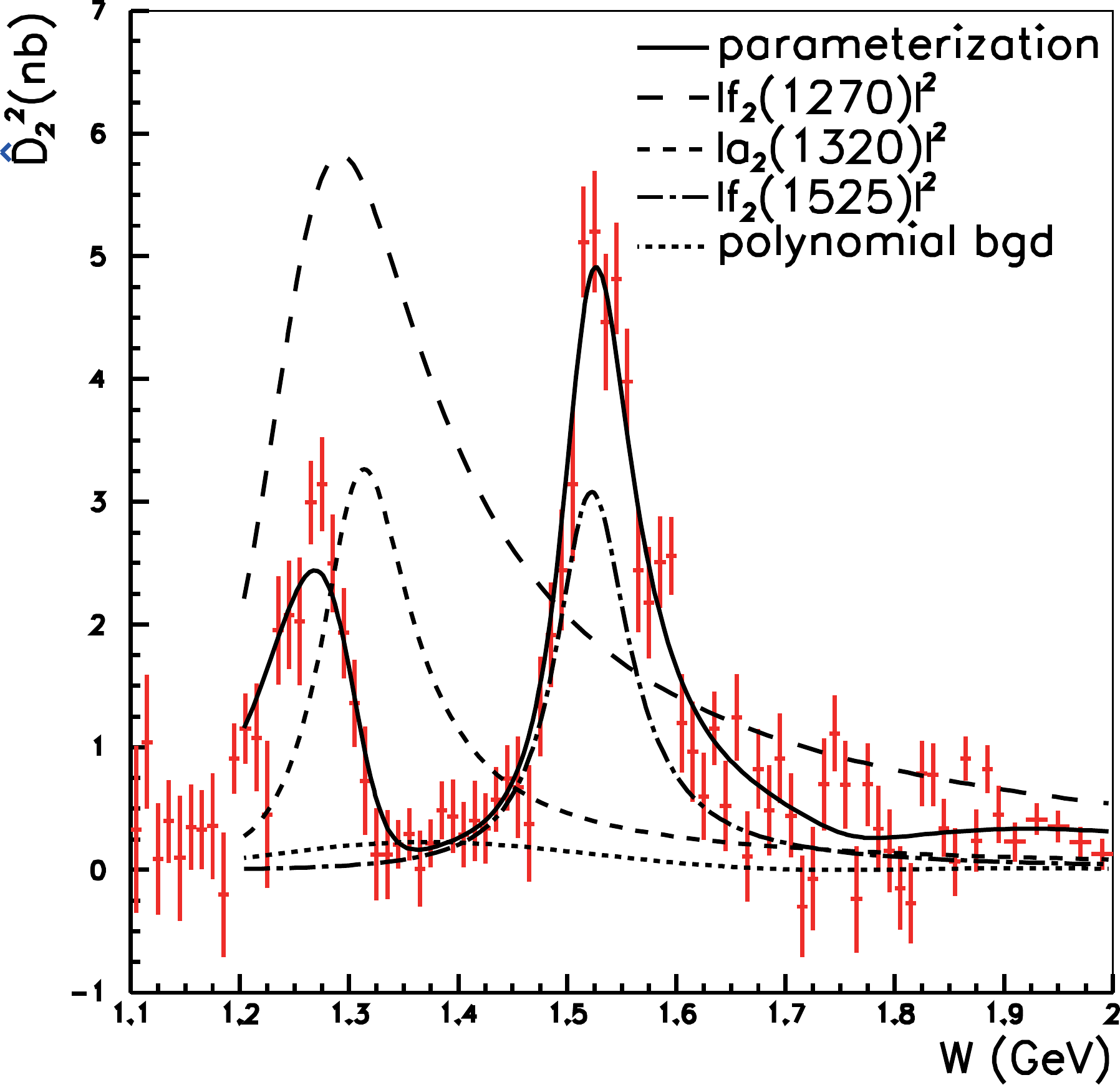,width=50mm}} 
  {\epsfig{file=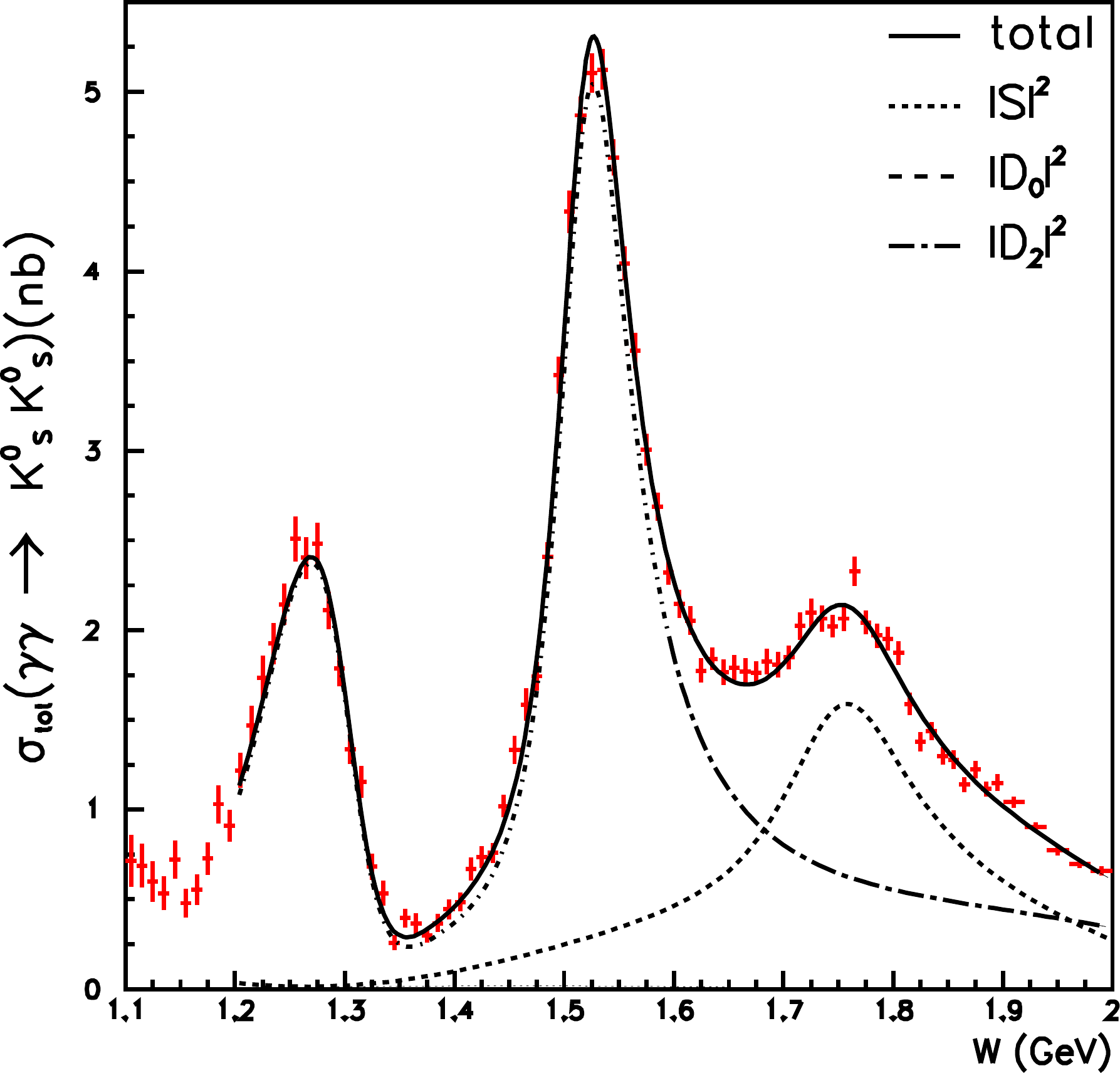,width=50mm}}
\caption{The fit-L of the $f_0(1710)$ fit
(solid line) superimposed on the spectrum of 
$\hat{S}^2$ (left top), $\hat{D}_0^2$ (left bottom)),
$\hat{D}_2^2$ (middle) and  
on the integrated cross section (for $|\cos \theta^*| \leq 0.8$) (right).
See the caption of Fig.~\ref{fig:f0Hsd02t} for the line 
convention (also shown in the legends).
}
\label{fig:f0Lsd02t}
\end{figure*}

\begin{figure}
 \centering
  {\epsfig{file=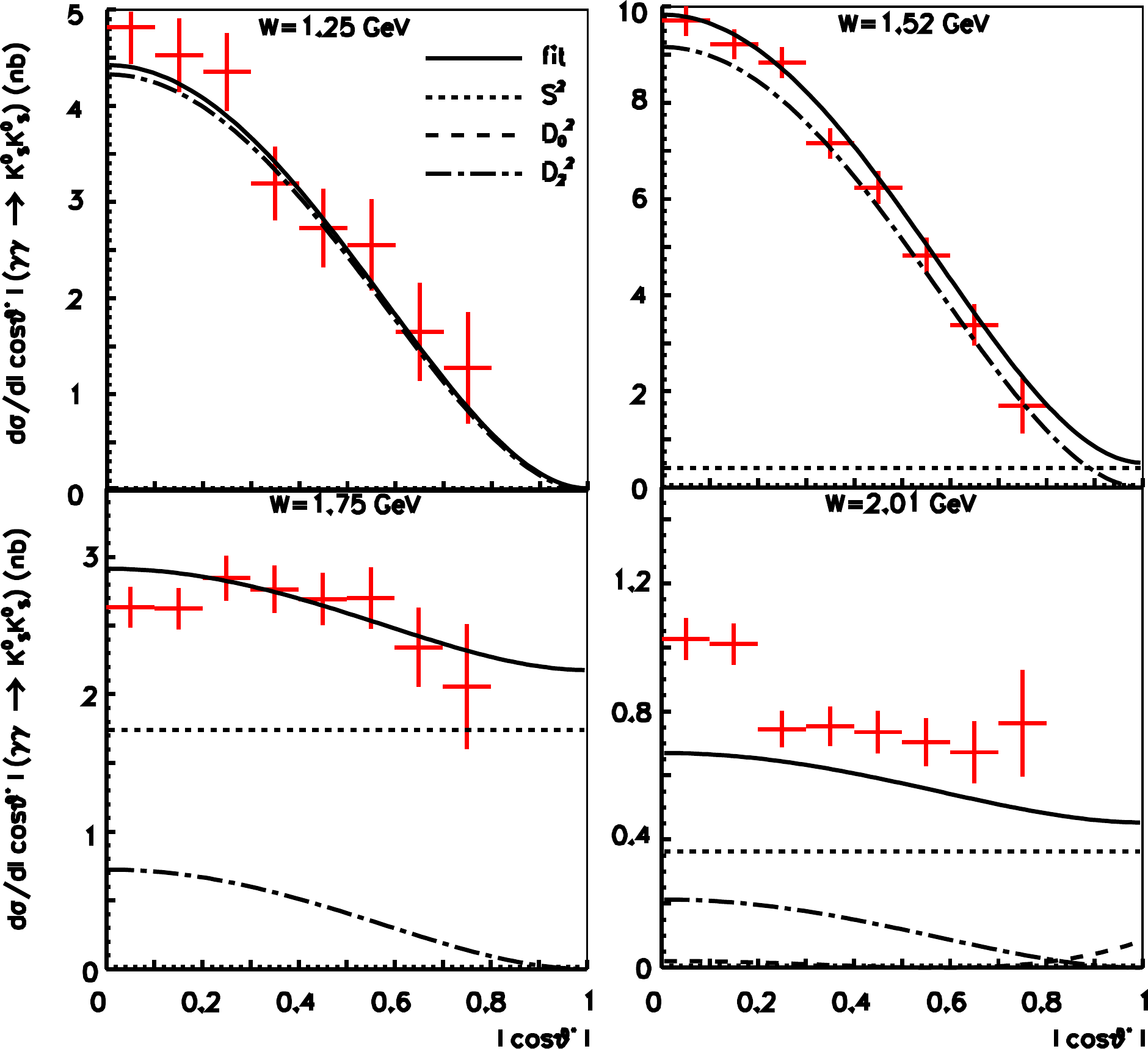,width=65mm}}
\caption{The fit-H of the $f_0(1710)$ fit
(solid line) on the differential cross section for
selected $W$ bins.
Contributions from $|S|^2$ (dotted line), $|D_0|^2$ (dashed line)
and $|D_2|^2$ (dot-dashed line) are also shown.
}
\label{fig:df0Hsel}
\end{figure}

\begin{figure}
 \centering
  {\epsfig{file=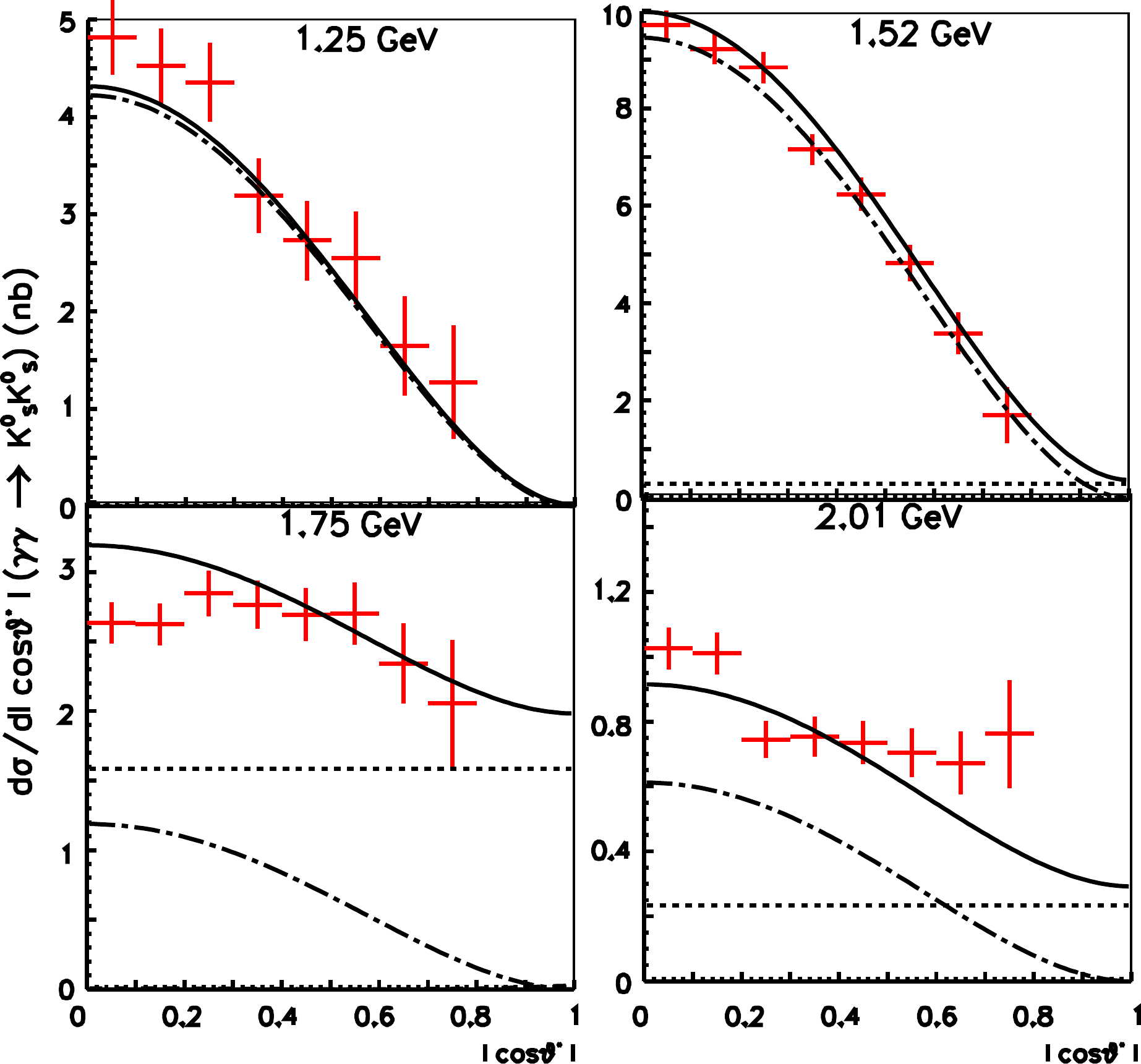,width=65mm}}
\caption{The fit-L of the $f_0(1710)$ fit
(solid line) superimposed on the differential cross section for
selected $W$ bins.
Contributions from $|S|^2$ (dotted line), $|D_0|^2$ (dashed line)
and $|D_2|^2$ (dot-dashed line) are also shown.
}
\label{fig:df0Lsel}
\end{figure}

A unique solution is obtained for each of the H and L solutions
(named ``fit-H'' and ``fit-L,'' where H and L stand 
for the H and L solutions of the $f_2'(1525)$ fit, respectively).
These solutions are summarized in Table~\ref{tab:fj17}.
Figures~\ref{fig:f0Hsd02t} and \ref{fig:f0Lsd02t} 
show the fitted results for fit-H and fit-L, respectively. 
Figures~\ref{fig:df0Hsel} and \ref{fig:df0Lsel} 
show fit-H and fit-L solutions superimposed on the 
differential cross section for selected $W$ bins.

\begin{figure*}
 \centering
  {\epsfig{file=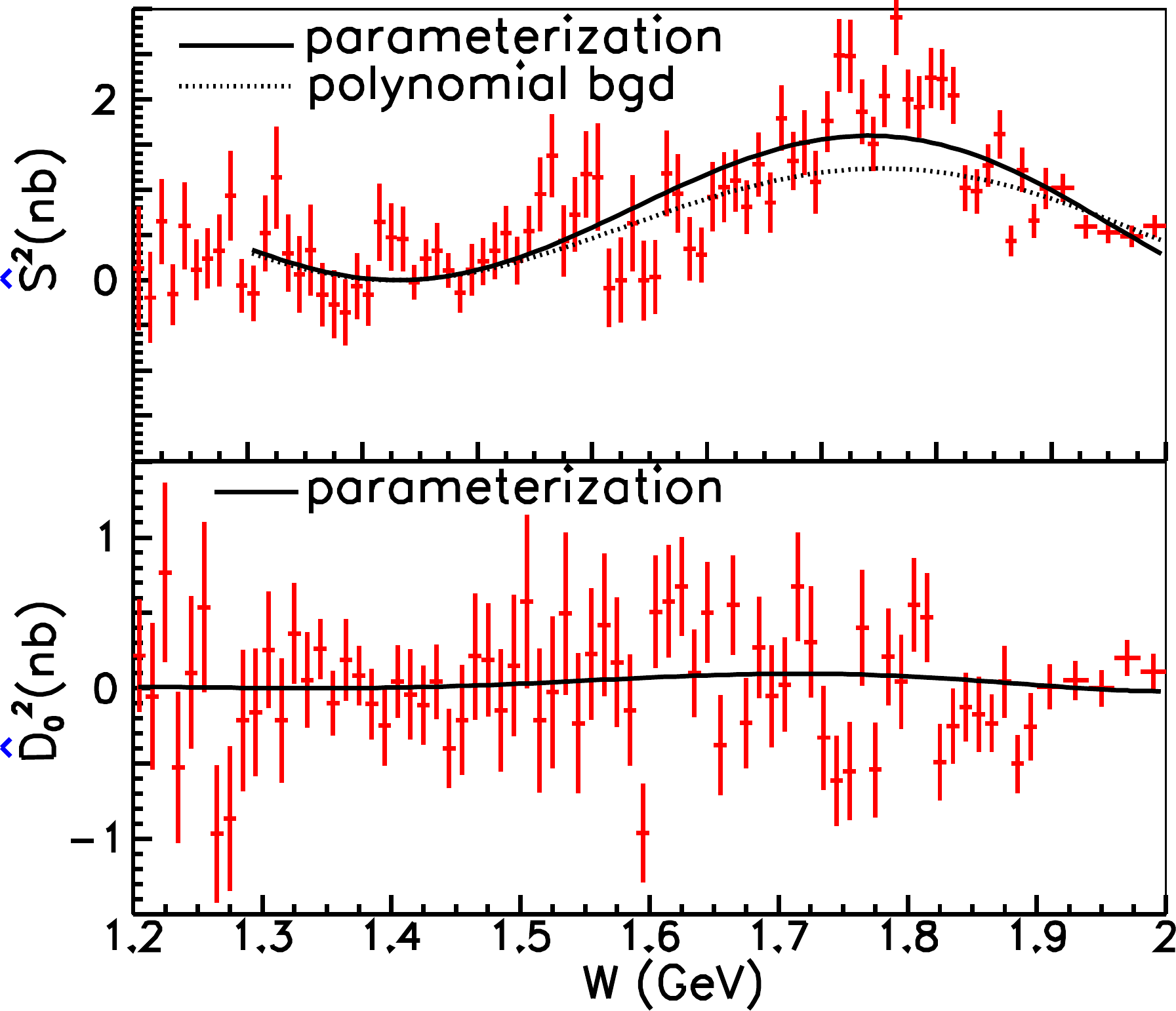,width=50mm}}
  {\epsfig{file=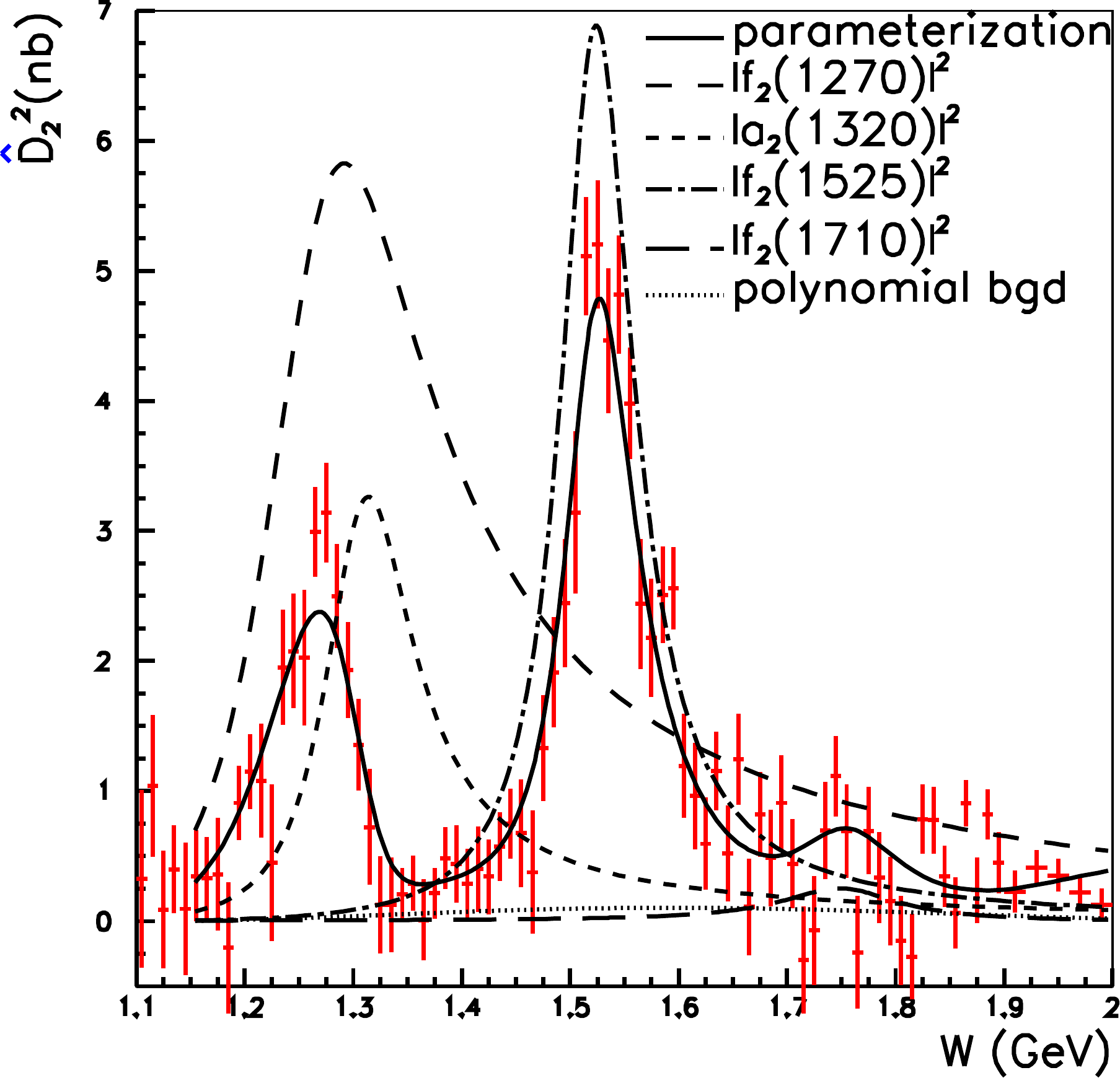,width=50mm}}
  {\epsfig{file=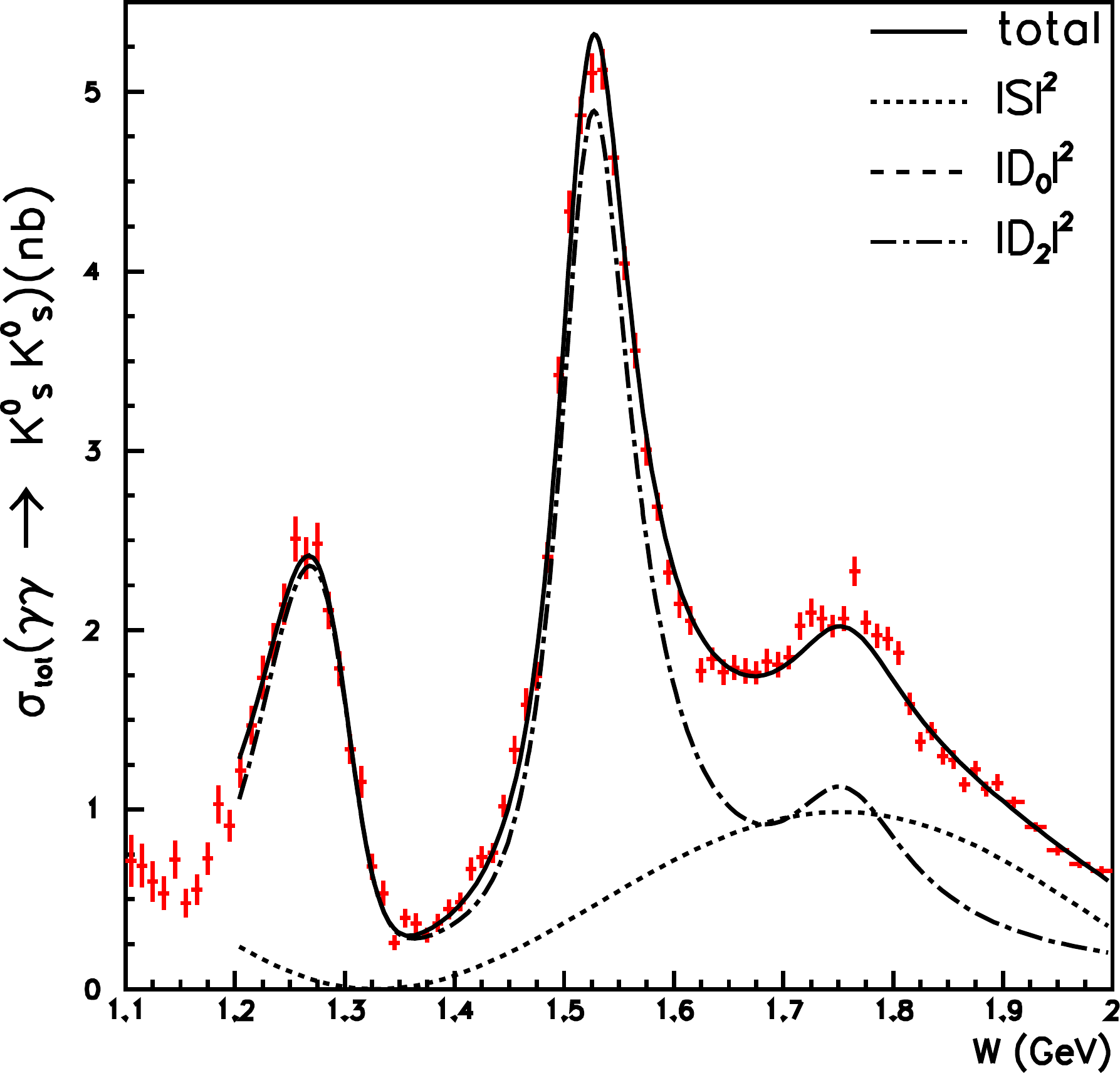,width=50mm}}
 \caption{The fit-H for the $f_2(1710)$ fit 
(solid line) superimposed on the spectrum of 
$\hat{S}^2$ (left top), $\hat{D}_0^2$ (left bottom),
$\hat{D}_2^2$ (middle) and
integrated cross section (for $|\cos \theta^*| \leq 0.8$) (right). 
Shown in the $\hat{D}_2^2$ ($\hat{S}^2$) spectrum are the fitted results
$f_2(1270)$ (long-dashed line),
$a_2(1320)$ (dashed line),
$f_2'(1525)$ (dot-dashed line), $f_2(1710)$ (very-long-dashed line)
and non-resonant background $|B_{D2}|^2$ (dotted line) 
($|B_S|^2$ (dotted line)).
In the integrated cross section, the fitted results of 
$|S|^2$ (dotted line), $|D_0|^2$ (dashed line) and
$|D_2|^2$ (dot-dashed line) are also shown.
}
\label{fig:f2Hsd02t}\end{figure*}

\begin{figure*}
 \centering
  {\epsfig{file=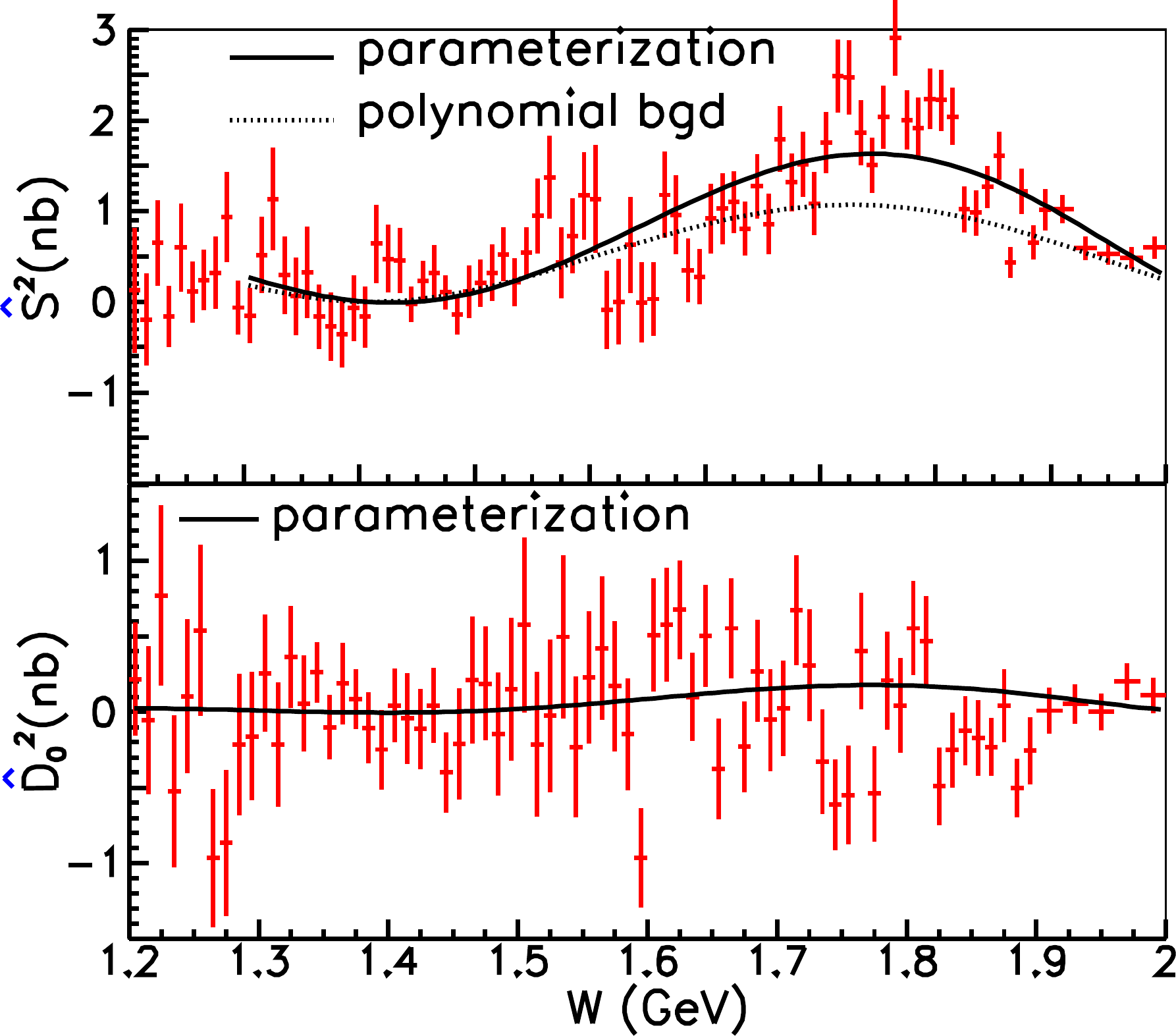,width=50mm}}
  {\epsfig{file=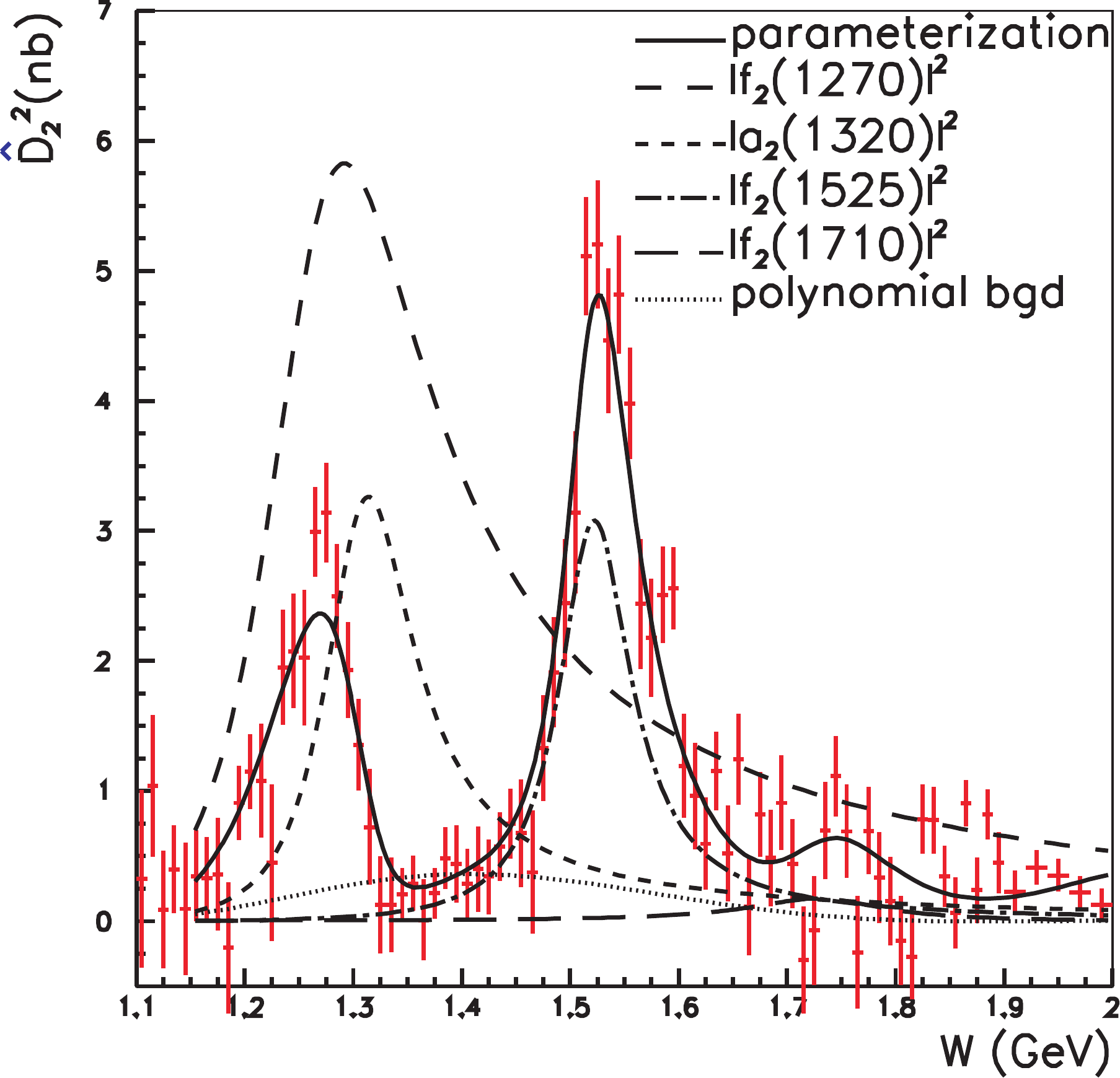,width=50mm}}
  {\epsfig{file=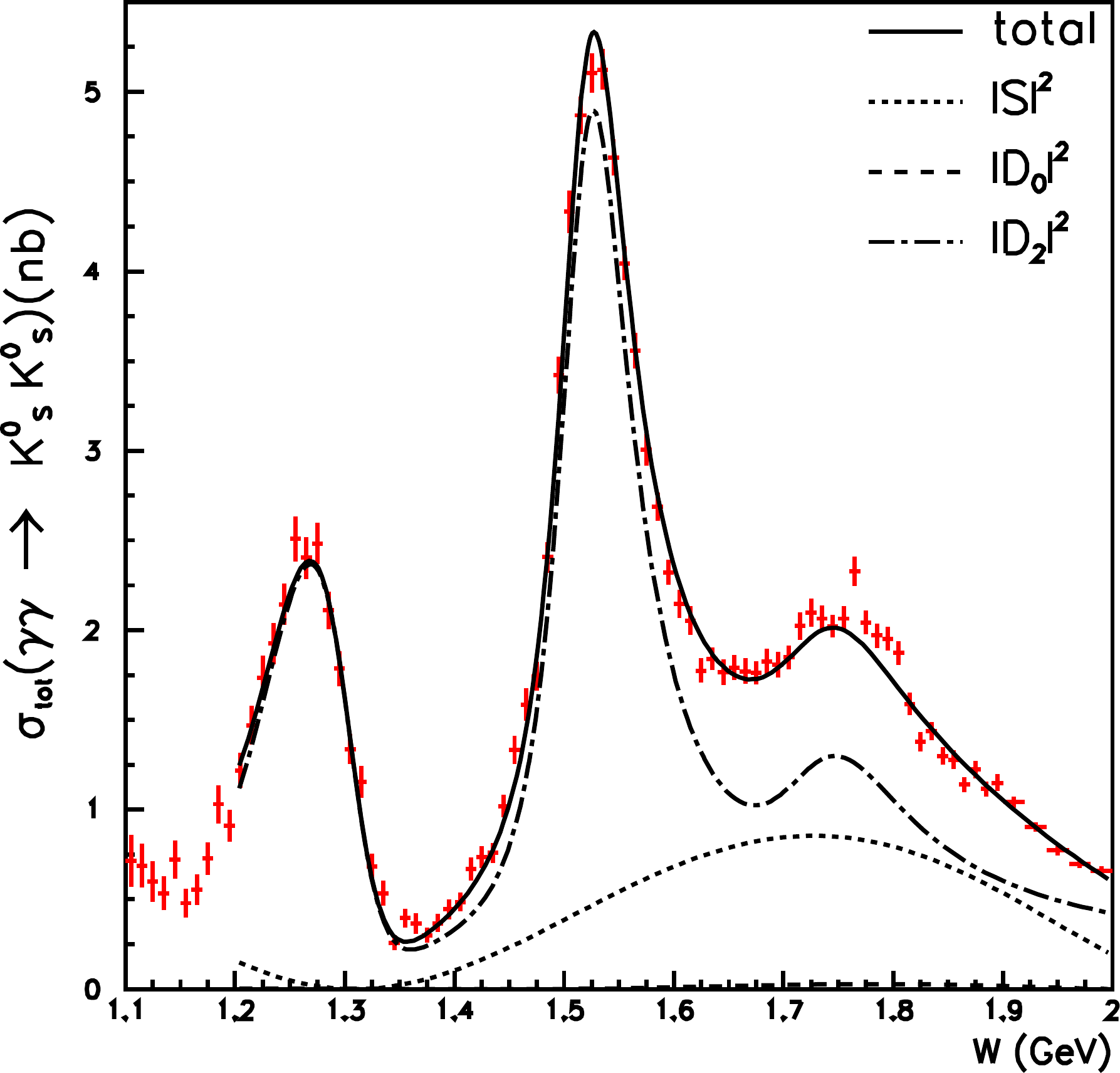,width=50mm}}
 \caption{The fit-L for the $f_2(1710)$ fit 
(solid line) superimposed on the spectrum of 
$\hat{S}^2$ (left top), $\hat{D}_0^2$ (left bottom),
$\hat{D}_2^2$ (middle) and
integrated cross section (for $|\cos \theta^*| \leq 0.8$) (right). 
See the caption of Fig.~\ref{fig:f2Hsd02t} for the line 
convention (also shown on plots).
}
\label{fig:f2Lsd02t}
\end{figure*}

We also study a case where the structure near $W=1.8$~GeV 
is assumed to be due to a tensor meson
(labeled the $f_2(1710)$, which can be either $a_2(1700)$
or $f_2(1810)$ (referred to as the ``$f_2(1710)$ fit'').
The contribution from tensor mesons may be suppressed due to
destructive interference between the $f_2(1810)$ and $a_2(1700)$;
this hypothesis could also be tested by analyzing $\gamma \gamma \to
K^+ K^-$ data.
A unique best fit with poor $\chi^2$ is obtained for 
the $f_2(1710)$ fit with either of the H and L solutions of 
the $f_2'(1525)$ fit.
Thus, the hypothesis of $J=2$ for the $f_J(1710)$ is disfavored
by the data.
Fitted values are summarized in Table~\ref{tab:fj17}.
Figures~\ref{fig:f2Hsd02t} and \ref{fig:f2Lsd02t} 
show the fitted results for the $f_2(1710)$ fit
for each of the H and L solutions. 

Furthermore, we fit the hypothesis where we assume no resonance near 
$W=1.8$~GeV. 
Three best fits are obtained for the hypothesis H
of the $f_2'(1525)$ fit with poor $\chi^2/ndf$: 
1264.5/589, 1265.3/589 and 1267.8/589.
One best fit is obtained for the L hypothesis with even worse
$\chi^2/ndf$ of 1349.8/589.
We conclude that our fit favors the presence
of the $f_0(1710)$ in our data.

Systematic uncertainties are estimated similarly to those for
the $f_2'(1525)$ fit.
In the $W$-range study, fits are performed in two fit regions: 
$1.12~\GeV \leq W \leq 1.92~\GeV$ and
$1.28~\GeV \leq W \leq 2.08~\GeV$.
For $W$-distortion, a study is performed by shifting the cross section by
$\pm 0.08 \times |d \sigma / d \Omega| (W - 1.6~\GeV)$;
for background waves, by changing each wave to
a first- or third-order polynomial;
and for the parameters of the $f_2'(1525)$, by shifting the
values by their MINOS errors.
The results for the systematic uncertainties are summarized in
Table~\ref{tab:sysf017}.
Table~\ref{tab:fj17} lists the results for the
$f_0(1710)$ fit (fit-H and fit-L).

\begin{center}
\begin{table*}
\caption{Fitted parameters for the $f_0(1710)$ fit and
$f_2(1710)$ fit. 
For the $f_0(1710)$ fit, the first errors
are statistical and the second systematic; they are 
summarized in Table~\ref{tab:sysf017}.
The parameters where the H and L solutions are combined
are also shown (explained in Sec.~\ref{sub:finloww}).}
\label{tab:fj17}
\begin{tabular}{l|cccc|cc} \hline \hline
Parameter 
& \multicolumn{4}{c|}{$f_0(1710)$ fit} 
& \multicolumn{2}{c}{$f_2(1710)$ fit} \\
& fit-H & fit-L & H,L combined & PDG & fit-H & fit-L \\
\hline
$\chi^2/ndf$ & 694.2/585 & 701.6/585 & -- & -- & 796.3/585 & 831.5/585\\
\hline
&&&&&&\\[-10pt]
Mass(${f_J})$ (MeV/$c^2$) & 
 $1750^{+5+29}_{-6-18}$ & $1749^{+5+31}_{-6-42}$ &
 $1750^{+6+29}_{-7-18}$ & $1720 \pm 6$ &
 $1750^{+6}_{-7}$ & $1729^{+6}_{-7}$ \\
$\Gamma_{\rm tot}({f_J})$ (MeV) &
 $138^{+12+96}_{-11-50}$ & $145^{+11+31}_{-10-54}$ &
 $139^{+11+96}_{-12-50}$ & $135 \pm 6$ &
 $132^{+12}_{-11}$ & $150 \pm 10$ \\
$\Gamma_{\gamma \gamma}\B(K \bar{K})_{f_J}$ (eV) &
 $12^{+3+227}_{-2-8}$ & $21^{+6+38}_{-4-26}$ &
 $12^{+3+227}_{-2-8}$ & unknown &
 $2.1^{+0.5}_{-0.3}$ & $1.6 \pm 0.2$ \\
\hline\hline
\end{tabular}
\end{table*}
\end{center}

\begin{table*}[h]
\caption{Systematic uncertainties for the $f_0(1710)$ fit.
The left (right) number in each row for each observable indicates
a positive (negative) deviation from the nominal values.}
\label{tab:sysf017}
\begin{center}
\begin{tabular}{l|rr|rr|rr|rr|rr|rr} \hline \hline 
& \multicolumn{6}{c|}{Fit-H} & \multicolumn{6}{c}{Fit-L} 
 \\ \cline{2-13}
Source & \multicolumn{2}{c|}{Mass} 
& \multicolumn{2}{c|}{$\Gamma_{\rm tot}$}
& \multicolumn{2}{c|}{$\Gamma_{\gamma \gamma} \B(K \bar{K})$}
& \multicolumn{2}{c|}{Mass} 
& \multicolumn{2}{c|}{$\Gamma_{\rm tot}$}
& \multicolumn{2}{c}{$\Gamma_{\gamma \gamma} \B(K \bar{K})$} \\
& \multicolumn{2}{c|}{~(MeV/$c^2$)~} 
& \multicolumn{2}{c|}{~~~~(MeV)~~~~} & \multicolumn{2}{c|}{(eV)}
& \multicolumn{2}{c|}{~(MeV/$c^2$)~} 
& \multicolumn{2}{c|}{~~~~(MeV)~~~~} & \multicolumn{2}{c}{(eV)} \\ \hline 
$W$-range & 21 & 0 & 0 & $-15$ & 0 & $-1$ & 16 & 0 & 5 & $-13$ & 6 & $-4$ \\
$W$ bias & 2 & $-2$ & 6 & $-5$ & 2 & $-1$ & 4 & 0 & 2 & $-7$ & 2 & $-4$ \\
Efficiency & 8 & $-4$ & 25 & $-30$ & 209 & 0 & 11 & $-5$ & 0 & $-28$ & 2 & $-12$ \\
Overall normalization & 4 & $-2$ & 9 & $-11$ & 1 & $-2$ & 7 & $-2$ & 4 & $-16$ & 5 & $-8$ \\
$|\cos \theta^*|$ bias & 0 & $-1$ & 3 & $-1$ & 1 & 0 & 4 & 0 & 2 & $-7$ & 2 & $-4$ \\ \hline
$B_S$ & 5 & $-7$ & 84 & $-9$ & 87 & $-2$ & 7 & 0 & 26 & $-3$ & 35 & $-2$ \\
$B_{D0}$ & 0 & 0 & 1 & 0 & 0 & 0 & 1 & 0 & 0 & $-2$ & 0 & $-1$ \\
$B_{D2}$ & 1 & 0 & 0 & $-1$ & 0 & 0 & 11 & $-37$ & 4 & $-11$ & 1 & $-2$ \\ \hline
Mass($f_2(1270)$) & 3 & $-1$ & 6 & $-6$ & 1 & $-1$ & 3 & 0 & 0 & $-4$ & 2 & $-3$ \\
$\Gamma_{\rm tot}(f_2(1270))$ & 0 & 0 & 1 & 0 & 0 & 0 & 2 & $-2$ & 5 & $-4$ & 4 & $-2$ \\
$\B(\gamma \gamma)(f_2(1270))$ & 7 & 0 & 12 & $-18$ & 2 & $-4$ & 6 & $-1$ & 0 & $-17$ & 5 & $-10$ \\
$r_R$ & 0 & 0 & 0 & 0 & 0 & 0 & 1 & $-1$ & 1 & 0 & 1 & 0 \\ \hline
Mass($a_2(1320)$) & 1 & 0 & 0 & $-2$ & 0 & 0 & 2 & 0 & 0 & $-2$ & 0 & $-1$ \\
$\Gamma_{\rm tot}(a_2(1320))$ & 2 & $-2$ & 7 & $-5$ & 2 & $-1$ & 3 & 0 & 2 & $-9$ & 3 & $-6$ \\
$\B(\gamma\gamma)(a_2(1320))$ & 1 & $-1$ & 2 & 0 & 1 & 0 & 2 & 0 & 0 & $-2$ & 0 & $-1$ \\ \hline
Mass($f_2'(1525)$) & 2 & $-2$ & 1 & 0 & 1 & 0 & 1 & $-1$ & 3 & $-4$ & 3 & $-3$ \\
$\Gamma_{\rm tot}(f_2'(1525))$ & 2 & $-2$ & 4 & $-3$ & 2 & $-1$ & 4 & 0 & 0 & $-4$ & 0 & $-2$ \\
$\B(\gamma\gamma)(f_2'(1525))$ & 14 & 0 & 0 & $-24$ & 0 & $-4$ & 14 & $-18$ & 14 & $-27$ & 9 & $-12$ \\ \hline
$\phi_{f_2'(1525)}$ & 4 & $-15$ & 33 & $-12$ & 22 & $-3$ & 4 & $-5$ & 0 & $-17$ & 3 & $-11$ \\
$\phi_{a_2(1320)}$ & 4 & $-1$ & 5 & $-8$ & 1 & $-2$ & 3 & 0 & 0 & $-4$ & 0 & $-2$ \\ \hline
Total & ~~29 & $-18$ & ~~96 & $-50$ & ~227 & $-8$ & ~~31 & $-42$ & ~~31 & $-54$ & ~~38 & $-26$ \\
\hline\hline
\end{tabular}
\end{center}
\end{table*}

\subsubsection{Final results for the region $W<2.0~\GeV$}
\label{sub:finloww}
As described above, we obtain two solutions referred to as H and L
for the $f_2'(1525)$ fit, and corresponding fits are performed in
the $f_0(1710)$ fit by fixing the $f_2'(1525)$ parameters to those of either 
the H or L solution.
Here, we combine solutions statistically to obtain final results. 

From each pair of solutions for an observable $x$, a probability 
density function (PDF) $P(x)$ is formed as the sum of asymmetric Gaussian 
functions that correspond to the two solutions with asymmetric errors.
These functions are weighted according to the $\chi^2$  differences 
between the two solutions.
The most probable value $x_f$ is the one that gives the maximum in $P(x)$.
Asymmetric statistical errors 
$\sigma^l$ and $\sigma^h$ are determined from a confidence interval
such that $P(x_f - \sigma^l) = P(x_f + \sigma^h)$ with
\begin{equation}
\int_{x_f - \sigma^l}^{x_f + \sigma^h} P(x) dx = 0.683 \; .
\end{equation}
The systematic uncertainty 
for observable $x$ is determined from the solution
with the largest deviation from $x_f$.
The final results thus obtained are listed
in Tables~\ref{tab:f2pfit} and \ref{tab:fj17}.

\subsection{Fitting the region $W > 2.0$~GeV}
\label{sub:above}
We investigate the structures around 2.3 and 2.6~GeV.
In fitting the region $2.0~\GeV \le W \le 3.0~\GeV$, we 
assume that the non-resonant backgrounds in the S, D$_0$, D$_2$, 
G$_0$ and G$_2$ waves obey a power law in $W$.
When we parameterize them using a
polynomial approximation as in Eq.~(\ref{eqn:para2}),
we obtain fits of poor quality.
We parameterize the backgrounds as
\begin{equation}
B_i = b_i \times \left( \frac{W}{W_0} \right)^{-c_i} e^{i \phi_i} ,
\label{eqn:bgdpow}
\end{equation}
where the index $i$ denotes S, D$_0$, D$_2$, G$_0$ or G$_2$ waves,
$W_0$ is chosen to be the lower boundary of the fitting region 
(nominally $W_0=2.0$~GeV), 
and $b_i$ and $c_i$ are the free parameters.
The phases of $B_S$ and $B_{D2}$ are chosen to be zero
as a reference for the other phases.
The parameters $b_i$ are set positive to resolve arbitrary
sign ambiguities.

We also investigate a possible contribution from the $J=4$ resonances.
Table~\ref{tab:param2} summarizes the parameters of
the $f_0(2200)$, $f_2(2300)$ and $f_4(2300)$ 
that are known to couple to
$K \bar{K}$~\cite{pdg2012}.

We allow $B_{G0}$ and/or $B_{G2}$ to be non-zero in 
Eq.~(\ref{eqn:bgdpow}).
We fit 13 assumptions for the structures
around 2.3 and 2.6~GeV that are
observed in the plot of the integrated cross section 
shown in Fig.~\ref{fig:ksksres}(a).
A fit performed assuming the presence of the $f_J(2200)$ 
($J=0,2,4$) and $f_{J'}(2500)$ 
($J'=0,2,4$) is referred to as an ``$f_J$-$f_{J'}$ fit.''
We also investigate hypotheses in which there are no resonances
(or only one) for the two structures.
These fits are referred to as ``no-resonance-''
(``only-$f_{J}$-'') fits, respectively.

When both $B_{G0}$ and $B_{G2}$ are allowed to be non-zero,
too many solutions
are obtained
because of the several combinations of interfering amplitudes
(not shown).
Thus, we focus on the hypothesis wherein only $B_{G2}$ is non-zero.
This choice is based on the idea that
the possible resonances $f_4(2200)$ and $f_4(2500)$
are included in the G$_2$ wave only because of helicity considerations.
A summary of the fitted results is given in Table~\ref{tab:higsum3}.
In this case, once again, some of the $f_J$-$f_{J'}$ fits give multiple 
solutions.
In some cases, one or more $c_i$ values in 
Eq.~(\ref{eqn:bgdpow}) assume unphysically large values.
Thus, we constrain the maximum values of $c_i$ to be 20.

A unique solution of relatively good quality is obtained for 
the $f_2$-$f_0$ fit, while other hypotheses yield larger values of 
$\chi^2/ndf$.
The $f_2$-$f_0$ fit is also favored for the case in which both $B_{G0}$ and 
$B_{G2}$ are assumed to be non-zero.
Thus, we conclude that the structure around 2.3~GeV 
is likely due to a tensor meson (referred to tentatively as 
$f_2(2200)$) and the one near 2.6~GeV is likely to be a scalar meson 
(possibly $f_0(2500)$).

\begin{figure}
 \centering
  {\epsfig{file=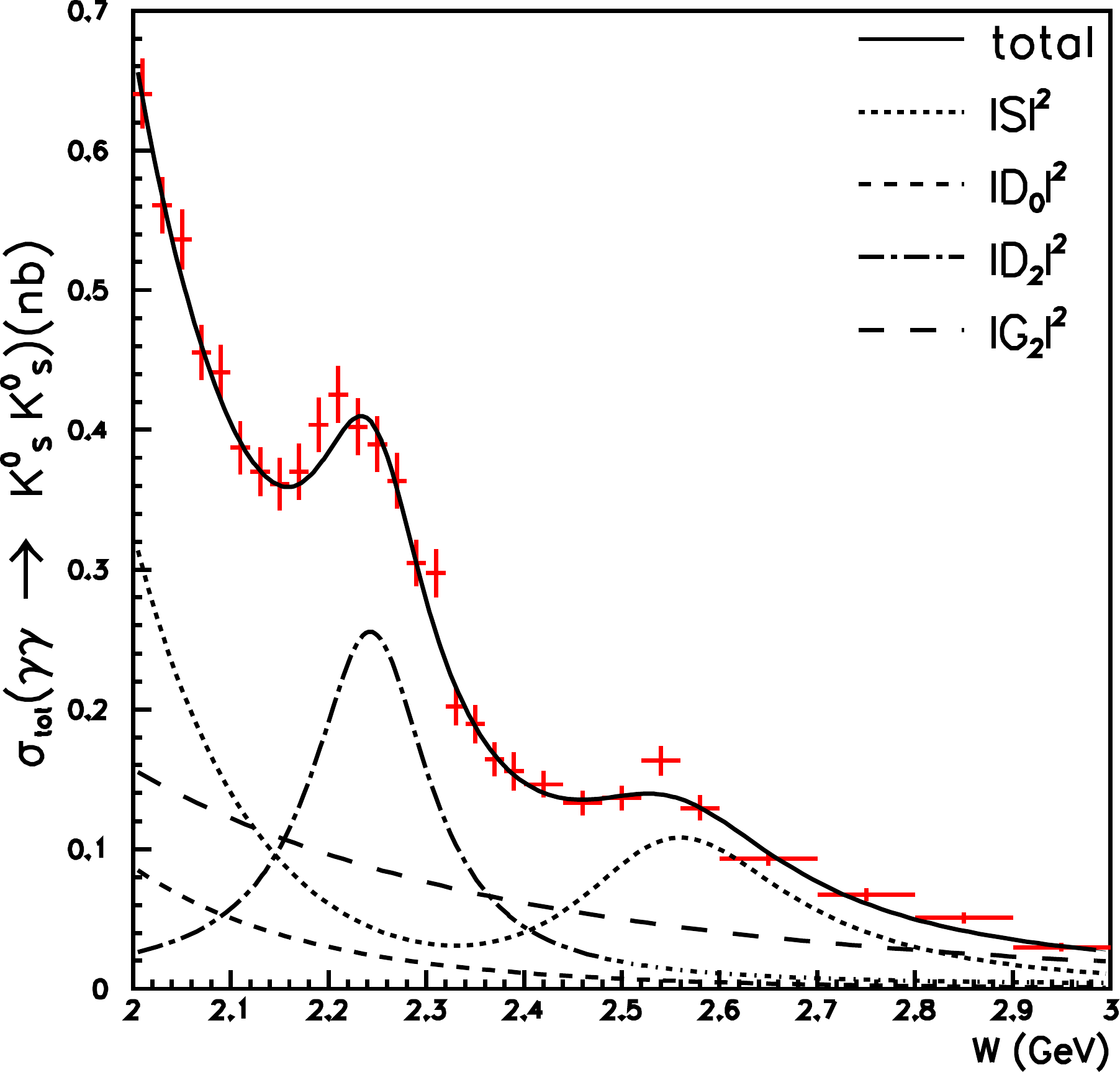,width=80mm}}
 \caption{Result of the $f_2$-$f_0$ fit (solid line)
superimposed on 
the integrated cross section (for $|\cos \theta^*| \leq 0.8$).
The fitted results of $|S|^2$ (dotted line),
$|D_0|^2$ (dashed line), $|D_2|^2$ (dot-dashed line) and
$|G_2|^2$ (long-dashed line) are also shown.
}
\label{fig:f2f0gt}
\end{figure}

\begin{figure}
 \centering
  {\epsfig{file=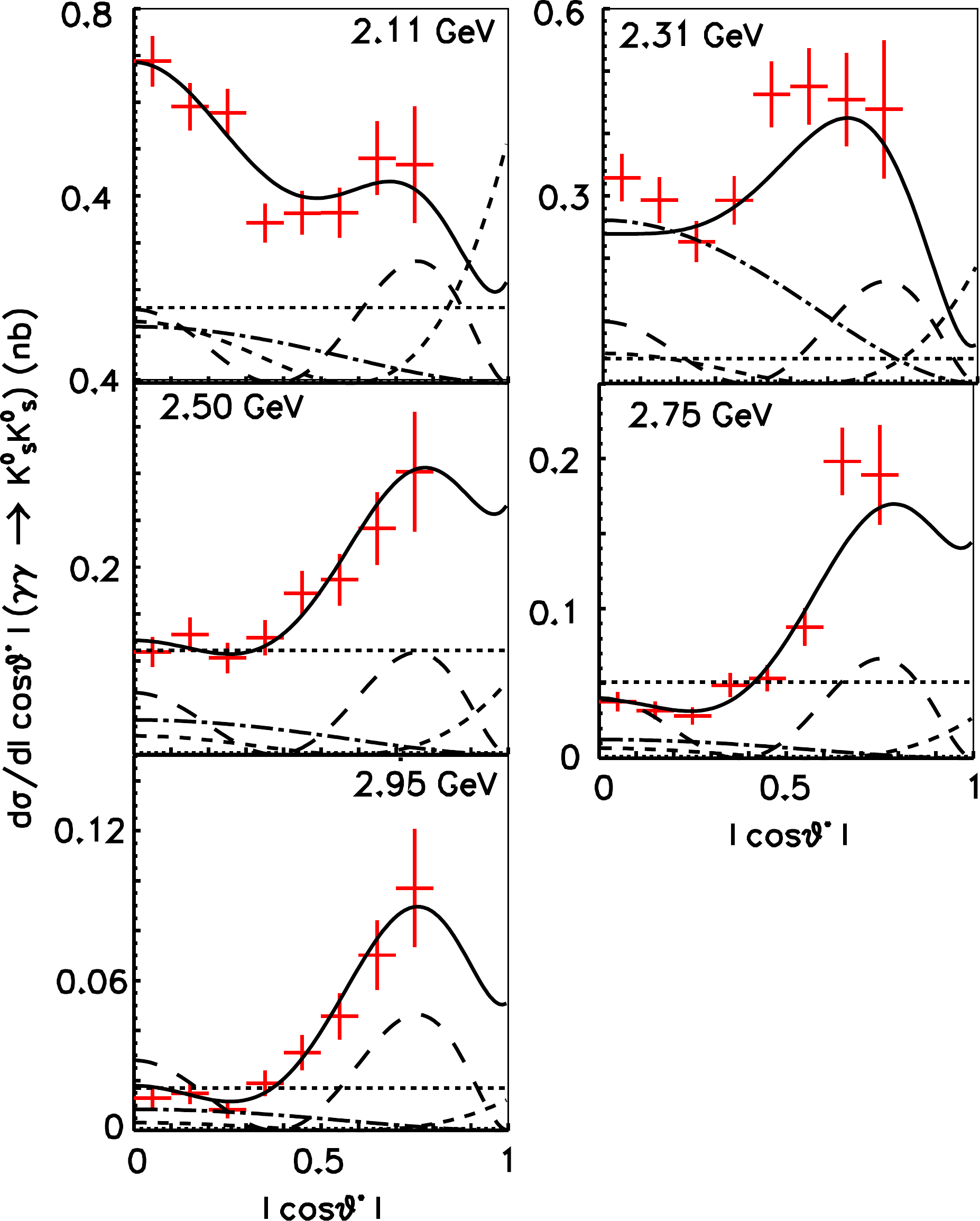,width=80mm}}
 \caption{Differential cross section and the fitted results of
the $f_2$-$f_0$ fit (solid line) at the $W$ bins indicated
in each panel.
The contributions of $|S|^2$ (dotted line), 
$|D_0|^2$ (dashed line), $|D_2|^2$ (dot-dashed line) and
$|G_2|^2$ (long-dashed line) are also shown.
}
\label{fig:f2f0dsel}
\end{figure}

The fitted values obtained from the $f_2$-$f_0$ fit are
summarized in Table~\ref{tab:f2f0}
for the mass, total width and $\Gamma_{\gamma \gamma}\B(K \bar{K})$
of the $f_2(2200)$ and $f_0(2500)$.
Figure
\ref{fig:f2f0gt} 
shows the fitted results for the $f_2$-$f_0$ fit superimposed on 
the integrated cross section.
Figure \ref{fig:f2f0dsel} shows the fitted results and contributions of
$|S|^2$, $|D_0|^2$, $|D_2|^2$, and $|G_2|^2$ to the
differential cross section in selected $W$ bins.
The systematic uncertainties shown in Table~\ref{tab:f2f0}
are estimated similarly to those described in Sec.~\ref{sub:below}.
To estimate the uncertainties from the background parameterization,
the background amplitudes are changed to $b_i ((W \pm 1~\GeV)/W_0)^{-c_i}$.
The results of the studies of systematic uncertainties are summarized in 
Table~\ref{tab:sysf2f0}.

\begin{center}
\begin{table}
\caption{Parameters of the $f_0(2200)$, $f_2(2300)$ 
and $f_4(2300)$~\cite{pdg2012}.}
\label{tab:param2}
\begin{tabular}{lccc} \hline \hline
Parameter  & $f_0(2200)$ & $f_2(2300)$ & $f_4(2300)$ \\ \hline
Mass (MeV/$c^2$ ) & $2189 \pm 13$ & $2297 \pm 28$ & $\sim 2300$\\
$\Gamma_{\rm tot}$ (MeV) & $238 \pm 50$ & $149 \pm 41$ & $250 \pm 80$\\
$f_J \to K \bar{K}$ & seen & seen & seen \\
$f_J \to \gamma \gamma$ & unknown & seen & unknown \\
\hline\hline
\end{tabular}
\end{table}
\end{center}

\begin{center}
\begin{table}
\caption{Summary of fitted results for $2.0~\GeV \le W \le 3.0~\GeV$
for 13 assumptions (with G$_2$ background).}
\label{tab:higsum3}
\begin{tabular}{c|ccc} \hline \hline
Assumption 
& No. of sol. & $\chi^2$ & $ndf$ \\ \hline
$f_0$-$f_0$ & 2 & 293.3, 293.9 & 214 \\
$f_0$-$f_2$ & 4 & 320.9, 321.9, 324.5, 327.6 & 214 \\
$f_0$-$f_4$ & 1 & 291.4 & 214 \\
\hline
$f_2$-$f_0$ & 1 & 228.3 & 214 \\
$f_2$-$f_2$ & 1 & 260.4 & 214 \\
$f_2$-$f_4$ & 1 & 323.6, 306.7 & 214 \\
\hline
$f_4$-$f_0$ & 1 & 411.6 & 214 \\
$f_4$-$f_2$ & 2 & 468.6, 472.1 & 214 \\
$f_4$-$f_4$ & 4 & 459.6, 464.1, 466.4, 467.5 & 214 \\
\hline
Only-$f_0$ & 1 & 390.0 & 218 \\
Only-$f_2$ & 1 & 323.6 & 218 \\
Only-$f_4$ & 1 & 518.7 & 218 \\
No resonances & 1 & 659.32 & 222 \\
\hline \hline
\end{tabular}
\end{table}
\end{center}

\begin{center}
\begin{table}
\caption{Parameters obtained from the $f_2$-$f_0$ fit.
The first errors
are statistical and the second systematic 
(summarized in Table~\ref{tab:sysf2f0}).}
\label{tab:f2f0}
\begin{tabular}{lcc} \hline \hline
Parameter & $f_2(2200)$ & $f_0(2500)$ \\ \hline
&& \\[-10pt]  
Mass (MeV/$c^2$) 
& $2243^{+7+3}_{-6-29}$ & $2539 \pm 14 {}^{+38}_{-14}$ \\
$\Gamma_{\rm tot}$ (MeV) 
& $145 \pm 12 {}^{+27}_{-34}$ & $274^{+77+126}_{-61-163}$ \\
$\Gamma_{\gamma \gamma}\B(K \bar{K})$ (eV) 
& $3.2^{+0.5+1.3}_{-0.4-2.2}$ & $40^{+9+17}_{-7-40}$ \\
\hline \hline
\end{tabular}
\end{table}
\end{center}

\begin{table*}[h]
\caption{Systematic uncertainties for the $f_2$-$f_0$ fit.
The left (right) number in each row for each observable indicates
a positive (negative) deviation from the nominal values.}
\label{tab:sysf2f0}
\begin{center}
\begin{tabular}{l|rr|rr|rr|rr|rr|rr} \hline \hline 
& \multicolumn{6}{c|}{$f_0(2200)$} & \multicolumn{6}{c}{$f_0(2500)$} 
 \\ \cline{2-13}
Source &\multicolumn{2}{c|}{Mass} & \multicolumn{2}{c|}{$\Gamma_{\rm tot}$}
& \multicolumn{2}{c|}{$\Gamma_{\gamma \gamma} \B(K \bar{K})$} 
& \multicolumn{2}{c|}{Mass} & \multicolumn{2}{c|}{$\Gamma_{\rm tot}$}
& \multicolumn{2}{c}{$\Gamma_{\gamma \gamma} \B(K \bar{K})$} \\
& \multicolumn{2}{c|}{(MeV/$c^2$)} & \multicolumn{2}{c|}{(MeV)} 
& \multicolumn{2}{c|}{(eV)}
& \multicolumn{2}{c|}{(MeV/$c^2$)} & \multicolumn{2}{c|}{(MeV)} 
& \multicolumn{2}{c}{(eV)} \\ \hline 
$W$-range & ~~~3 & ~~$-3$ & ~~26& ~$-14$ & ~1.3 & ~$-0.7$ & ~~~6& ~$-11$ & ~101& ~$-89$ & ~~10& ~~$-5$ \\
$W$ bias & 0& $ 0$ & 0& $ 0$ & 0.1 & $-0.1$ & 0& $ 0$ & 8& $-7$ & 1& $-1$ \\
Efficiency & 0& $ 0$ & 1& $-1$ & 0.3 & $-0.3$ & 0& $ 0$ & 9& $-9$ & 4& $-4$ \\
Overall normalization & 0& $ 0$ & 0& $ 0$ & 0.1 & $-0.1$ & 0& $ 0$ & 0& $-1$ & 2& $-2$ \\
$|\cos \theta^*|$ bias & 0& $ 0$ & 0& $ 0$ & 0.1 & $-0.1$ & 0& $ 0$ & 0& $-2$ & 2& $-2$ \\
\hline
$B_S$ & 0& $-15$ & 0& $-24$ & 0.0 & $-1.6$ & 25& $-1$ & 0& $-105$ & 0& $-39$ \\
$B_{D0}$ & 0& $-15$ & 0& $-25$ & 0.0 & $-1.5$ & 24& $ 0$ & 0& $-108$ & 0& $-39$ \\
$B_{D2}$ & 0& $ 0$ & 0& $ 0$ & 0.0 & $ 0.0$ & 0& $ 0$ & 0& $ 0$ & 0& $ 0$ \\
$B_{G0}$ & 0& $-19$ & 7& $ 0$ & 0.1 & $ 0.0$ & 0& $-9$ & 0& $-62$ & 0& $-3$ \\
$B_{G2}$ & 0& $-1$ & 0& $ 0$ & 0.0 & $ 0.0$ & 1& $ 0$ & 0& $-4$ & 0& $-1$ \\\hline
Total & 3& $-29$ & 27& $-37$ & 1.3 & $-2.3$ & 35& $-14$ & 101& $-186$ & 11& $-56$ \\ 
\hline\hline
\end{tabular}
\end{center}
\end{table*}


\subsection{Discussion of the results of the resonance study}
In this subsection, we summarize the fitted results and
discuss their implications.
First, the destructive interference between the $f_2(1270)$
and $a_2(1320)$~\cite{lipkin} is confirmed with high accuracy; 
the relative phase, $\phi_{a2}$, is
measured to be $\left( 172.6^{+6.0+12.2}_{-0.7-7.0}\right)^{\circ}$, where
the first error is statistical and the second systematic.
The mass, total width and $\Gamma_{\gamma \gamma} \B(K \bar{K})$
of the $f_2'(1525)$ are measured to be 
$1525.3^{+1.2+3.7}_{-1.4-2.1}~\MeV/c^2$,
$82.9^{+2.1+3.3}_{-2.2-2.0}~\MeV$ and
$48^{+67}_{-8}{}^{+108}_{-12}~\eV$, respectively.
The systematic uncertainty of $\Gamma_{\gamma \gamma} \B(K \bar{K})$
is fairly large.
Nevertheless, this is the first measurement of this
parameter that includes
the interference with a non-resonant amplitude.

The structure near 1.6~GeV is attributed
to a scalar meson and is interpreted to be the $f_0(1710)$.
To obtain the significance for the assignment of the $f_0(1710)$ 
over that of the $f_2(1710)$,
fits are performed for each of the sources of systematic uncertainty 
for the two hypotheses
and the minimum $\chi^2$ difference is identified among these fits.
It is found to be 63.3,  which corresponds to a
significance of 7.9$\sigma$ favoring the  $f_0(1710)$.

A similar study is performed for the $f_J(2200)$ hypothesis by comparing
the values of $\chi^2_{\rm min}$ of the $f_2$-$f_0$ fit and 
$f_0$-$f_0$ fit for each source of systematic uncertainty.
We obtain a minimum $\chi^2$ difference to be of 11.3, corresponding to 
a $3.4\sigma$ significance.
For the  $f_0(2500)$, 
the $f_2$-$f_0$ fit gives the best $\chi^2$;
the next best, the $f_2$-$f_2$ fit, yields a $4.3\sigma$ significance.
Thus, while we cannot make definitive assignments about 
the spins of the $f_J(2200)$ and $f_{J'}(2500)$, $J=2$ and $J'=0$
hypotheses, respectively, are favored.

The values of $\Gamma_{\gamma \gamma} \B(K \bar{K})$ 
for the $f_0(1710)$, $f_2(2200)$ and $f_0(2500)$
are estimated for the first time and are found to be
$12^{+3+227}_{-2-8}~\eV$, 
$3.2^{+0.5+1.3}_{-0.4-2.2}~\eV$ and 
$40^{+9+17}_{-7-40}~\eV$, respectively.
Each value could provide
important information on the constituents of the corresponding resonance.
For example, the $f_0(1710)$ is identified as an unmixed scalar glueball
according to a coupled-multi-channel analysis~\cite{albaladejo}. 
However, the $f_0(1710)$ is unlikely to be a glueball candidate
because the observed value of 
$\Gamma_{\gamma \gamma} \B(K \bar{K})$,
combined with the implied value of $\Gamma_{\gamma \gamma} \B(\pi \pi)$
($\simeq \Gamma_{\gamma \gamma} \B(K \bar{K})$
for the flavor-SU(3)-symmetric decay of a glueball)
would indicate a large two-photon width, contrary to the expectation 
of much less than 1~eV for a glueball 
(see, e.g., Refs.~\cite{scalar, scalar2, scalar3, scalar4}).
Therefore, we conclude that the $f_0(1710)$ is a resonance with
a large $s \bar{s}$ admixture.

The measured mass of the $f_2(2200)$, $2243^{+7+3}_{-6-29}~\MeV/c^2$,
is close to that of the $f_J(2220)$ and smaller than
that of the $f_2(2300)$~\cite{pdg2012}.
The $f_J(2220)$ is usually assumed to be a glueball candidate
due to the small value of its total width ($23^{+8}_{-7}~\MeV$).
The structure found by Belle in the $\gamma \gamma \to K^+ K^-$
reaction around 2.3~GeV~\cite{kabe} is interpreted as the $f_2(2300)$
by PDG~\cite{pdg2012}.
The measured total width of the $f_2(2200)$,  
$145 \pm 12 {}^{+27}_{-34}$~MeV, is much wider than that
of the $f_J(2220)$ and is similar to that of the $f_2(2300)$.

The $f_0(2500)$, whose mass and width are found to be
$2539 \pm 14 {}^{+38}_{-14}~\MeV/c^2$ and
$274^{+77+126}_{-61-163}~\MeV$, respectively,
is the first scalar to be identified in this mass region~\cite{pdg2012};
this observation invites an independent confirmation.

\section{Derivation of charmonium contribution}
\label{sec:charm}
We present our estimates of the $\chi_{c0}$ and $\chi_{c2}$ contributions
by measuring the yields of the fit components in the region 
$|\cos \theta^*|<0.5$ and $2.8~{\rm GeV} < W < 4.0$~GeV 
(Fig.~\ref{fig:kskschic}). 
We use only samples with $|\cos \theta^*|<0.5$ to enhance 
the fraction of the charmonium components while disentangling
them from the continuum contribution.
We derive $\Gamma_{\gamma\gamma}(R){\cal B}(R \to \ks \ks)$ for 
these charmonium states. 
We also search for a possible contribution from a higher-mass charmonium 
in the 3.6 -- 4.0 ~GeV/$c^2$ region.

We assume the angular distributions for the  $\chi_{c0}$ and $\chi_{c2}$
components to be flat and proportional to $\sin^4 \theta^*$ (from
pure helicity-2~\cite{kabe2}), respectively,
to derive $\Gamma_{\gamma\gamma}{\cal B}$ from the yield in
$|\cos \theta^*|<0.5$.
We discuss the effect of interference with the continuum.

\subsection{Evaluation of parameters for the  $\chi_{cJ}$ charmonia}
\label{sub:evalu}
\begin{figure}
\centering
\includegraphics[width=7cm]{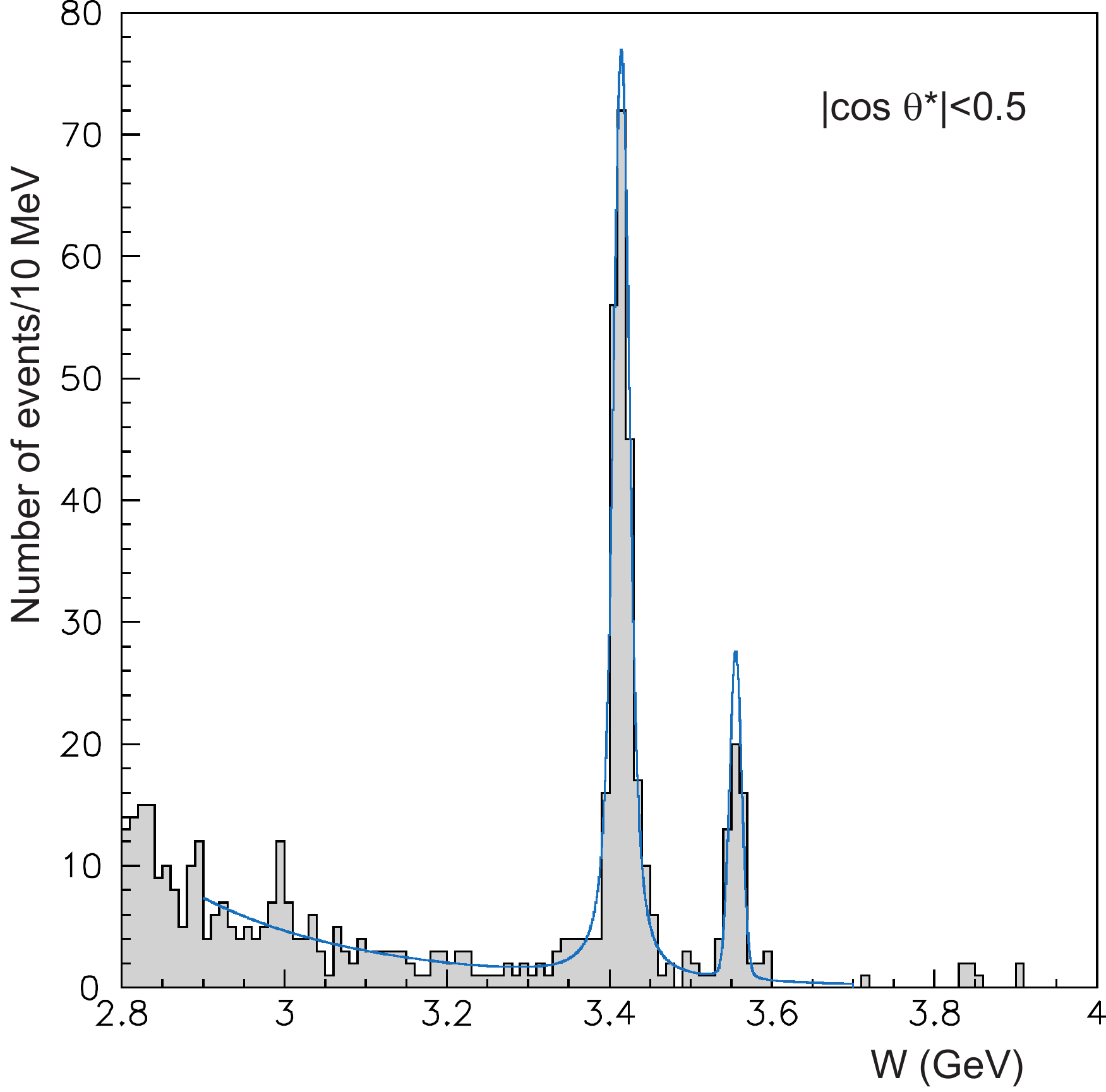}
\centering
\caption{ $W$ distribution of the candidate events for  
$2.8~\GeV < W < 4.0~\GeV$
and  $|\cos \theta^*|<0.5$. 
The two distinct peaks are from the 
$\chi_{c0}$ and $\chi_{c2}$. The curve is a fit without interference.
 }
\label{fig:kskschic}
\end{figure}

The peak structures observed in the yield distribution 
for $3.3~\GeV < W < 3.6~\GeV$ (Fig.~\ref{fig:kskschic})
are from charmonium production:
$\gamma \gamma \to \chi_{c0}$, $\chi_{c2} 
\to \ks \ks$. 
We fit the distribution to contributions from the $\chi_{c0}$ and
$\chi_{c2}$ as well as a smooth continuum component
represented by
\begin{eqnarray}
Y(W) &=& |\sqrt{\alpha kW^{-\beta}}+e^{i\phi}\sqrt{N_{\chi_{c0}}}
{\rm BW}_{\chi_{c0}}(W)|^2 + \nonumber \\
&& N_{\chi_{c2}}|{\rm BW}_{\chi_{c2}}(W)|^2 + \alpha (1-k)W^{-\beta},
\label{eqn:chicj}
\end{eqnarray}
in the $W$ and $|\cos \theta^*|$ ranges 2.9 -- 3.7~GeV and
$|\cos \theta^*| < 0.5$, respectively,
where ${\rm BW}_{\chi_{cJ}}(W)$ is a Breit-Wigner function for 
the charmonium amplitude, which is proportional to 
$1/(W^2-M_{\chi_{cJ}}^2-iM_{\chi_{cJ}} \Gamma_{\chi_{cJ}})$
and is normalized by $\int |{\rm BW}_{\chi_{cJ}}(W)|^2dW=1$.
The component $\alpha W^{-\beta}$ corresponds to the contribution from 
the continuum, with a fraction $k$ that interferes with the 
$\chi_{c0}$ amplitude with a relative phase $\phi$.  

It is impossible to determine the interference parameters for the
 $\chi_{c2}$ by any fits because of its smaller intrinsic 
width compared to the mass resolution. 
We fit the $\chi_{c2}$ yield ($N_{\chi_{c2}}$) with a function in which
no interference term is included, as shown by Eq.~(\ref{eqn:chicj});
later, we estimate the maximum
effects from the interference term when we evaluate the uncertainty
for the two-photon decay width of $\chi_{c2}$.

All parameters except the width of the $\chi_{c2}$ are free.
The $\chi_{c2}$ width is fixed to 2.0~MeV,
which is smaller than the estimated mass resolution of $\sim 7$~MeV.
Smearing effects due to a non-zero mass resolution 
are taken into account in the fit, using a
Gaussian function with $\sigma = 7.0$~MeV.
The small $W$ dependence of the product
of the efficiency and luminosity function, 
$\mp 0.95\%$ for a change in $W$ of $\pm 10$~MeV, is
folded in the $\chi_{c0}$ resonance curve.

A binned maximum likelihood method is applied with
the bin width $\Delta W = 10$~MeV. 
We examine two cases: with and without the interference
of the $\chi_{c0}$. 
We could not determine the $k$ parameter; that is, any $0 < k \leq 1$
gives exactly the same fit quality for the
constructive ($\phi \approx \pi/2$) and
destructive ($\phi \approx 3\pi/2$) interference cases.
Therefore, the statistical error range for 
the yield of $\chi_{c0}$ corresponds to the full range
of the interference assumption $0 < k \leq 1$.

The maximum effect of the interference of $\chi_{c2}$ with the
continuum component is calculated from Eq.~(\ref{eqn:chiint})
because we cannot measure it from the line shape of the charmonium,
so we include its maximum possible range
in the statistical error.
The number of $\chi_{c2}$ events that is proportional
to the square of the resonance amplitude is converted from 
the observed number $N_{\chi_{c2}}$ to that with 
the maximum interference effect $N'_{\chi_{c2}}$ 
using the relation
\begin{equation}
\sqrt{N'_{\chi_{c2}}} = \sqrt{\frac{\pi \Gamma n_I}{2} + N_{\chi_{c2}}}
\pm \sqrt{\frac{\pi \Gamma n_I}{2}},
\label{eqn:chiint}
\end{equation}
where $\Gamma$ and $n_I$ are the total width of the $\chi_{c2}$
and the observed yield density of the continuum component
per unit energy in the $W$ range around the $\chi_{c2}$, 
respectively.

The fitted results are summarized in Table~\ref{tab:charm1}, where
${\cal L}$ is the likelihood value.
The normalization $N_{\chi_{c0}}$ in Eq.~(\ref{eqn:chicj}) 
is proportional to the square of the resonance amplitude,
even when the interference is assumed.
The yields from the fits are translated to products 
of the two-photon decay width and the branching fraction, 
$\Gamma_{\gamma \gamma}(\chi_{cJ}){\cal B}(\chi_{cJ} \to \ks \ks)$, 
shown in Table~\ref{tab:charm2}.

To estimate the systematic uncertainties associated with the choice of 
the signal shape approximation, we vary their shape parameters.
We change the $W$ resolution from 7 to 8~MeV and modify the term 
in the denominator of the Breit-Wigner function, from
$-iM\Gamma$ to $-i W \Gamma$. 
The observed changes of the central values of the $\chi_{c0}$ and $\chi_{c2}$ 
yields are less than 3\%. 
This is because the
$\chi_{c0}$ and $\chi_{c2}$ contributions are well separated from
each other, and the continuum contribution is very small
around the charmonium peaks.
The systematic uncertainty is thus dominated by the
contributions from the 
efficiency and luminosity function, and is about 10\% in total. 
The systematic uncertainties for
$\Gamma_{\gamma\gamma}{\cal B}$ are shown in Table~\ref{tab:charm2}. 

The present results are consistent with and supersede
those from the previous measurements~\cite{chen, pdg2012}.
The interference effect was not taken into account
in the previous Belle result.

\begin{center}
\begin{table*}
\caption{Results of the fits (see the text) to obtain the charmonium
contributions with and without interference effects. 
Errors are statistical only. Logarithmic likelihood
($\ln {\cal L}$) values are only meaningful when comparing two or more
fits.}
\label{tab:charm1}
\small
\begin{tabular}{c|ccccc}
\hline
\hline
Interference & $N_{\chi_{c0}}$ & $k$ & $\phi$ &
$N_{\chi_{c2}}$ & $-2\ln{\cal L}/ndf$ \\
\hline
&&&&&\\[-10pt]
Not included & $248.3^{+17.9}_{-17.2}$ & (0.0, by def.) & $-$ & $53.0^{+8.1}_{-7.4}$ & $57.34/73$\\
Included & $266 \pm 53$ & any of 0--1 & two sols. 
& $53^{+14}_{-12}$ & $57.22/71$\\
\hline
\hline
\end{tabular}
\end{table*}
\end{center} 
\ \\
\begin{center}
\begin{table*}
\caption{Products of the two-photon decay width
and the branching fraction for the two charmonia, 
where $\Gamma_{\gamma \gamma}{\cal B}(\chi_{cJ})$
is the abbreviation for
$\Gamma_{\gamma \gamma}(\chi_{cJ}){\cal B}(\chi_{cJ} \to \ks\ks)$.
Mass and width parameters determined by the present fits
are also presented. 
Comparisons with the previous Belle results~\cite{chen} and the 
PDG2012~\cite{pdg2012} values are also shown.
The first and second errors (if given) 
are statistical and systematic,
respectively.}
\label{tab:charm2}
\small
\begin{tabular}{c|cc|cccc}
\hline
\hline
Interference & $\Gamma_{\gamma \gamma}{\cal B}(\chi_{c0})$ & 
$\Gamma_{\gamma \gamma}{\cal B}(\chi_{c2})$  & Mass($\chi_{c0})$ 
& Width($\chi_{c0}$) & Mass($\chi_{c2})$ & Width($\chi_{c2})$ \\
  & (eV) & (eV) & (MeV/$c^2$) & (MeV) & (MeV/$c^2$)  & (MeV) \\
\hline
&&&&&&\\[-10pt]
Not included & $8.09 \pm 0.58 \pm 0.83$  & $0.268^{+0.041}_{-0.037} \pm 0.028$ &  
$3414.8 \pm 0.9$ & $13.2 \pm 2.1$ & $3555.4 \pm 1.3$ & (2.0, fix)   \\
Included & $8.7 \pm 1.7 \pm 0.9$ & $0.27^{+0.07}_{-0.06} \pm 0.03$ & 
$3414.6 \pm 1.1$ & $13.2 \pm 2.1$ & $3555.4 \pm 1.3$ & (2.0, fix)
 \\
\hline
Belle 2007 & $7.00 \pm 0.65 \pm 0.71$ & $0.31 \pm 0.05 \pm 0.03$ & - & - & - & -\\
PDG 2012 & $7.3 \pm 0.5$ & $0.297 \pm 0.026$ & $3414.75 \pm 0.31$ & $10.4 \pm 0.6$ & $3556.20 \pm 0.09$ & $1.98 \pm 0.11$
\\
\hline
\hline
\end{tabular}
\end{table*}
\end{center} 
\normalsize

\subsection{Possible higher-mass charmonium states}
\label{sub:himasscc}
 We could expect a contribution from the possible higher-mass charmonium
states, $\chi_{cJ}(2P)$ ($J=0$, 2) in the $W$ region above 3.8~GeV. 
The $\chi_{c2}(2P)$ has been
found near 3927~MeV/$c^2$ in the two-photon process~\cite{pdg2012} in its
decay to the $D\bar{D}$ state,
but the $\chi_{c0}(2P)$ has not yet been observed in this decay mode.
Although no theoretical
predictions are available for the branching fractions 
${\cal B}(\chi_{cJ}(2P) \to \ks \ks)$, a yield of a few events is 
expected if $\Gamma(\chi_{cJ}(2P) \to \ks \ks) \approx
\Gamma(\chi_{cJ}(1P) \to \ks \ks)$ and postulated or observed
values for $\Gamma_{\rm tot}$ and $\Gamma_{\gamma\gamma}$ for such states
are taken.

As seen in Fig.~\ref{fig:kskschic}, we find 8 events in the $W$
region between 3.7 and 4.0~GeV, consistent with
5.2 events expected in the region 
from the extrapolated continuum background determined by the fit 
below 3.7~GeV (with a continuum yield density of
$dY/dW = 59.2(W/3.5\GeV)^{-13.5}~[\GeV^{-1}]$, including interference).
In the $W$ region between 3.80 and 3.95~GeV,
where we expect the presence of contributions from the
two higher-mass charmonium states, 7 events are observed, while
only 2.3 events are expected from the continuum.
The probability for this observation ($p$-value) is 0.9\%. 

We evaluate an upper limit for 
$\Gamma_{\gamma \gamma}(\chi_{c2}(2P)){\cal B}(\chi_{c2}(2P) \to \ks \ks)$.
We find 2 events in the $\chi_{c2}(2P)$ mass region, 3.879 -- 3.975~GeV/$c^2$,
which is defined by $M \pm 2 \Gamma$ using the known mass and total
width~\cite{pdg2012}. 
We adopt $N^{UL}_{\chi_{c2}(2P)} = 5.32$ as
the upper limit of the yield with a 90\% confidence level (CL) for the
contribution of the $\chi_{c2}(2P)$, assuming no
background contribution for a conservative limit and based on the
Poisson distribution with this mean value giving a 10\% probability
for two or fewer observed events. 
This translates into an upper limit for the product of the two-photon 
decay width and the branching fraction of the $\chi_{c2}(2P)$ of
$\Gamma_{\gamma \gamma}(\chi_{c2}(2P)){\cal B}(\chi_{c2}(2P) \to \ks \ks) 
< 0.064~{\rm eV\ (90\%\ CL)}$ without interference.
This upper limit takes into account the uncertainty of the efficiency by
increasing the limit by one standard deviation.

The $X(3915)$ found in the $\gamma \gamma \to X(3915) \to \omega J/\psi$
process~\cite{x3915be} has been confirmed
and its spin-parity is assigned to be $J^P=0^+$~\cite{x3915ba}. 
Assigning this resonance to be the $\chi_{c0}(2P)$ state together with
the values of mass and total width given by the most recent tabulation by 
PDG~\cite{pdg2013} ($M=3918.4 \pm 1.9$~MeV/$c^2$ and 
$\Gamma = 20 \pm 5$~MeV), 
we extract the upper limit of 
$\Gamma_{\gamma \gamma}(\chi_{c0}(2P)){\cal B}(\chi_{c0}(2P) \to \ks \ks) 
< 0.49~{\rm eV\ (90\%\ CL)}$;
the same two events that are found for the analysis of the 
$\chi_{c2}(2P)$ in the $M \pm 2 \Gamma$ region 
correspond to $N^{UL}_{\chi_{c0}(2P)} = 5.32$.

\subsection{Search for the decay $\eta_c \to \ks \ks$}
The decay $\eta_c \to \ks \ks$ violates both $P$ and $CP$ invariance.
We search for this decay mode in the present data. 
Copious production of the $\eta_c$ in two-photon collisions 
has been established in several decay modes by previous 
measurements~\cite{pdg2012}.

A small peak-like structure near 2.99~GeV seen in 
Fig.~\ref{fig:kskschic} is not statistically significant and corresponds to a 
fluctuation at the $1.7\sigma$ level, which is evaluated from
the difference between log-likelihoods for the fits
without and with a contribution of
the $\eta_c$, taking into account the interference effect
that is described below.
We thus set the upper limit of the branching fraction for 
this decay mode.

 We fit the event distribution in the range 
2.8~GeV$< W < 3.3$~GeV with a function similar to Eq.~(19) 
in which the $\chi_{c0}$ contribution
is replaced by that of the $\eta_c$ and the $\chi_{c2}$ term is not
included. We fix the mass and width of the $\eta_c$ to be 2981~MeV/$c^2$
and 30~MeV, respectively. The best fit without interference gives
$N_{\eta_c} = 5.4 \pm 5.0$. 
This is consistent with zero. 
We determine the 90\% CL upper limit with the 
$N_{\eta_c}^{\rm UL}$ value that corresponds to the $(1.64)^2$ worse 
log-likelihood $-2 \ln {\cal L}$ than that of the best fit.

\begin{figure}
\centering
\includegraphics[width=8cm]{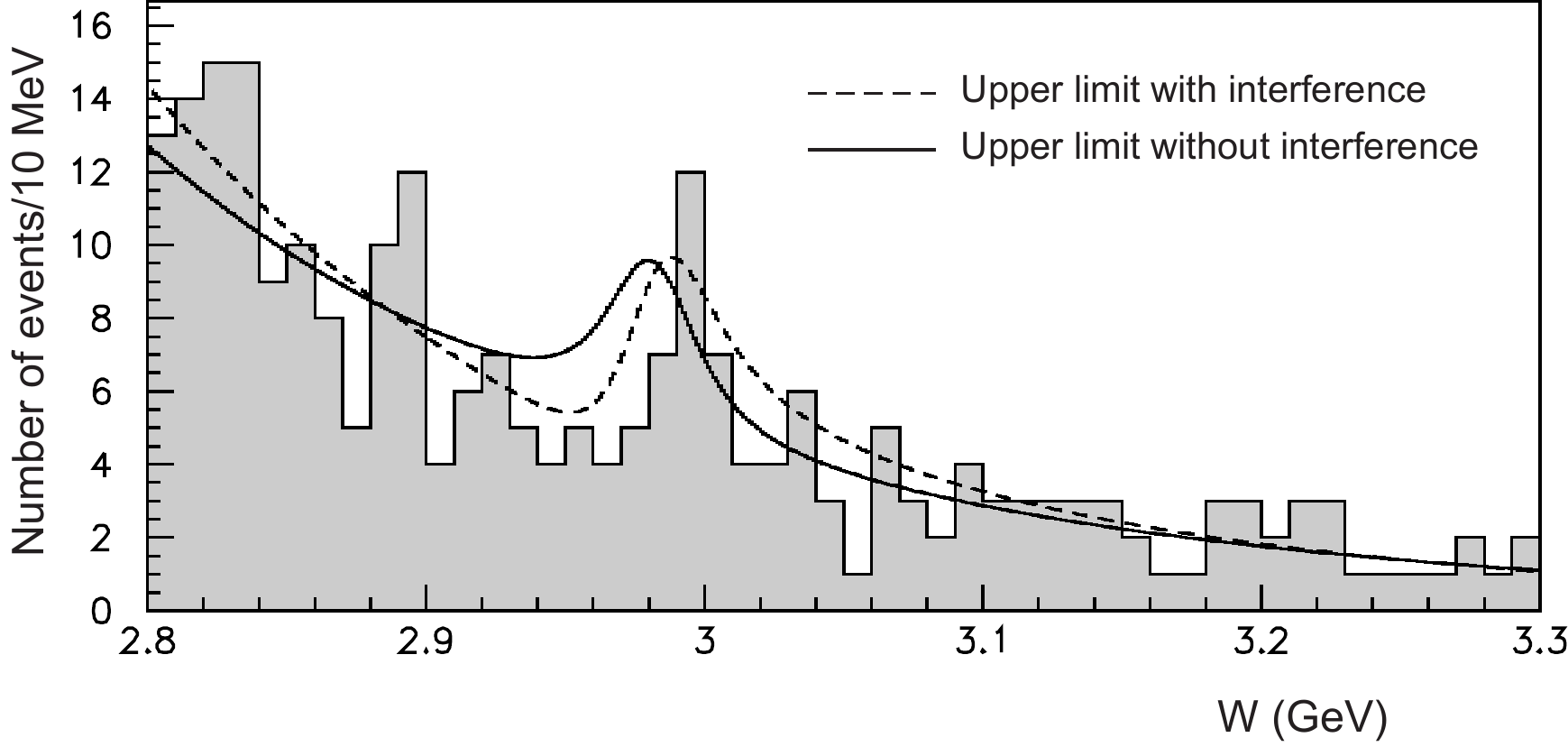}
\centering
\caption{Experimental event distribution in the range 
$2.8~\GeV < W < 3.3~\GeV$ and $|\cos \theta^*| < 0.5$ 
and the results of the fits used to estimate the upper limits for 
$\eta_c \to \ks \ks$ in cases with (dashed) and without (solid) interference.
 }
\label{fig:etac_limit}
\end{figure}

We take into account uncertainties in the mass, width and the mass 
resolution associated with our measurement, and repeat the fit by adjusting 
these parameters by $\pm 2$~MeV/$c^2$,  $\pm 4$~MeV
and in $5--7$~MeV, respectively, and choose the most conservative
upper limit. 
The obtained upper limit is $N_{\eta_c}^{\rm UL} = 15$ 
($N_{\eta_c}^{\rm UL} = 85$) without (with) interference.
The curves describing the results of the fits used to estimate the
upper limits as described are shown in Fig.~\ref{fig:etac_limit}.

The 90\% CL upper limits for 
$\Gamma_{\gamma\gamma}(\eta_c){\cal B}(\eta_c \to \ks \ks)$
and ${\cal B}(\eta_c \to \ks \ks)$ are summarized in Table~\ref{tab:charm3};
for the latter, $\Gamma_{\gamma\gamma}(\eta_c) = 5.3 \pm 0.5$~keV is
used~\cite{pdg2012}. These upper limits take into account the 
uncertainties from systematic error of the measurement and the 
$\Gamma_{\gamma\gamma}(\eta_c)$ value by shifting the limits by 
a ratio corresponding to $1 \sigma$ in the direction of increased values.

\begin{center}
\begin{table}
\caption{Upper limits for products of the two-photon decay width
and the branching fraction for
the $\eta_c \to \ks\ks$ decay, 
where  $\Gamma_{\gamma \gamma}{\cal B}(\eta_c)$
stands for $\Gamma_{\gamma \gamma}(\eta_c){\cal B}(\eta_c \to \ks\ks)$.
}
\label{tab:charm3}
\begin{tabular}{c|ccc}
\hline
\hline
Interference & $\Gamma_{\gamma \gamma}{\cal B}(\eta_c)$  & 
${\cal B}(\eta_c \to \ks\ks)$ &  \\
\hline
Not included & < 0.29 eV & < $5.6 \times 10^{-5}$ & 90\% CL \\  
Included & < 1.6 eV & < $3.2 \times 10^{-4}$ & 90\% CL \\  
\hline
\hline
\end{tabular}
\end{table}
\end{center} 
\normalsize

\section{QCD studies in the high-energy region}
\label{sec:qcd}
In this section, the cross-section behavior is studied and compared
with predictions from QCD-based models and calculations in the region 
$W>2.6~\GeV$.
First, we compare the differential cross section with 
the $1/\sin^4 \theta^*$ dependence.
Then the $W^{-n}$ behavior of the integrated cross section is examined.

\subsection{Angular dependence of the differential cross section}
\label{sub:angdep}
We compare the angular dependence of the differential
cross section with the $1/\sin^4 \theta^*$ dependence,
which is claimed by the handbag model~\cite{handbag}. 
Earlier Belle measurements for this process supported
such a dependence in the $W$ region between 2.4 and 3.3~GeV for
$|\cos \theta^*|<0.6$~\cite{chen}. 

\begin{figure}
\centering
\includegraphics[width=7.7cm]{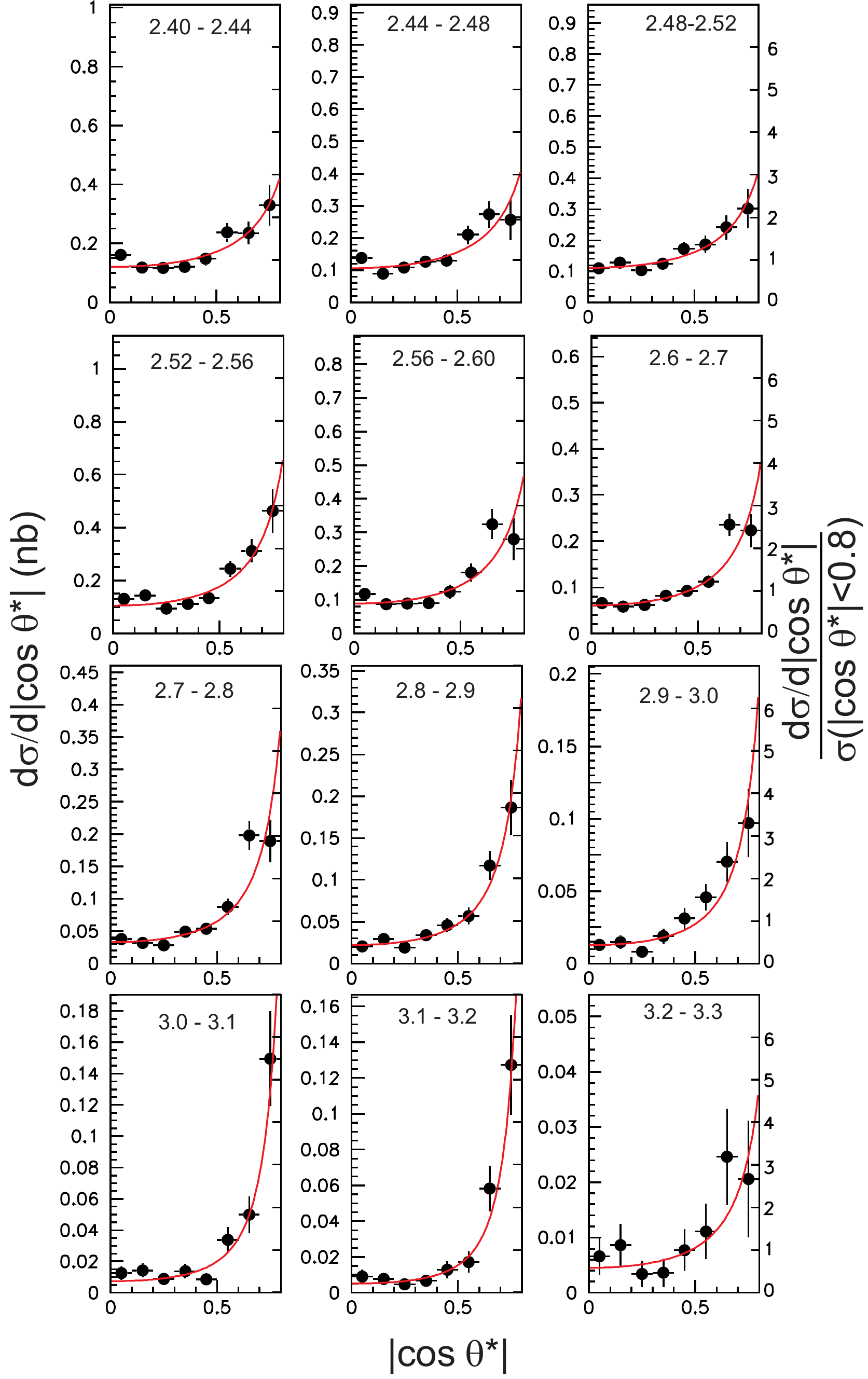}
\centering
\caption{
Data for the $\cos \theta^*$ dependence of the 
differential cross section and the results of the
fits performed with the function proportional to
$1/\sin^\alpha \theta^*$ (solid curve). 
The numbers in each panel show the $W$ region in GeV.
The left (right) vertical scale of each subfigure corresponds to 
the absolute scale (normalized in such a way that the average is 1.25,
as described in the text)
of the differential cross section.
}
\label{fig:angqcd1}
\end{figure}

\begin{figure}
\centering
\includegraphics[width=7cm]{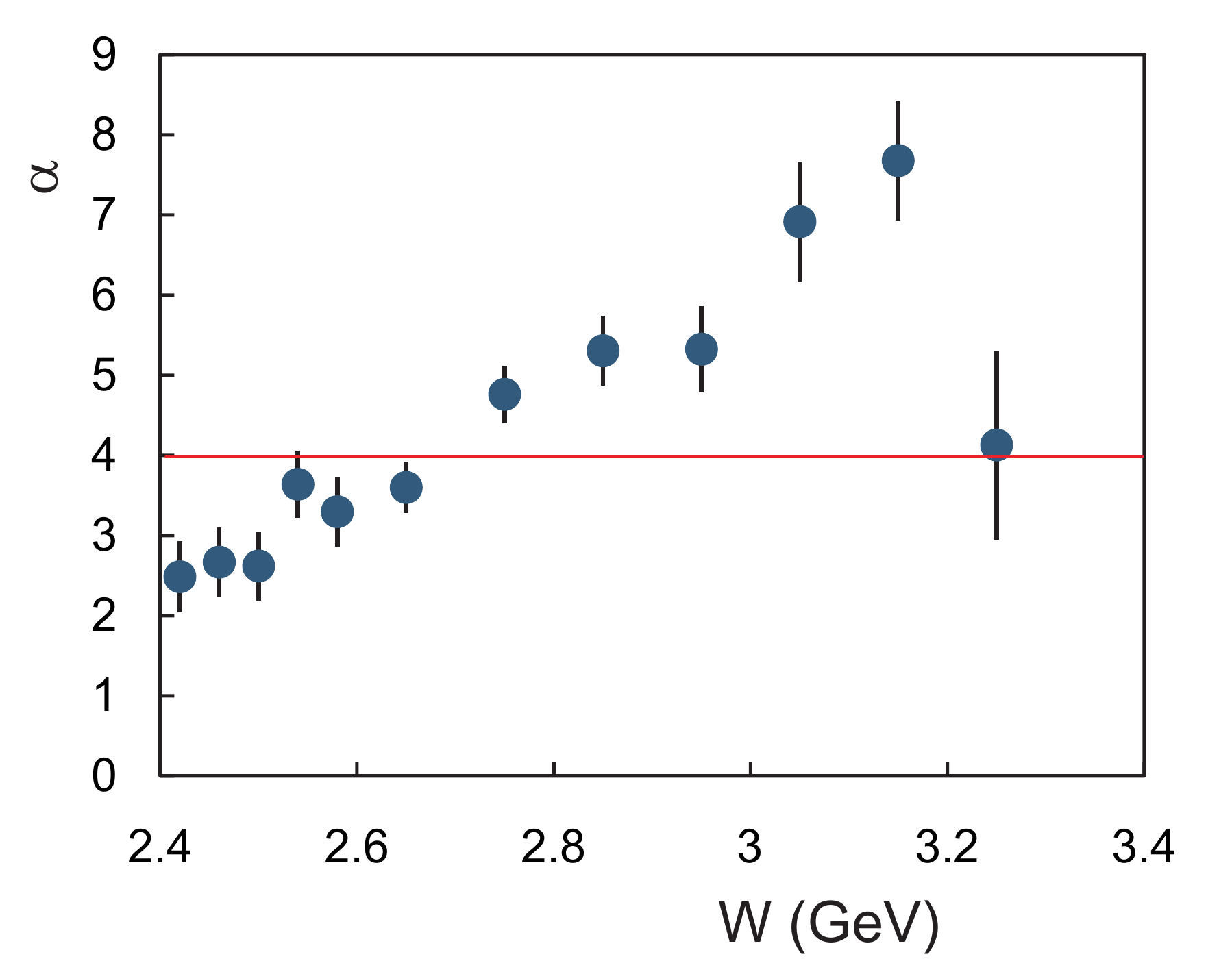}
\centering
\caption{$W$ dependence of the parameter $\alpha$, which
characterizes the angular dependence of the differential
cross section. 
The horizontal line at $\alpha=4$ corresponds to 
the claim of the handbag model (see the text).
}
\label{fig:angqcd2}
\end{figure}

To make a quantitative statement about the behavior of
the cross section, we fit the differential cross section 
using the approximation $A / \sin^\alpha \theta^*$, \textit{i.e.},
\begin{equation}
\frac{d\sigma}{d|\cos \theta^*|} = \frac{A}{\sin^\alpha \theta^*}
\label{eqn:angdist}
\end{equation}
in each $W$ bin.
We summarize the fitted results for the 12 regions in 
Fig.~\ref{fig:angqcd1}, where the right scales are 
differential cross sections normalized to the integrated
cross section in the range $|\cos \theta^*|<0.8$ (that gives 
the average 1/0.8 = 1.25). 
This scale is added to improve the visibility of the plots for 
different $W$ bins.
The $\chi^2/ndf$ values obtained from the fits are
between 3/6 and 19/6.
The obtained $W$ dependence of the parameter $\alpha$ is shown in
Fig.~\ref{fig:angqcd2}. 
The parameter $\alpha$ is found to be above 4 
for the $W$ range between 2.7 and 3.3~GeV,
but no tendency toward 4 
is observed in the high-energy part of the $W$ region. 
We note that we find a resonance-like 
contribution considered to be a scalar at around 2.5~GeV,
as described in Sec.~\ref{sub:above},
which could affect the $W$ dependence of $\alpha$ 
in the region around 2.4 -- 2.7~GeV.

Information on the meson ($M$) distribution amplitude (DA) $\phi_M$ 
can be obtained
by comparing the observed angular dependence to that of the theoretical
calculation~\cite{bl}; the angular dependence of the data
is steeper and more forward-peaked,
which indicates that the DA is flatter than assumed.

The function proportional to 
$1/\sin^4 \theta^* + b \cos^2 \theta^*$
that has been applied in our analysis of the $\gamma \gamma
\to \pi^0\pi^0$ process yields fits of poor quality in this
study, as the rise of the $\cos^2 \theta^*$ term for the forward 
angles is insufficient to describe the trend observed in data. 

\subsection{$W$ dependence}
\label{sub:wdep}
\begin{figure}
\centering
\includegraphics[width=8cm]{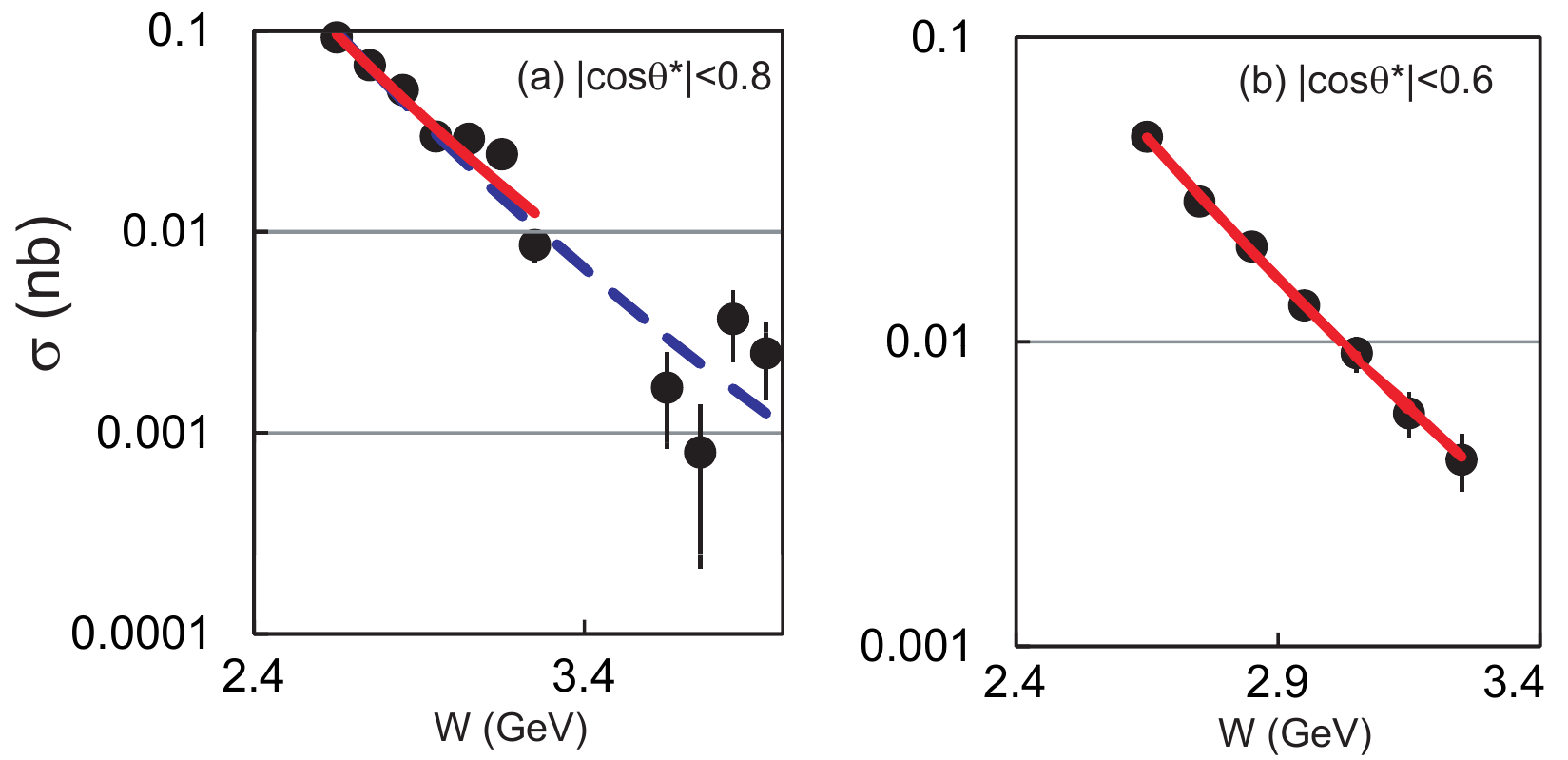}
\centering
\caption{ Results for the cross section integrated
over $|\cos \theta^*|$ regions (a) below 0.8 and (b) below 0.6. 
The $W$ dependence is
fitted to $W^{-n}$ in the different $W$ regions:
2.6 -- 4.0~GeV excluding 3.3 -- 3.6~GeV
(dashed line) and 2.6 -- 3.3~GeV (solid line).
}
\label{fig:wpower}
\end{figure}

The $W$ dependence of the cross section integrated over the angle
provides important information about the mechanism of the exclusive
meson-pair production. 
We fit the cross section with
\begin{equation}
\sigma (|\cos \theta^*|<0.8) = a W^{-n} \; ,
\label{eqn:wdist}
\end{equation}
for the $W$ region 2.6 -- 4.0~GeV, excluding 3.3 -- 3.6~GeV.
We exclude the region below 2.6~GeV because a resonance-like
contribution is found there. 
We obtain $n= 11.0 \pm 0.4$. 
This result is shown by the dashed line in Fig.~\ref{fig:wpower}(a). 
The error is statistical only.

We also try the fits for the narrower $W$ region, 
2.6 -- 3.3~GeV, for $\sigma(|\cos \theta^*|<0.8)$ and 
 $\sigma(|\cos \theta^*|<0.6)$, and obtain
$n= 10.0 \pm 0.5$ and $n= 11.8\pm 0.6$, respectively.

In our previous work, we obtained $n = 10.5 \pm 0.6$
for $W = 2.4 - 4.0$~GeV excluding 3.3 -- 3.6~GeV
and $|\cos \theta^*|<0.6$~\cite{chen}. 
The present analysis in this region yields $n=10.8 \pm 0.2$.
We quote this number only for verification,
as we now know that it includes a resonance-like 
contribution around 2.5~GeV.
These results are summarized in Table~\ref{tab:wdep}.

We estimate the systematic uncertainty for the $n$ measurement 
as follows:
since the overall normalization error does not affect the 
determination of $n$, we consider the $W$-dependent distortion 
effect only.
As in the resonance studies, we assume  $\pm 4\%$ distortions
at the two ends of the fit range and continuous variations
between them. 
The distortion for a $W$ range changes the $n$ value with
$\Delta n \approx \log 1.08/\log(W_2/W_1)$, where $W_1$ and $W_2$
delimit the fit region (chosen to be 2.65~GeV and 3.25~GeV, 
respectively). 
We obtain the estimated systematic uncertainty $\Delta n = 0.4$.

The slope parameter $n$ that ranges between 10 and
11 for the present process is larger 
than 6 and 7--8 that are 
predicted~\cite{bl} and observed~\cite{nkzw}, respectively,
for the $\pi^+\pi^-$ and $K^+K^-$ processes.
For the process $\gamma \gamma \to \ks \ks$, as discussed 
in Refs.~\cite{chernyak, chernyak2},
the coefficient of the leading-term amplitude is much smaller than
that of the non-leading term.
Therefore, at this energy the $W$ dependence of the cross section is
mainly determined by that of the non-leading terms.
In the $W$ region measured in this experiment and 
including a non-leading term, the perturbative QCD
predicts $n=10$~\cite{chernyak2}, 
which is in reasonable agreement
with our measurement.

\begin{center}
\begin{table*}
\caption{Results for the slope parameter $n$ from the
power fit $\sigma \sim W^{-n}$ for $\gamma \gamma \to \ks \ks$
in different fit ranges.
The result from the previous work~\cite{chen} 
is also shown. The first and second errors are statistical and systematic,
respectively.}
\label{tab:wdep}
\small
\begin{tabular}{cccc}
\hline
\hline
$W$ range (GeV) & $|\cos \theta^*|$ range & $n$ & Note \\
\hline
$2.6 - 4.0$ (excluding $3.3 - 3.6$) & $<0.8$ & $11.0 \pm 0.4 \pm 0.4$ & \\
$2.6 - 3.3$ & $<0.8$ & $10.0 \pm 0.5 \pm 0.4$ & \\
$2.6 - 3.3$ & $<0.6$ & $11.8 \pm 0.6 \pm 0.4$ & \\
\hline
$2.4 - 4.0$ (excluding $3.3 - 3.6$) & $<0.6$ & $10.5 \pm 0.6 \pm 0.5$ &
Belle 2007\\
\hline
\hline
\end{tabular}
\end{table*}
\end{center} 

\section{Summary and conclusion}
\label{sec:summary}
We have measured the cross section for the process
$\gamma \gamma \rightarrow \ks \ks$ for $1.05~\GeV \le W \le 4.00~\GeV$
with the Belle detector at the asymmetric-energy KEKB collider.
The data sample of 972~fb$^{-1}$ is 
three orders of magnitude larger
than in the previous measurements~\cite{tasso1, pluto, cello, tasso2, L3}.
The differential cross section is measured up to 
$|\cos \theta^*| = 0.8$,
which allows high-sensitivity studies of the amplitudes.

In our study, the differential cross section has been fitted
to obtain information on partial waves.
The obtained spectra of $\hat{S}^2$, $\hat{D}_0^2$ and $\hat{D}_2^2$
indicate the presence of the $f_0(1710)$, $f_J(2200)$ and 
$f_{J'}(2500)$ in addition to the well known $f_2(1270)$, 
$a_2(1320)$ and $f_2'(1525)$.
Then fits to the differential cross section are performed by 
assuming possible resonances in the partial waves. 

First, we perform a fit in the region $1.15~\GeV \le W \le 1.65~\GeV$ to
determine the parameters of the $f_2'(1525)$ as well as the
relative phase between the $f_2(1270)$ and $a_2(1320)$.
Two solutions are obtained and combined statistically.
The phase difference between the $a_2(1320)$ and $f_2(1270)$
is measured to be $\left( 172.6^{+6.0+12.2}_{-0.7-7.0}\right)^{\circ}$,
confirming the destructive interference between the two mesons
and agreeing with theoretical predictions~\cite{lipkin}.
The mass, total width and $\Gamma_{\gamma \gamma}\B(K \bar{K})$ 
of the $f_2'(1525)$ are measured to be
$1525.3^{+1.2+3.7}_{-1.4-2.1}$~MeV/$c^2$,
$82.9^{+2.1+3.3}_{-2.2-2.0}$~MeV
and $48^{+67+108}_{-8-12}$~eV, respectively.
Note that no interference effect was taken into account in the previous 
measurements~\cite{tasso1, pluto, cello, L3}.

Evidence for the existence of 
the $f_0(1710)$, $f_2(2200)$ and $f_0(2500)$
in this channel is obtained.
Masses (widths) of these resonances are measured to be
$1750^{+6+29}_{-7-18}$, 
$2243^{+7+3}_{-6-29}$,
$2539 \pm 14 {}^{+38}_{-14}~\MeV/c^2$ 
($139^{+11+96}_{-12-50}$, 
$145 \pm 12 {}^{+27}_{-34}$,
$274^{+77+126}_{-61-163}~\MeV$), respectively.
Their $\Gamma_{\gamma \gamma}\B(K \bar{K})$ 
values are measured for the first time to be
$12^{+3+227}_{-2-8}$, 
$3.2^{+0.5+1.3}_{-0.4-2.2}$, 
$40^{+9+17}_{-7-40}~\eV$, respectively.

We conclude that the $f_0(1710)$ and $f_2(2200)$ are unlikely to be
glueballs because their total widths and $\Gamma_{\gamma \gamma}\B(K \bar{K})$ 
values are much larger than those expected for a pure glueball state.

Analyses in the region $W>2.6~\GeV$ are updated;
parameters of the $\chi_{c0}$ and $\chi_{c2}$ and 
the exponents $\alpha$ and $n$ in $(\sin \theta^*)^{-\alpha}$
and $W^{-n}$ describing the angular and $W$
behavior of the cross section are extracted from data.
The value of $\alpha$ does not show the tendency toward 4
observed in our previous work where the available angular
region is limited to $|\cos \theta^*|<0.6$~\cite{chen}.
The fitted value of $n=11.0 \pm 0.4 \pm 0.4$ is much larger than the
QCD asymptotic prediction of 6 or 7~\cite{bl} but agrees fairly
well with $n=10$ 
predicted by a qualitative QCD estimate~\cite{chernyak2}.
For the process $\gamma \gamma \to \ks \ks$,
according to Refs.~\cite{chernyak, chernyak2},
the $W$ dependence of the cross section is
determined by that of the non-leading term
in the $W$ region measured by this experiment;
the coefficient of the leading term amplitude is much smaller than
that of the non-leading term.
The results are consistent with the previous analyses~\cite{chen}
with improved statistics and supersede the measurements for the 
cross section, the $\chi_{cJ}(1P)$ parameters and the slope 
parameter $n$.

We provide upper limits for the decay of the $\chi_{cJ}(2P)$,
$\Gamma_{\gamma \gamma}(\chi_{c2}(2P)){\cal B}(\chi_{c2}(2P) \to \ks \ks) 
< 0.064$~eV and $\Gamma_{\gamma \gamma}(\chi_{c0}(2P)){\cal B}(\chi_{c0}(2P) \to \ks \ks) < 0.49$~eV 
at 90\%\ CL, where the $\chi_{c0}(2P)$ coincides with the former
$X(3915)$~\cite{pdg2012, pdg2013}.
A new upper limit for the branching fraction of
the $P$- and $CP$-violating decay $\eta_c \to \ks \ks$ is obtained to be
$3.2 \times 10^{-4}$ ($5.6 \times 10^{-5}$) at 90\% CL 
with (without) the interference effect.

\acknowledgments
%

We are grateful to V. Chernyak for fruitful discussions.
We thank the KEKB group for the excellent operation of the
accelerator; the KEK cryogenics group for the efficient
operation of the solenoid; and the KEK computer group,
the National Institute of Informatics, and the 
PNNL/EMSL computing group for valuable computing
and SINET4 network support.  We acknowledge support from
the Ministry of Education, Culture, Sports, Science, and
Technology (MEXT) of Japan, the Japan Society for the 
Promotion of Science (JSPS), and the Tau-Lepton Physics 
Research Center of Nagoya University; 
the Australian Research Council and the Australian 
Department of Industry, Innovation, Science and Research;
Austrian Science Fund under Grant No. P 22742-N16;
the National Natural Science Foundation of China under
contract No.~10575109, 10775142, 10875115 and 10825524; 
the Ministry of Education, Youth and Sports of the Czech 
Republic under contract No.~MSM0021620859;
the Carl Zeiss Foundation, the Deutsche Forschungsgemeinschaft
and the VolkswagenStiftung;
the Department of Science and Technology of India; 
the Istituto Nazionale di Fisica Nucleare of Italy; 
the BK21 and WCU program of the Ministry Education Science and
Technology, National Research Foundation of Korea Grant No.\ 
2010-0021174, 2011-0029457, 2012-0008143, 2012R1A1A2008330,
BRL program under NRF Grant No. KRF-2011-0020333,
and GSDC of the Korea Institute of Science and Technology Information;
the Polish Ministry of Science and Higher Education and 
the National Science Center;
the Ministry of Education and Science of the Russian
Federation and the Russian Federal Agency for Atomic Energy;
the Slovenian Research Agency;
the Basque Foundation for Science (IKERBASQUE) and the UPV/EHU under 
program UFI 11/55;
the Swiss National Science Foundation; the National Science Council
and the Ministry of Education of Taiwan; and the U.S.\
Department of Energy and the National Science Foundation.
This work is supported by a Grant-in-Aid from MEXT for 
Science Research in a Priority Area (``New Development of 
Flavor Physics''), and from JSPS for Creative Scientific 
Research (``Evolution of Tau-lepton Physics'').



\begin{thebibliography}{99}
\bibitem{pw} We denote individual partial
waves by roman letters and parameterized waves by italic.
\end{thebibliography}
\end{document}